\documentclass{aastex631}
\usepackage{amsmath,amssymb}
\usepackage{graphicx,subfigure}
\usepackage{comment}

\usepackage[T1]{fontenc}
\usepackage[utf8]{inputenc}
\usepackage{lmodern}

\begin{document}

\title{Turbulent infall onto class 0 disks as cause of CAI brief condensation episode in the solar system}

\author{Jiachen Zheng}
\affiliation{School of Physics and Astronomy, Beijing Normal University, Beijing, 100875, P.R.China}

\author[0000-0002-3641-6732]{Xing Wei}
\affiliation{School of Physics and Astronomy, Beijing Normal University, Beijing, 100875, P.R.China}
\correspondingauthor{Xing Wei}
\email{xingwei@bnu.edu.cn}

\author{Hongping Deng}
\affiliation{Shanghai Astronomical Observatory, Chinese Academy of Sciences, Shanghai 200030, P.R.China}

\author[0000-0002-9408-2857]{Wenrui Xu}
\affiliation{Center for Computational Astrophysics, Flatiron Institute, New York 10010, USA}

\author{Douglas N. C. Lin}
\affiliation{Department of Astronomy and Astrophysics, University of California, Santa Cruz, CA, USA}
\affiliation{Institute for Advanced Studies, Tsinghua University, Beijing, P.R.China}

\begin{abstract}
Calcium-aluminum-rich inclusions (CAIs) in carbonaceous chondritic meteorites are the oldest relics in the solar system. Notably, their radiogenic age feature a brief (100 kyr) condensation episode. In contrast, the reservoirs of the short-lived isotopes in CAIs, presumably supernovae or asymptotic giant stars, pollutes star-forming regions in giant molecular cloud complexes (GMC) over much longer (Myr) duration.  Through a series of numerical simulations, we show here 
the possibility that, within an extended region (2$\sim$3 AU), nearly all ``pre-solar'' CAI-loaded grains in the infall clouds were sublimated and re-condensed during the early ($ \lesssim 10^5$ yr) infall and formation of class-0 disks. We adopt a set of initial conditions from a previous hydrodynamic simulation of the collapse of GMC and the formation of young stellar clusters.   We analyze the evolution of the disk's thermal distribution and dynamical structure resulting from the interaction between circumstellar disks and infalling gas. Our follow-up simulations, with much higher resolution, show significant and rapid changes in the disk orientation and morphology due to the dynamic infall of external streamers. Warps and global spiral density waves commonly appear.  They lead to 
intense dissipation which heats the gas to sufficiently high temperature to sublimate prior-generation CAIs.  This solid-to-gas phase transition is followed by subsequent cooling and re-condensation.  The CAI contained in the meteorites today could be the relics of the last episode of major infall onto class 0 disks.
\end{abstract}

\keywords{accretion --- solar system --- circumstellar disk}

\section{Introduction} \label{sec:intro}

With the continuous development of multi-wavelength astronomical observation techniques and the improvement of related theoretical models and numerical simulations, the evolutionary sequence of low-mass stars (such as the Sun) and their systems during formation has been largely clarified. Based on existing theoretical models and observational statistical data, we can roughly infer how stellar systems similar to the Solar System form, from the gravitational collapse of molecular clouds to the accretion of planetesimal-sized bodies. (For example, \citealp{Cameron1995}; 
\citealp{Shu1987}; \citealp{Andre2000}; \citealp{Alexander2001}.) However, many details of this
process remain unclear, and we still require quantitative information on the early 
chronology of Solar-System formation. Chondrite meteorites serve as the 
crucial source of data we need. These primitive meteorites are formed from 
preplanetary grains and dust that were condensed, collided, coagulated, and compressed 
into rocks during the early epochs of the solar system's formation. Despite these
violent processes, these entities retain their aggregate nature and the unique 
characteristics of the diverse particles that make them up. The ``chondrules'' in 
these chondritic meteorites typically refer to millimeter-sized solidified molten 
droplets, mainly composed of magnesium-iron silicate minerals and glass.  They 
may also contain metals or sulfides (i.e., samples with nearly solar elemental 
compositions, undifferentiated). Thus, chondrites can be considered as a sample 
of data from the Class 0/1 stages of stellar formation. These chondrites contain 
many short-lived radionuclides (SLRs), which have half-lives on the order of tens 
of millions of years or less, with time scales comparable to those of protostar 
and protostellar disk formation. These isotopes have long since decayed from 
their original abundances, but their relative abundances can nonetheless be used 
as high-precision timers for the Solar System. Moreover, the origin of these 
short-lived radionuclides plays a critical role in understanding the formation 
mechanisms of the Solar System. The earliest discovery was made by \cite{Jeffery1961}, 
who found an excess of the daughter product \(^{129}\mathrm{Xe}\) that was correlated 
with the relative abundance of iodine, revealing the presence of its parent nuclide \(^{129}\mathrm{I}\). Later, in the 1970s, the discovery of a significant excess of \(^{26}\mathrm{Mg}\) in the calcium-aluminum-rich inclusions (CAIs) of the Allende 
meteorite (\citealp{gray1974excess}; \citealp{lee1974}) showed that the degree of 
\(^{26}\mathrm{Mg}\) excess was correlated with the Al/Mg ratio in the CAI mineral 
separates \citep{lee1976demonstration}, suggesting the it in situ decay of 
\(^{26}\mathrm{Al}\).

CAIs exhibit significant diversity in composition, mineralogy, structure, and size. They are the oldest known objects, having formed in the infant solar nebula, and their isotopic composition preserves nucleosynthetic signatures from the pre-solar nebula. Primarily composed of oxides and silicates of calcium, aluminum, magnesium, and titanium, CAIs are notably deficient in volatile elements such as iron, alkali metals, and water. These minerals include many phases that are also found in synthetic high-temperature ceramics and phase equilibrium diagrams relevant to ceramic systems. CAIs are consistently enriched in certain lithophile trace elements such as scandium, yttrium, zirconium, hafnium, the rare earth elements, and certain siderophile trace elements like those in the platinum group. Enrichment factors for these elements typically range from 10× to 100× relative to average solar system abundances (CI chondrites) (\citealp{grossman1980refractory}; \citealp{ireland1988trace}). These elements, along with the major components mentioned above, are all refractory, meaning they and their compounds have exceptionally high vaporization temperatures. As a result, CAIs are commonly referred to as ``refractory inclusions." 
The mineralogy of CAIs is strikingly similar to the phases predicted to condense from a hot solar vapor during cooling \citep{yoneda1995condensation}, indicating formation under extremely high-temperature conditions in the early solar system. Chronological constraints indicate that the primary formation of CAIs occurred over a relatively short interval, on the order of $\sim10^{4}$–$10^{5}$ yr (\citealp{macpherson2005calcium}; \citealp{chaussidon2015timing}). Many igneous CAIs record partial or complete melting and, in some cases, evidence for subsequent reheating and chemical modification. These later processing events span a longer evolutionary timescale, extending up to $\sim1$–2 Myr after initial CAI formation, reflecting the age offsets between CAIs and chondrules and the development of secondary mineral assemblages rather than the duration of CAI formation itself \citep{macpherson2005calcium}. We note that not all CAIs experienced such extended reprocessing: so-called primitive CAIs preserve their initial formation signatures and define time zero, whereas other CAIs record varying degrees of secondary thermal and chemical modification over longer timescales \citep{macpherson2012well}.
While CAIs formed in high-temperature regions of the inner protoplanetary disk, their present-day distribution among chondrite groups requires efficient early transport and selective retention \citep{jacquet2024early}. In particular, CAIs are strongly enriched in carbonaceous chondrites (the CC reservoir) but are comparatively scarce in non-carbonaceous chondrites (the NC reservoir), implying large-scale radial transport followed by limited communication between inner and outer disk regions (\citealp{jacquet2024early}; \citealp{marschall2023inflationary}). Recent models suggest that early viscous expansion of a compact disk can transport CAIs outward to several astronomical units, whereas subsequent pressure maxima or dust traps—possibly associated with early giant-planet growth—may inhibit inward drift and exchange between the NC and CC reservoirs, thereby preserving the observed compositional dichotomy (\citealp{jacquet2024early}; \citealp{marschall2023inflationary}).
Therefore, CAIs can be regarded as sensitive probes of the earliest stages of Solar-System evolution, recording both a brief high-temperature formation episode in the inner disk and the dynamical processes that governed material transport and reservoir separation in the young protoplanetary disk.

In the early stages of CAI research, Shu proposed the X-wind model, suggesting that CAIs could be formed when solid bodies were lifted by the aerodynamic drag of a magnetocentrifugally driven wind from relatively cool, shaded regions of a disk close to the star into the heat of direct sunlight \citep{shu1996toward}. This innovative hypothesis provided the first  
comprehensive attempt to tie together a global and coherent picture of star and planet formation. However, with more precise measurements and calculations of the \(^{26}\mathrm{Al}/^{27}\mathrm{Al}\) ratio in CAIs from carbonaceous chondrite meteorites (CCMs), it was determined that the age uncertainty of normal CAIs is approximately 0.1 million years(\citealp{jacobsen200826al}; \citealp{connelly2012absolute}). These results limit the plausibility of the X-wind model as a source of early solar \(^{26}\mathrm{Al}\) through solar irradiation, given that the timescales of solar flare events are much longer than the narrow age distribution of CAI formation.

Recent studies \citep{forbes2021solar}, particularly in regions such as Ophiuchus and the Upper Scorpius association, highlight the pervasive presence of ${}^{26}$Al in star-forming environments and its likely inheritance by pre-solar molecular cloud material. These results suggest that multiple supernova events over tens of millions of years contributed to the enrichment of ${}^{26}$Al in the precursor materials of the Solar System. This long-term enrichment process establishes the initial radiogenic inventory of CAI precursors but does not, by itself, determine the timing or mechanism of CAI formation.
The formation of CAIs instead requires a subsequent, localized high-temperature event that produced the final melting, cooling, and solidification of these refractory inclusions, thereby resetting the Al--Mg chronometer(\citealp{grossman2002zoned}; \citealp{grossman2002formation}; \citealp{lodders2003solar}). Several physical mechanisms have been proposed for such a short-lived heating episode within the protoplanetary disk. These include heating during episodes of rapid accretion \citep{bell1994ApJ, bell1995ApJ, hartmann1996fu}, although the temperatures inferred for typical FU Orionis outbursts may not be sufficient to affect regions extending to several astronomical units. Heating associated with a nearby supernova explosion at distances of order $\sim 0.1$~pc has also been proposed (\citealp{zwart2018consequences}; \citealp{zwart2019formation}), but such events require fine-tuned timing and geometry and are expected to be rare \citep{gritschneder2012}. Another possibility is transient shock heating arising from turbulent accretion and disk reorientation \citep{bate2010chaotic}, which could, in principle, produce a significant but short-lived temperature increase over extended regions of the disk. These scenarios remain under investigation as potential explanations for the brief, high-temperature event responsible for the final formation and isotopic resetting of CAIs.

To evaluate the plausibility of disk-driven heating reaching CAI sublimation conditions, we provide a simple estimate of the midplane temperature in the inner disk based on viscous heating and radiative cooling balance.The vertically integrated viscous heating rate is given by
\begin{equation}
Q^{+} = \nu \Sigma \left( r \frac{d\Omega}{dr} \right)^2 = \frac{9}{4}\,\nu\,\Sigma\,\Omega_K^2 ,
\end{equation}
where $\nu$ is the kinematic viscosity, $\Sigma$ is the surface density, and
$\Omega_K=(G M_\ast/r^3)^{1/2}$ is the Keplerian angular velocity. While this formulation formally assumes a viscous description of angular momentum transport, we note that in sufficiently massive, self-gravitating disks, the transport is expected to be dominated by global gravitational instabilities rather than effective local viscosity \citep{lodato2005testing}.  Nevertheless, we provide this order of magnitude estimation, which may partly inform us of the total energy dissipated. 
In the optically thick limit, radiative cooling is described by diffusion. The radiative flux from each side of the disk surface is $\sigma T_{\rm eff}^4$, so the total cooling rate is
\begin{equation}
Q^{-} = 2\sigma T_{\rm eff}^4 ,
\end{equation}
where $T_{\rm eff}$ is the effective temperature and $\sigma$ is the
Stefan--Boltzmann constant. The midplane temperature $T_m$ is related to
$T_{\rm eff}$ through the gray atmosphere approximation,
\begin{equation}
T_m^4 \simeq \frac{3}{4}\,\tau\,T_{\rm eff}^4 ,
\end{equation}
where $\tau$ is the one-sided Rosseland mean optical depth measured from the
disk midplane to a single disk surface.
Assuming steady-state accretion, the thermal balance condition is given by
\begin{equation}
    \nu \Sigma \left( r \frac{d\Omega}{dr} \right)^2 = \frac{9}{4}\,\nu\,\Sigma\,\Omega_K^2 = \frac{8}{3}\,\frac{\sigma T_m^4}{\tau},
\end{equation}
where $\nu$ is the viscosity, $\Sigma$ is the surface density, $\Omega$ is the angular velocity, $\sigma$ is the Stefan--Boltzmann constant, $T_m$ is the midplane temperature, and $\tau$ is the optical depth. Assuming a steady-state disk with torque balance $\nu \Sigma \approx \dot{M}_\ast/3\pi$, this yields
\begin{equation}
    T_m = \left( \frac{9 G M_\ast \dot{M}_\ast \tau}{32 \pi \sigma r^3} \right)^{1/4}.
\end{equation}
Adopting representative values $M_\ast = 1\,M_\odot$, $\dot{M}_\ast = 10^{-5}\,M_\odot/\mathrm{yr}$, $r = 2\,\mathrm{AU}$, and $\tau = 700$ (which is the typical Rosseland mean optical depth measured in the inner disk midplane in our simulations; see the simulations for more details), we obtain $T_m \approx 1350\,\mathrm{K}$. This temperature is close to the sublimation threshold of CAI grains ($\sim 2000$\,K). Here we would like to clarify that this estimate is intended only to provide a rough indication of the temperature scale. Our numerical results are obtained from self-consistent hydrodynamical simulations and do not rely on the validity of the viscous \(\alpha\)-prescription. In addition, recent observations of the Class I protostar HOPS-315 in the Orion B molecular cloud indicate the presence of a so-called “thermostat” region at $\sim$2.2 AU from the central star. This region corresponds to a high-pressure disk midplane zone in which temperature-dependent silicate vaporization and recondensation provide a negative thermal feedback, regulating the midplane temperature near the silicate sublimation threshold of $\approx1300$K \citep{mcclure2025refractory}.

The rest part of this paper is organized as follows. We present the numerical methods in our hydrodynamical simulations in Section \ref{sec:method}. The results are shown in Section \ref{sec:results}. The conclusions and discussions are presented in Section \ref{sec:conclusions}.

\section{Methods} \label{sec:method}

The calculations were performed with the GIZMO code in the Meshless Finite Mass (MFM) mode \citep{Hopkins2015}, which is a Godunov-type Lagrangian scheme. By its Lagrangian nature, it is ideal for simulating the gravitational collapse of the nebular gas and naturally enhances the numerical resolution in the formed circumstellar disk \citep{fu2025formation}. Compared to the 
traditional smoothed particle hydrodynamics (SPH) methods, MFM updates fluid variables by solving for flux across effective cell interfaces with a Riemann solver and does not require the use of artificial viscosity, resulting in better shock capturing \citep{Hopkins2015} and significantly reduced numerical dissipation \citep{deng2017convergence,deng2019local}. It also exhibits excellent angular momentum conservation, which is essential for accurate modeling of warp disks induced by misaligned infall. For example, MFM method was able to capture the parametric instability in warp disks which is challenge even for state of the art grid codes \citep{deng2021parametric,deng2022non}. In our study, angular momentum transport and shock heating are key processes. As mentioned above, our code offers distinct advantages in these aspects, providing a more reliable numerical foundation for our research. 

\subsection{Initial conditions} \label{subsec:IC}

To evaluate the effects of chaotic infall, we utilised the results from Matthew Bate's population synthesis study of circumstellar disks (\citealp{Bate2012}; \citealp{Bate2019}). This study involves hydrodynamic simulations of star cluster formation, resulting in a large sample of stellar disk systems, including various young circumstellar disks.  In Bate’s simulations,  sink particles were introduced to represent the central stars, thus avoiding extremely small time steps in the stellar core regions, which would otherwise make direct computation of such small-scale structures impractical \citep{bate1995modelling}. 
More recent cluster formation simulations, such as those from the STARFORGE\footnote{\url{https://starforge.space/whatis.html}} project \citep{grudic2021starforge}, offer increasingly detailed characterizations of star-forming environments and disk assembly. These datasets are not yet publicly available, and are therefore not included here.
From the formed star cluster, we selected solar-mass stars and traced their evolution back to the onset of sink particle formation, which corresponds to the Class 0 stage of protostellar evolution. The Class 0 phase is characterized by ongoing gravitational collapse, high mass infall rates from the surrounding envelope, and the early assembly of a rotationally supported disk, with a substantial fraction of the system mass still residing in the infalling envelope. The physical conditions at this stage naturally involve strong, time-variable accretion flows and non-axisymmetric infall, making it particularly suitable for investigating the effects of chaotic infall on the early thermal and dynamical evolution of circumstellar disks. 
We stress that the use of cluster formation simulations here is primarily motivated by the need to adopt more realistic initial and boundary conditions, rather than to imply that such an environment is required for the infall scenario explored in this work. In particular, cluster simulations self-consistently capture turbulent fragmentation, filamentary accretion, and dynamical interactions, which give rise to time-dependent and anisotropic infall histories that are difficult to prescribe in idealized isolated setups. Similar variability in accretion may also arise in other star formation environments (e.g., filament-fed collapse), and thus our conclusions do not depend on the system forming in a cluster.
We therefore adopt the disk and envelope properties at this stage as the initial conditions for zoom-in simulations. We emphasize that SPH calculations are used only to construct the initial conditions, while the subsequent evolution presented in this work is entirely performed using the MFM method. We selected the accretion particles and the surrounding gas particles within a $2000\,\text{AU}$ radius, and used the particle splitting scheme in the GIZMO code to enhance their resolution for the simulation's initial conditions. The particles, along with their mass, position coordinates, velocity, and specific internal energy $u_i$ (which tracks the thermal energy of each Lagrangian gas particle), were refined through multiple iterations of particle splitting to reach the required resolution of 10 million particles (for resolution testing, see Appendix A). We performed a short relaxation of the high-resolution system to remove the discretisation noise introduced by the particle splitting process. In our simulations, we adopt the background temperature of 10 K as specified in the Bate simulation data. This temperature has a minor impact on our simulations, as the thermal evolution in our calculations does not reach this minimum temperature. In addition, Our analysis focuses on the inner disk within about 5 AU. At the high accretion rates adopted here, viscous heating dominates over protostellar irradiation in the optically thick disk midplane at these radii, as shown in previous irradiated disk models. Stellar irradiation primarily affects disk surface layers and larger radii, which are not relevant for the present study.(e.g.\citealp{chambers2009analytic}; \citealp{bitsch2013stellar}; \citealp{harsono2015volatile})

We adopt non-periodic boundary conditions in GIZMO, such that gas is free to leave the computational domain without reflection or imposed constraints.
Our simulations also employ sink particles with an accretion radius $R_{\text{sink}} = 0.5\,\text{AU}$. Gas that falls within this radius is accreted by the sink particle if it is gravitationally bound and has specific angular momentum below that required to form a circular orbit at $R_{\text{sink}}$.  Thanks to the discard of artificial viscosity and good conservation properties, we can resolve flow down to the sink radius \citep{deng2017convergence}. We note that the chosen sink radius is much smaller than the disk region we focus on (see Figure \ref{fig:initial}). The code units we use are 1 solar mass and 1 AU with the gravitational constant $G=1$. In GIZMO, the effective mass resolution is set by the kernel mass. With a gas particle mass of
$1.12\times10^{-7}\,M_\odot$ and a target neighbor number of 200 (Wendland C4 kernel), the effective mass resolution of our simulations is $M_{\rm eff}\approx2.23\times10^{-5}\,M_\odot$ ($\sim7.4\,M_\oplus$).

\subsection{Radiative transfer} \label{subsec:Radiative}

Radiative transfer plays a critical role in hydrodynamical simulations, particularly for young self-gravitating circumstellar disks. The calculations presented in this work do not employ a simple barotropic equation of state. Although a barotropic equation of state can adequately describe the thermal evolution during the early stages preceding disk formation, it becomes unsuitable once the circumstellar disk has formed. At this stage, the temperatures are generally higher than those predicted by a barotropic relation \citep{Xu2021}. 
To more realistically model the thermal evolution of circumstellar disks, we adopt an approximate radiative cooling method proposed by \cite{Young2024}, while leaving more computationally expensive full radiation hydrodynamic simulations \citep[e.g.,][]{Ni2025} for future work. 
This approach follows the polytropic cooling framework, in which each gas element is assumed to be embedded within a spherically symmetric pseudo-cloud. Under this assumption, the radiative cooling rate can be estimated from a local approximation of the optical depth, $\tau \sim \kappa \Sigma$, where $\Sigma$ is an effective column density determined from local thermodynamic quantities.
Specifically, this approach employs a hybrid column density estimator that combines the gravitational-potential-based method of \cite{Stamatellos2007} with the pressure-gradient-based method of \cite{Lombardi2015}. In the Stamatellos method, the column density is estimated from the local gravitational potential, which sets a characteristic length scale for the surrounding gas distribution. This approach performs well in quasi-spherical geometries, such as collapsing cores or dense clumps, but is known to systematically overestimate the column density in disc-like structures.
In contrast, the Lombardi method estimates the column density using the local pressure gradient, which provides a measure of the pressure scale height, $H_{P,i} = P_i / |\nabla P_i|$, leading to $\Sigma \sim \rho H_P$. This makes it better suited for disc geometries with strong vertical stratification. However, the method becomes unreliable in regions where the pressure gradient is small, such as near the disk midplane or at the centers of quasi-spherical clumps.
In regions where pressure gradients become small, such as the disk midplane or the centers of quasi-spherical clumps, the estimator naturally reduces to the Stamatellos formalism, thereby avoiding the large scatter and breakdown of the Lombardi method in these limits. Conversely, in disk-like geometries with significant pressure gradients, the Lombardi estimator is preferentially selected, substantially reducing the systematic overestimation of column density inherent to the Stamatellos approach in such regions. The hybrid scheme dynamically selects the more appropriate estimator based on the local flow properties, thereby reducing the systematic biases present in either method alone and enabling a consistent treatment of cooling across the entire system.
As discussed in \cite{Young2024}, this approach remains an approximation to full radiative diffusion in regions of very high optical depth; in such regions, the cooling efficiency may be modestly underestimated, leading to a slight overestimate of the midplane temperature. The mean column density for particle $i$ is given by
\begin{equation}
\bar{\Sigma}_i = \zeta' \rho_i \left( H_{P,i}^{-2} + H_{s,i}^{-2} \right)^{-1/2},\label{eq1}
\end{equation}
which the pressure scale height based on the particle's pressure gradient $H_{P,i} = \frac{P_i}{\rho_i \lvert \boldsymbol{a}_{h,i} \rvert}$, and the self-gravity scale height based on the particle's gravitational potential $H_{s,i} = \frac{\zeta_n}{\zeta'} \left[ \frac{-\psi_i}{4 \pi G \rho_i} \right]^{1/2}$. Here, \(\rho_i\) is the density, \(P_i\) is the pressure, \(a_{h,i}\) is the hydrodynamic acceleration, \(\psi_i\) is the gravitational potential, and \(G\) is the gravitational constant. The coefficients \(\zeta'\) and \(\zeta_n\) are dimensionless constants. \(\zeta_n\) depends on the polytropic index \(n\) (with \(\zeta_2 = 0.368\) for \(n = 2\)), while \(\zeta' = 1.014\), and is nearly independent of \(n\) (\citealp{Stamatellos2007} \citealp{Forgan2009}). Based on these parameters, we can derive the radiative cooling rate
\begin{equation}
\left. \frac{du_i}{dt} \right|_{\mathrm{rad}} = \frac{4\sigma \left[ T_0^4 - T_i^4 \right]}{\bar{\Sigma}^2 \bar{\kappa}_i(\rho_i, T_i) + \kappa_i^{-1}(\rho_i, T_i)},\label{eq2}
\end{equation}
with $u_i$ is the particle's specific internal energy.
For this calculation, the temperature floor is set to $T_0 = 10\text{K}$, representing the background radiation. The opacity $\kappa$ follows the expression provided by \cite{Bell&Lin1994} for temperatures above 1000K, and following \cite{Xu2023}
\begin{equation}
\kappa(T) =
\begin{cases} 
\left( \frac{T}{100\,\mathrm{K}} \right)^2 \, \mathrm{cm}^2\,\mathrm{g}^{-1} & (T < 100\,\mathrm{K}) \\[10pt]
1 \, \mathrm{cm}^2\,\mathrm{g}^{-1} & (T \geq 100\,\mathrm{K})
\end{cases}\label{sec3}
\end{equation}
for temperatures below 1000K. Replace the particle's temperature \(T_i\) with the equilibrium temperature \(T_{\mathrm{eq},i}\) in equation \eqref{eq2}. If the computed \(T_{\mathrm{eq},i}^{4} \le T_0^{4}(r_i)\), we simply take \(T_{\mathrm{eq},i}^{4} = T_0^{4}(r_i)\). The corresponding equilibrium internal energy is \(u_{\mathrm{eq},i} = u(T_{\mathrm{eq},i}, \rho_i)\). Subsequently, the thermal timescale
\begin{equation}
t_{\text{therm},i} = \left( u_{\text{eq},i} - u_i \right) 
\left[ 
\left. \frac{du_i}{dt} \right|_{\text{hydro}} + 
\left. \frac{du_i}{dt} \right|_{\text{rad}} 
\right]^{-1}\label{eq4}
\end{equation}
is derived. Using this, the cooling rate is calculated and expressed as
\begin{equation}
\left. \frac{du_i}{dt} \right|_{\text{cool}} = \frac{1}{\delta t} 
\left[ 
u_i \exp\left(-\frac{\delta t}{t_{\text{therm},i}}\right) + 
u_{\text{eq},i} \left( 1 - \exp\left(-\frac{\delta t}{t_{\text{therm},i}}\right) \right) - u_i
\right].\label{eq5}
\end{equation}
Here $\delta t$ denotes the simulation time step.
The energy equation is then updated via
\begin{equation}
u_i(t + \delta t) = u_i(t) + \delta t \left. \frac{du_i}{dt} \right|_{\text{cool}}.\label{eq6}
\end{equation}

\section{Results} \label{sec:results}

\begin{figure}
\centering
\includegraphics[width=0.45\textwidth]{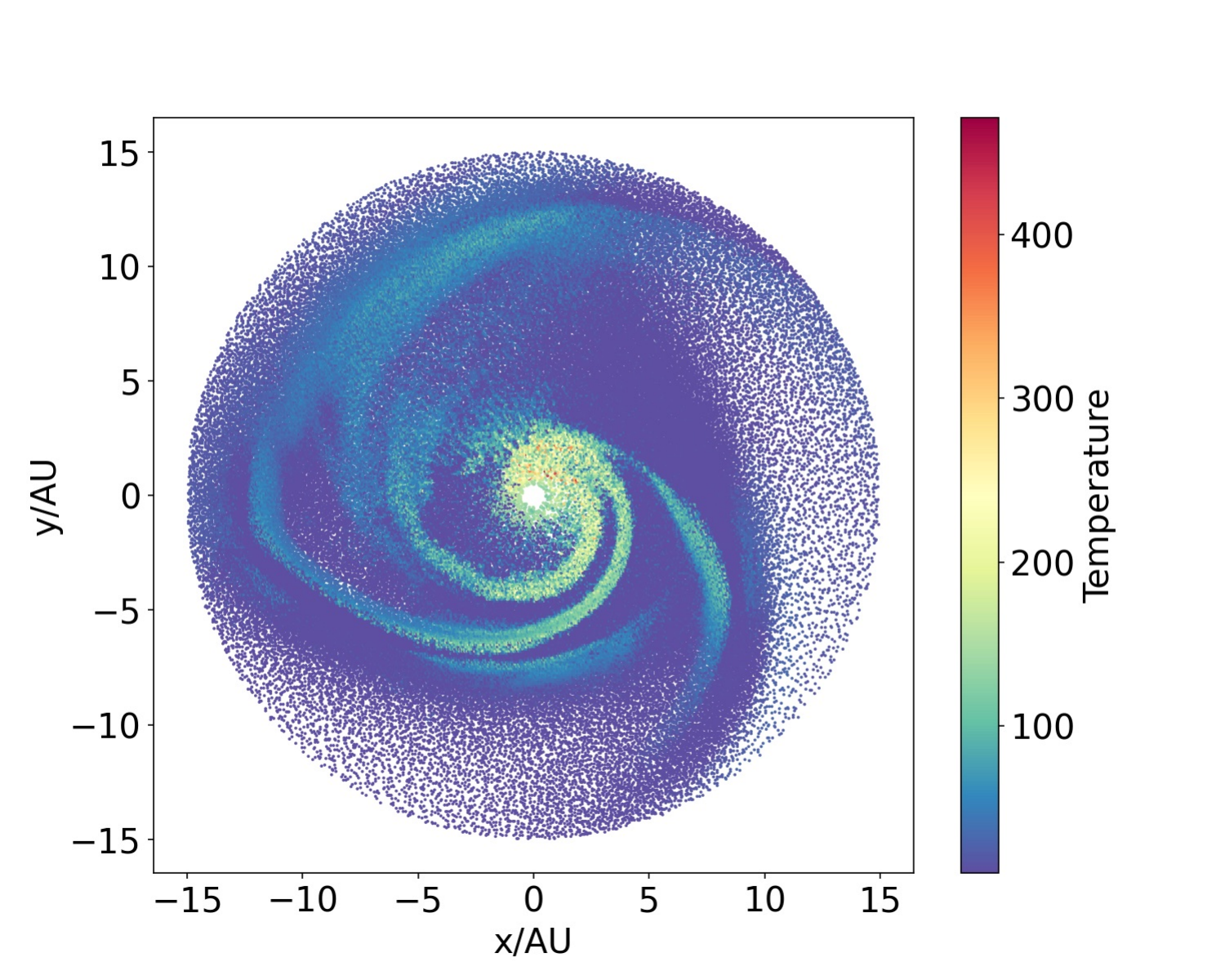}
\includegraphics[width=0.45\textwidth]{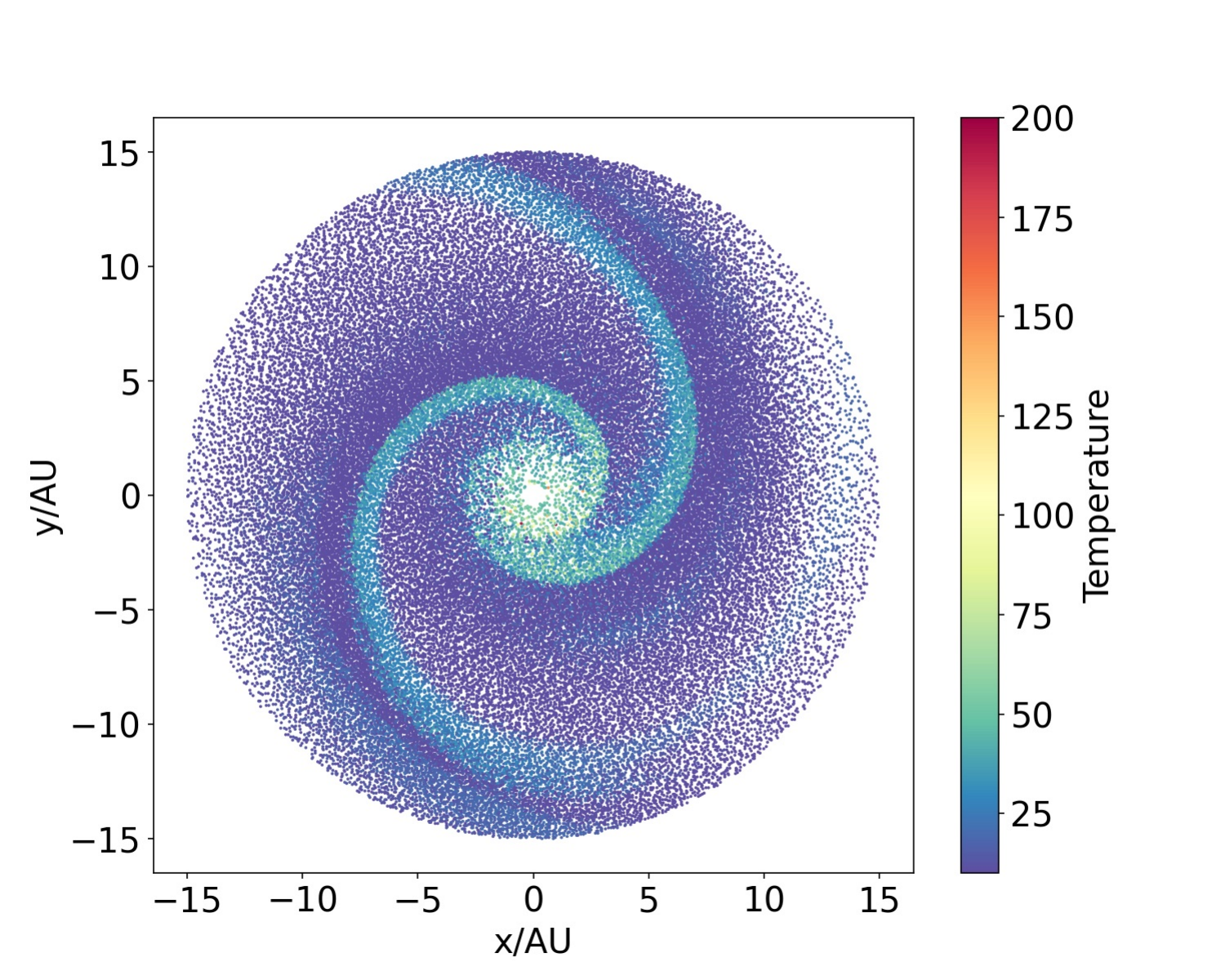}
\caption{These panels highlight the initial temperature distributions in the vicinity of the newborn stars. The slices have a thickness of 1 AU and extend to radii $<$30 AU. The left panel shows data for Run 1, and the right panel for Run 2. 
\label{fig:initial}}
\end{figure}

We conducted two representative simulations: Run 1, which features a rapid infall phase, and Run 2, which exhibits a relatively calm evolution. The initial density and temperature distribution near the midplane of the disks for both systems are shown in Fig.~\ref{fig:initial}. The two systems have total masses of $1\,M_\odot$ and $0.5\,M_\odot$ within a radius of $2000\,\text{AU}$, and their initial central stellar masses are $0.06\,M_\odot$ and $0.03\,M_\odot$, respectively. The corresponding accretion rates onto the star (i.e., the mass growth rate of the central sink particle) are approximately $1 \times 10^{-5}\,M_\odot\,\mathrm{yr}^{-1}$ and $1 \times 10^{-6}\,M_\odot\,\mathrm{yr}^{-1}$. The disk region is defined as the area where the rotational velocity satisfies $v_\phi > 0.5\,v_{\rm K}$ and the density satisfies $\rho > 10^{-3}\,\rho_{\rm max}$, where $v_{\rm K}$ is the local Keplerian velocity and $\rho_{\rm max}$ is the maximum density in the simulation domain.
In the original Bate simulations, these two protostars accrete steadily at the above rates. In this work, we focus on modeling the early-stage thermal evolution during the initial phases of this accretion process.

\subsection{Run 1 with infall} \label{subsec:run1}
\subsubsection{Global system evolution}
\begin{figure*}
\centering
\includegraphics[width=0.32\textwidth]{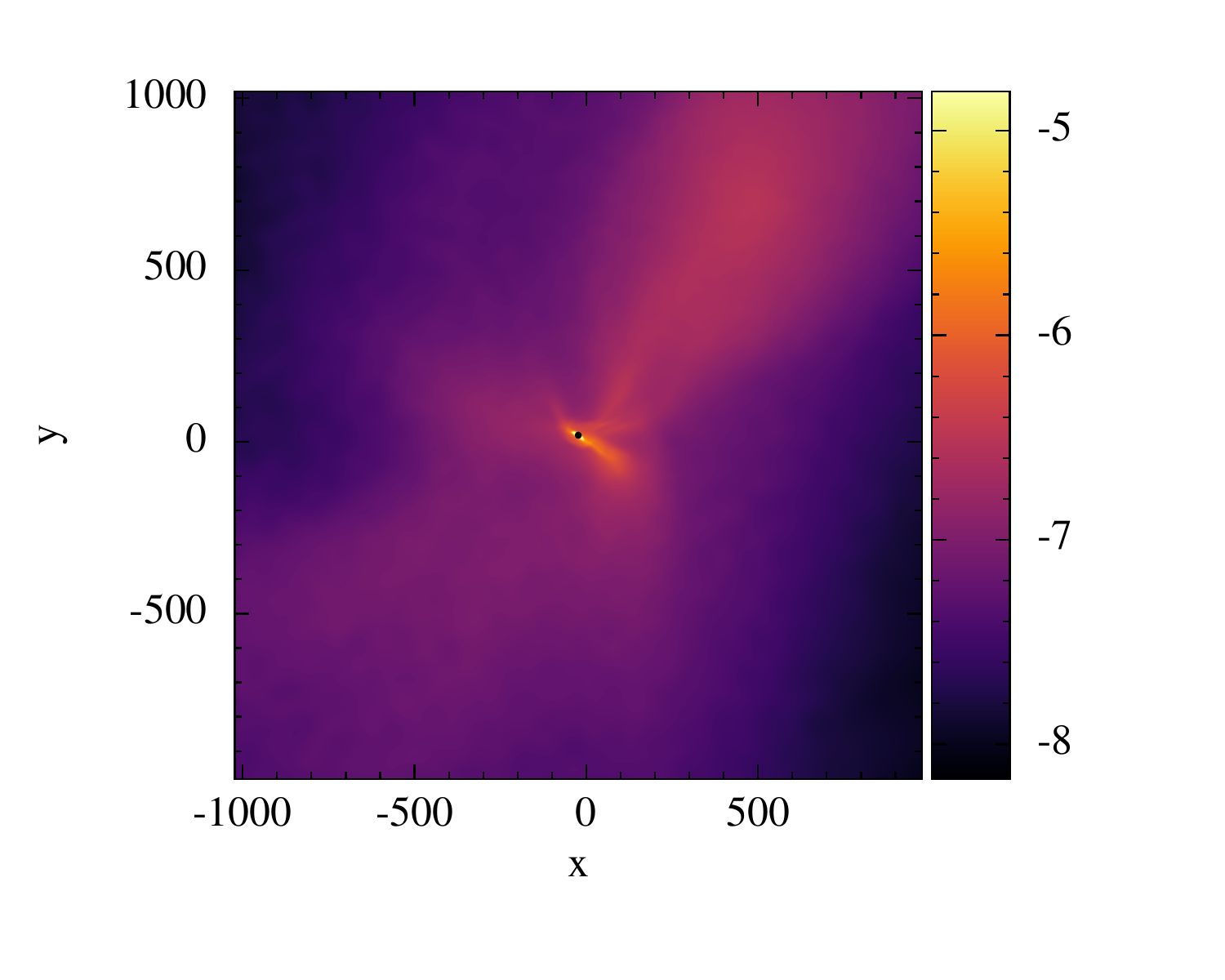}
\includegraphics[width=0.32\textwidth]{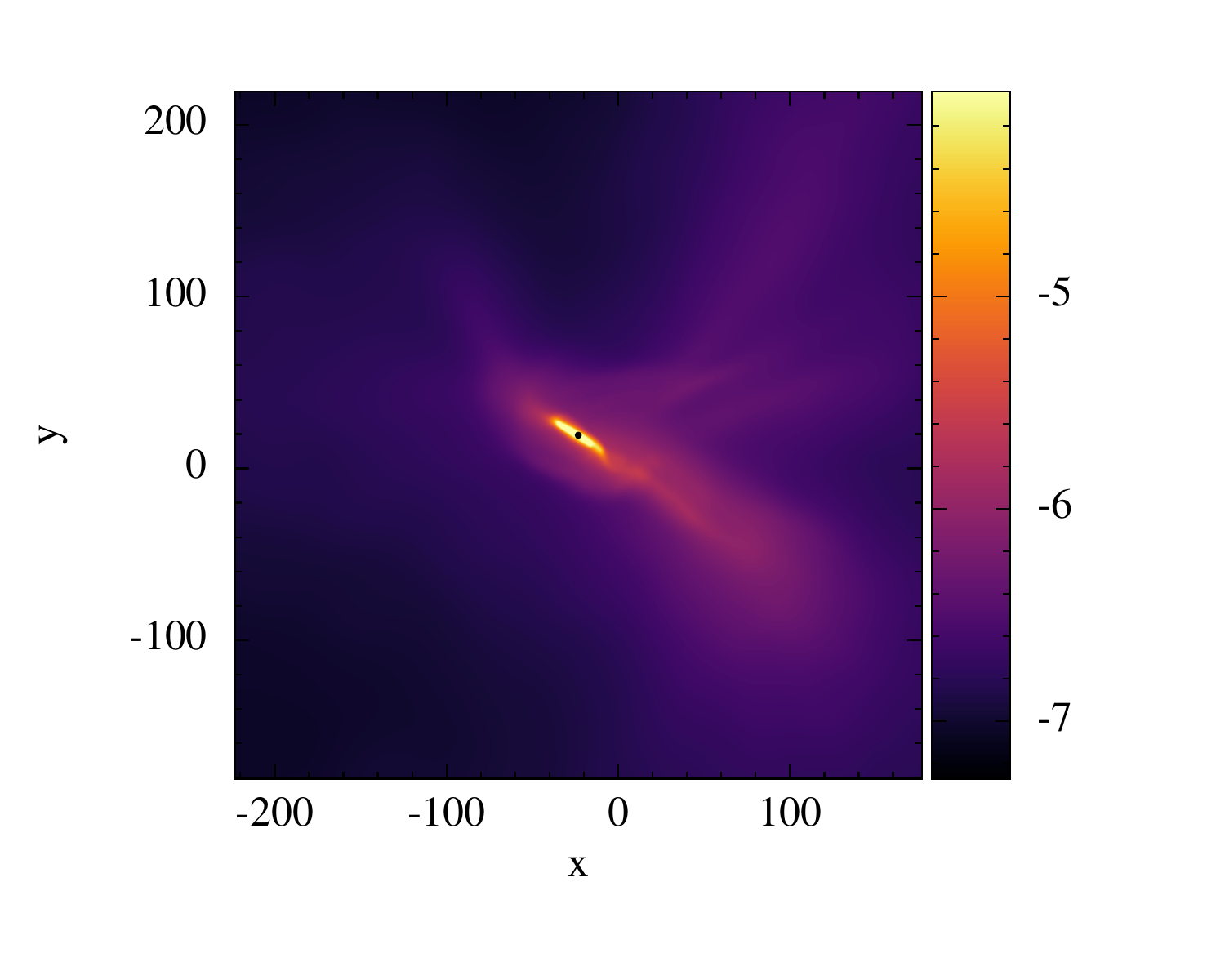}
\includegraphics[width=0.32\textwidth]{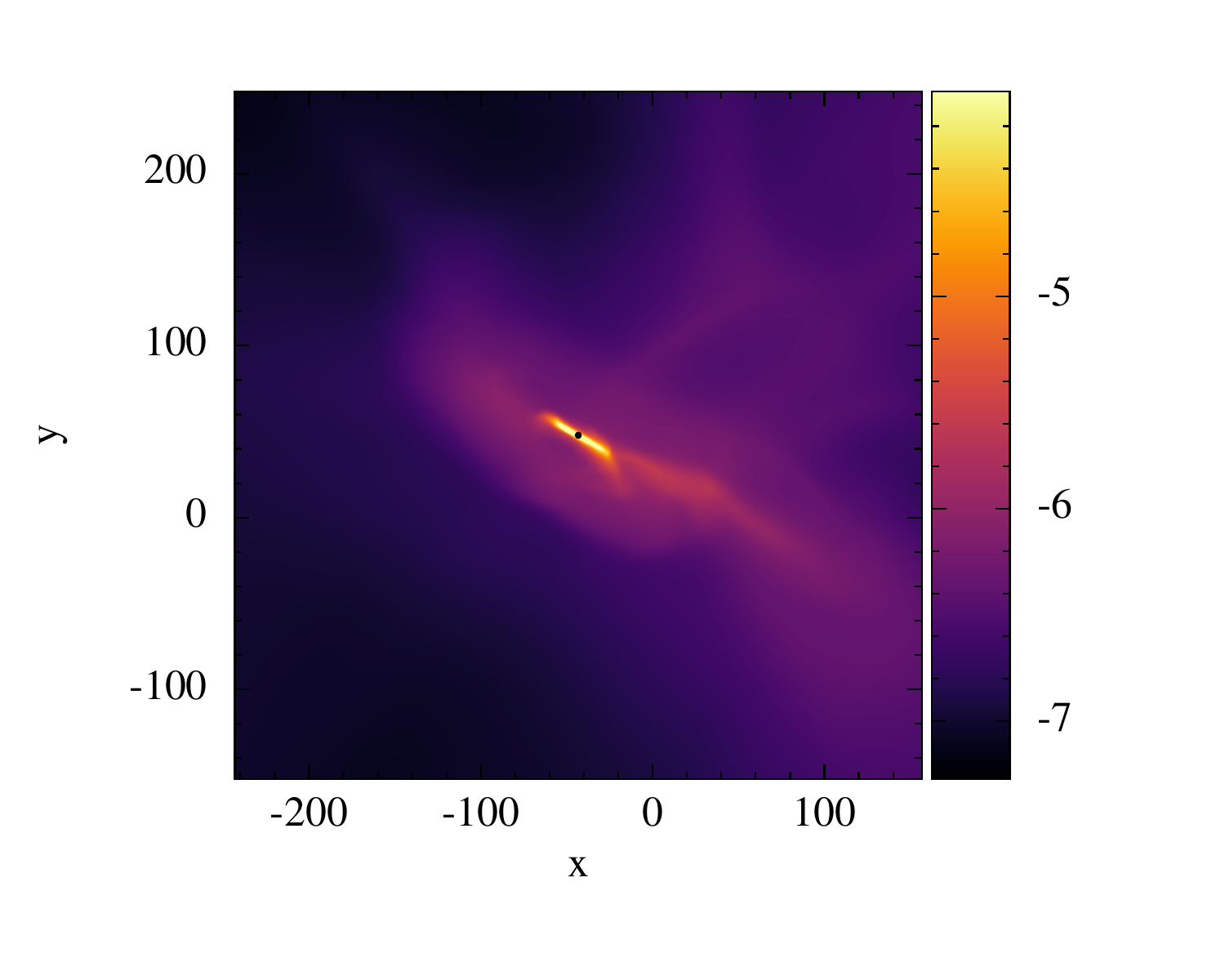}
\includegraphics[width=0.32\textwidth]{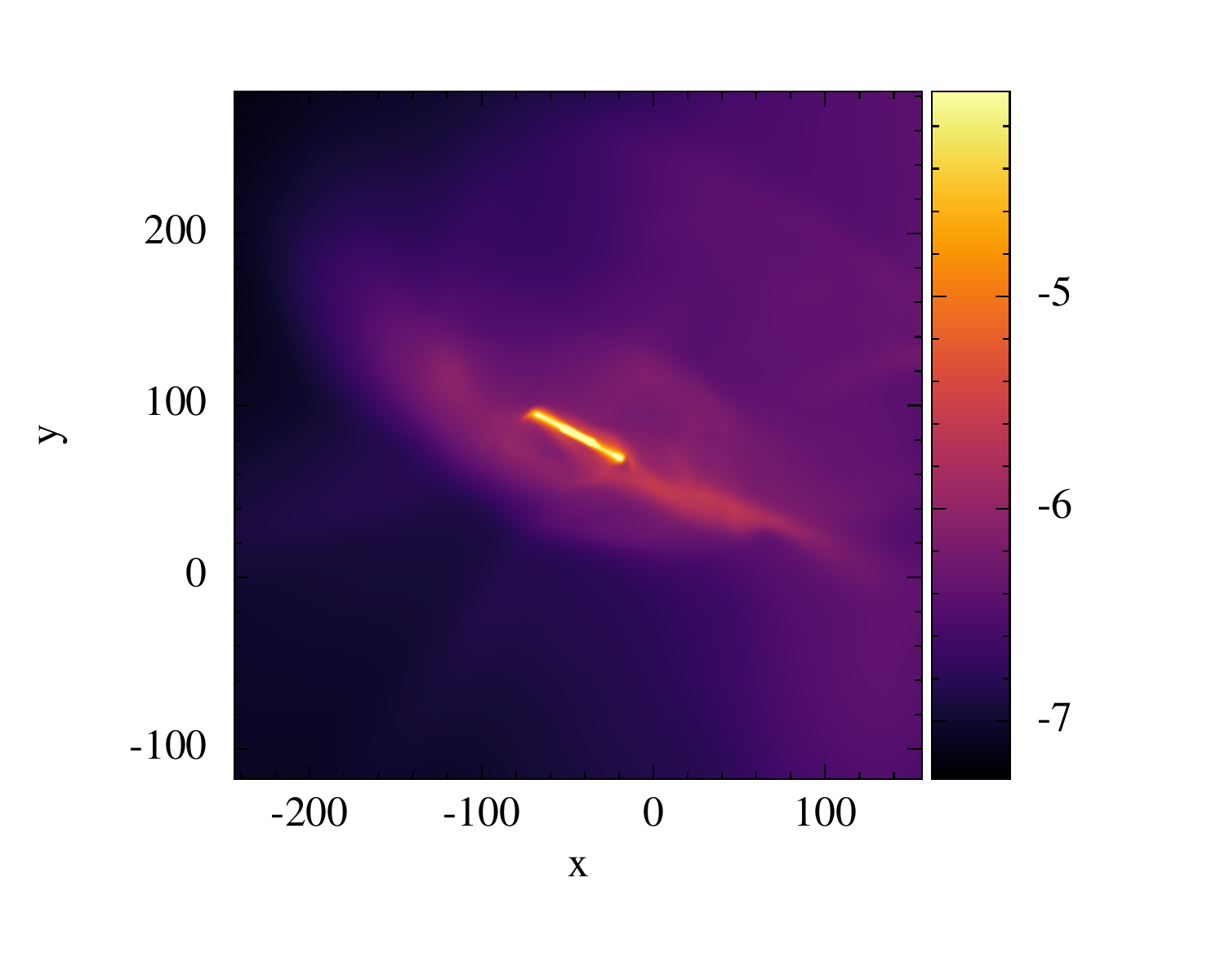}
\includegraphics[width=0.32\textwidth]{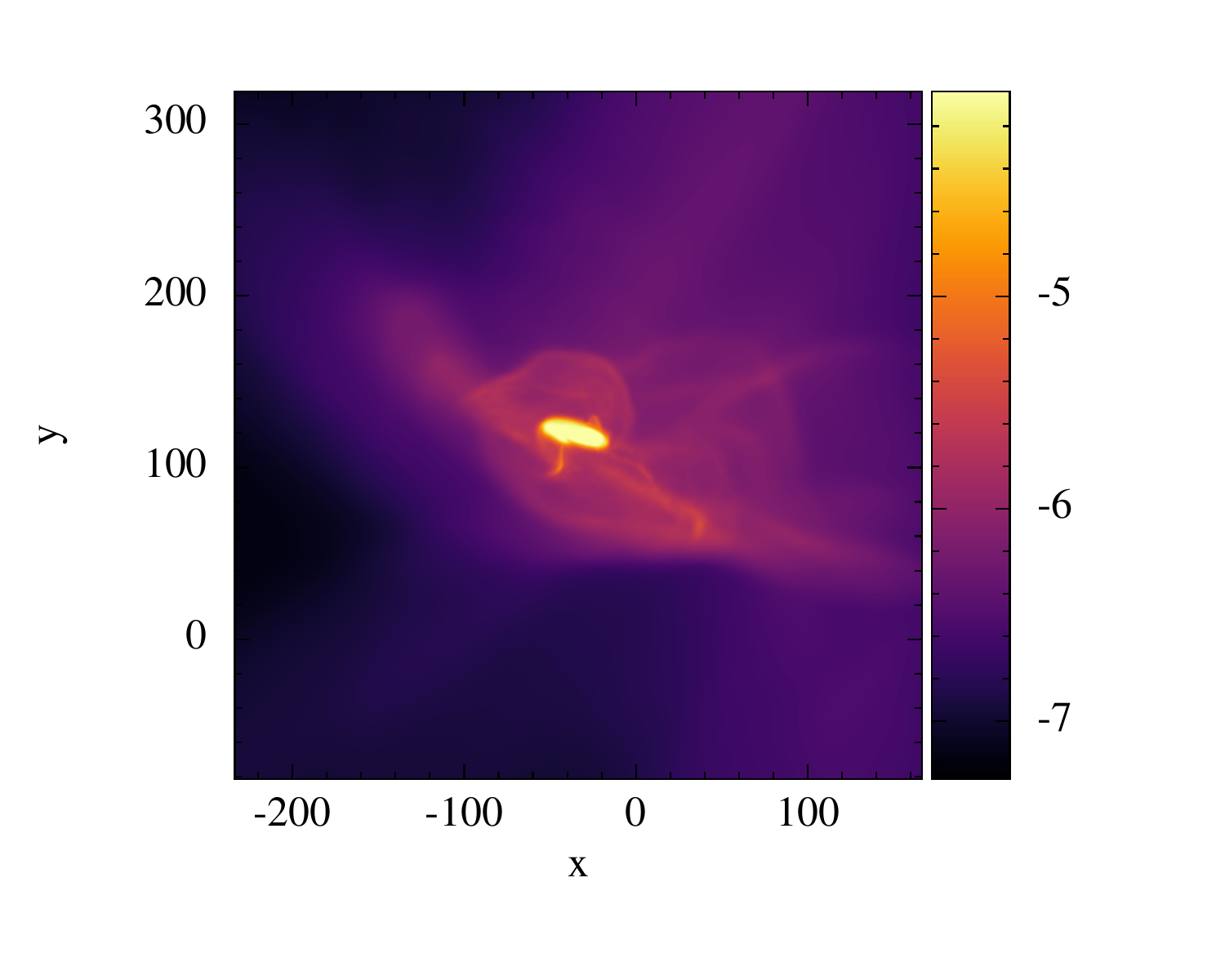}
\includegraphics[width=0.32\textwidth]{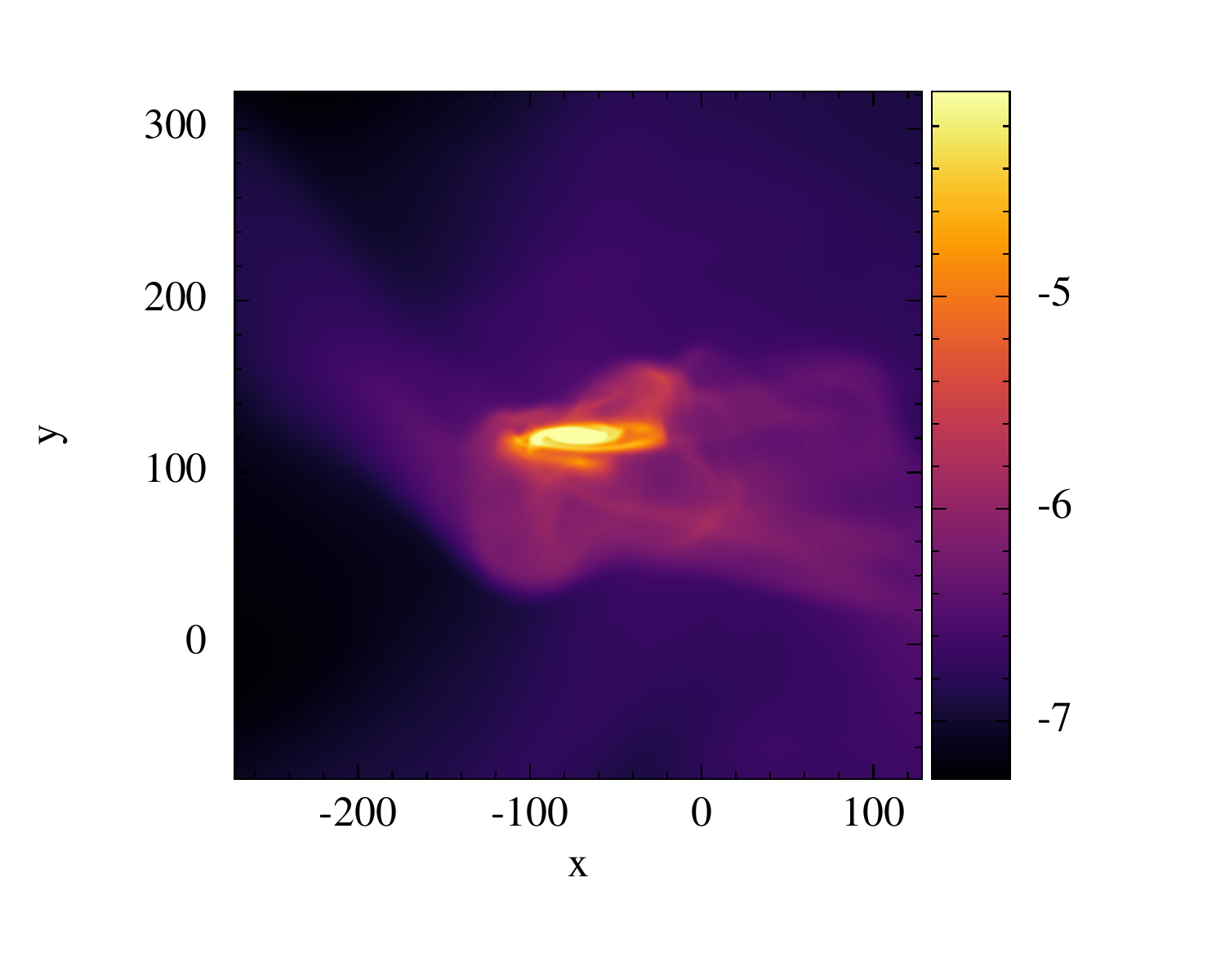}
\includegraphics[width=0.32\textwidth]{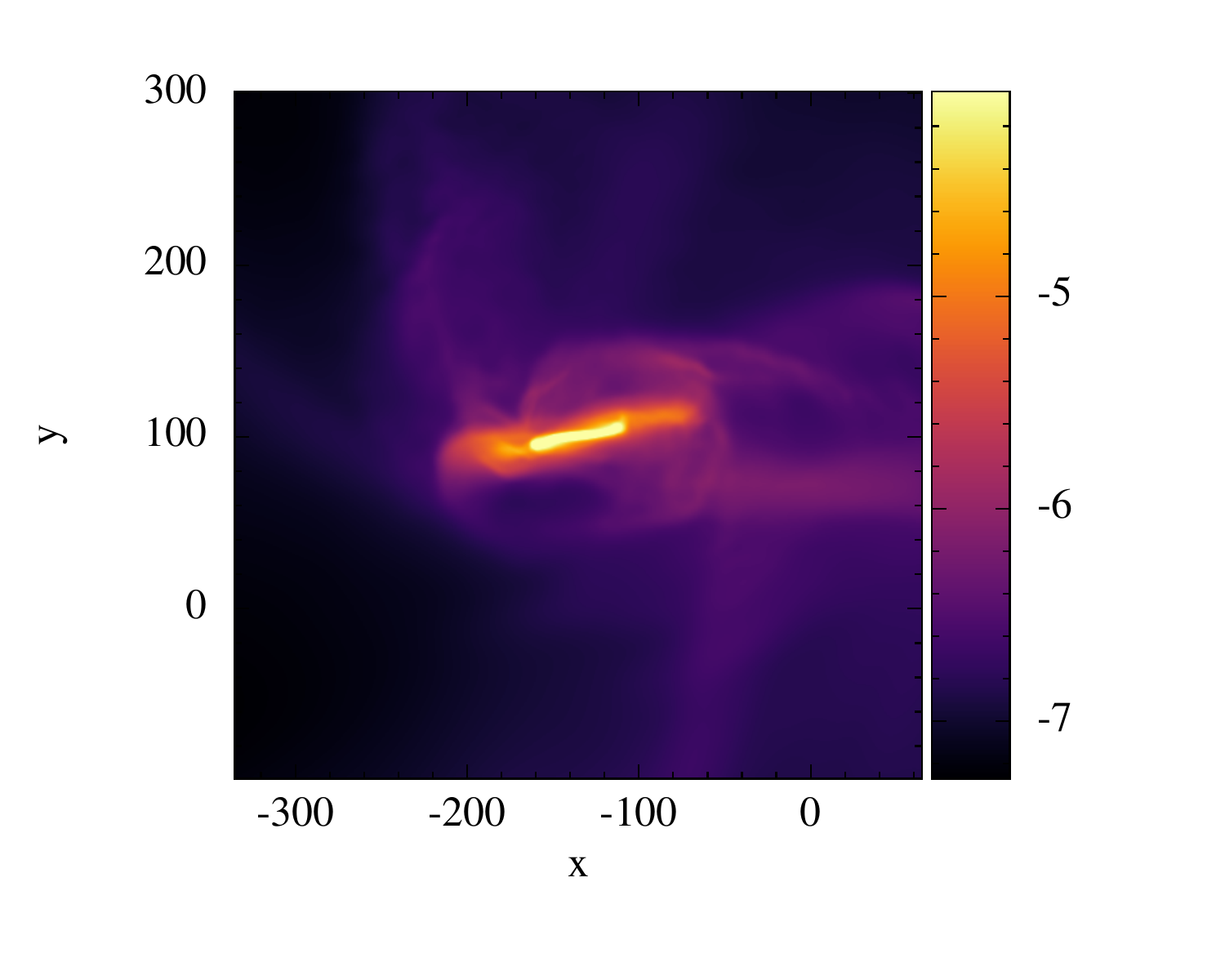}
\includegraphics[width=0.32\textwidth]{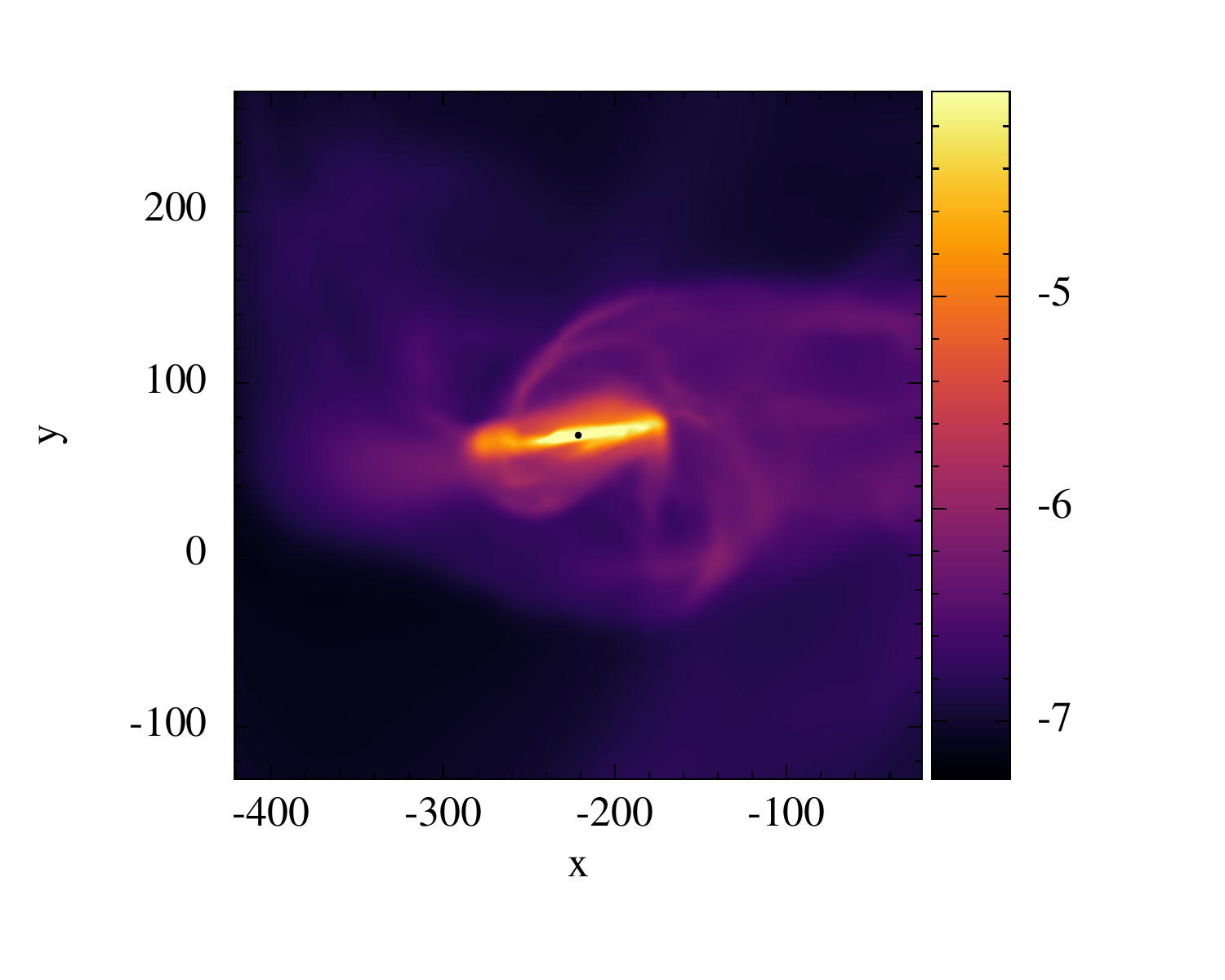}
\includegraphics[width=0.32\textwidth]{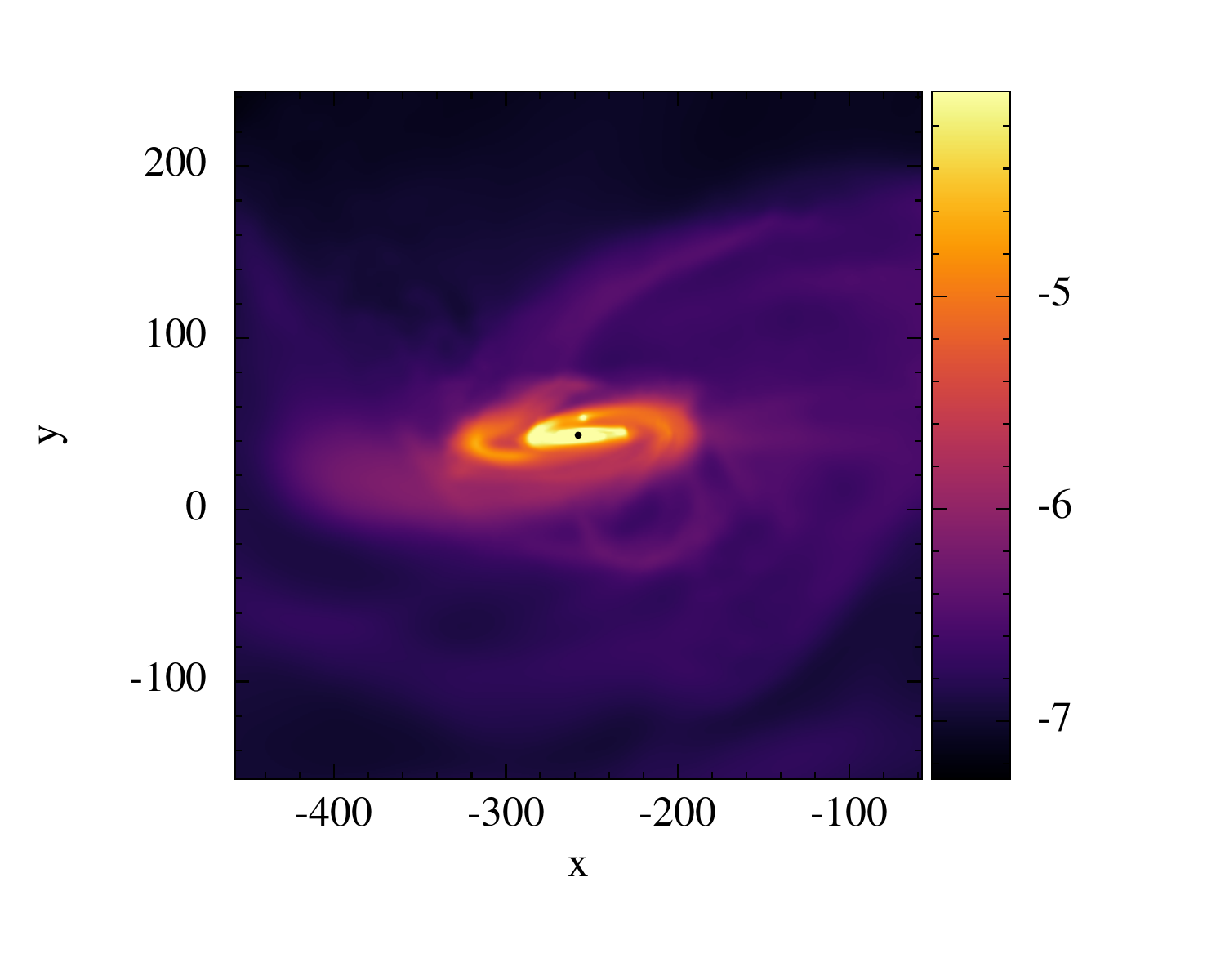}
\caption{The figures show snapshots of the system's global evolution, arranged in reading order from left to right, top to bottom. The first panel in the top left show the initial density distributions of Run 1 with the box size of 2000 AU. while the subsequent eight images cover the equally separated (by 2000 years) episodes from 1000 to 15,000 years. The box size for the last eight snapshots is 400 AU. Each box's x-y coordinates drift with the host star at its center. The color-bar indicates the column density (in unit of $M_\odot/\mathrm{AU}^2$) 
obtained by integrating the gas density along the $z$-axis.
\label{fig:general}}
\end{figure*}

\begin{figure}
\centering
\includegraphics[width=0.45\textwidth]{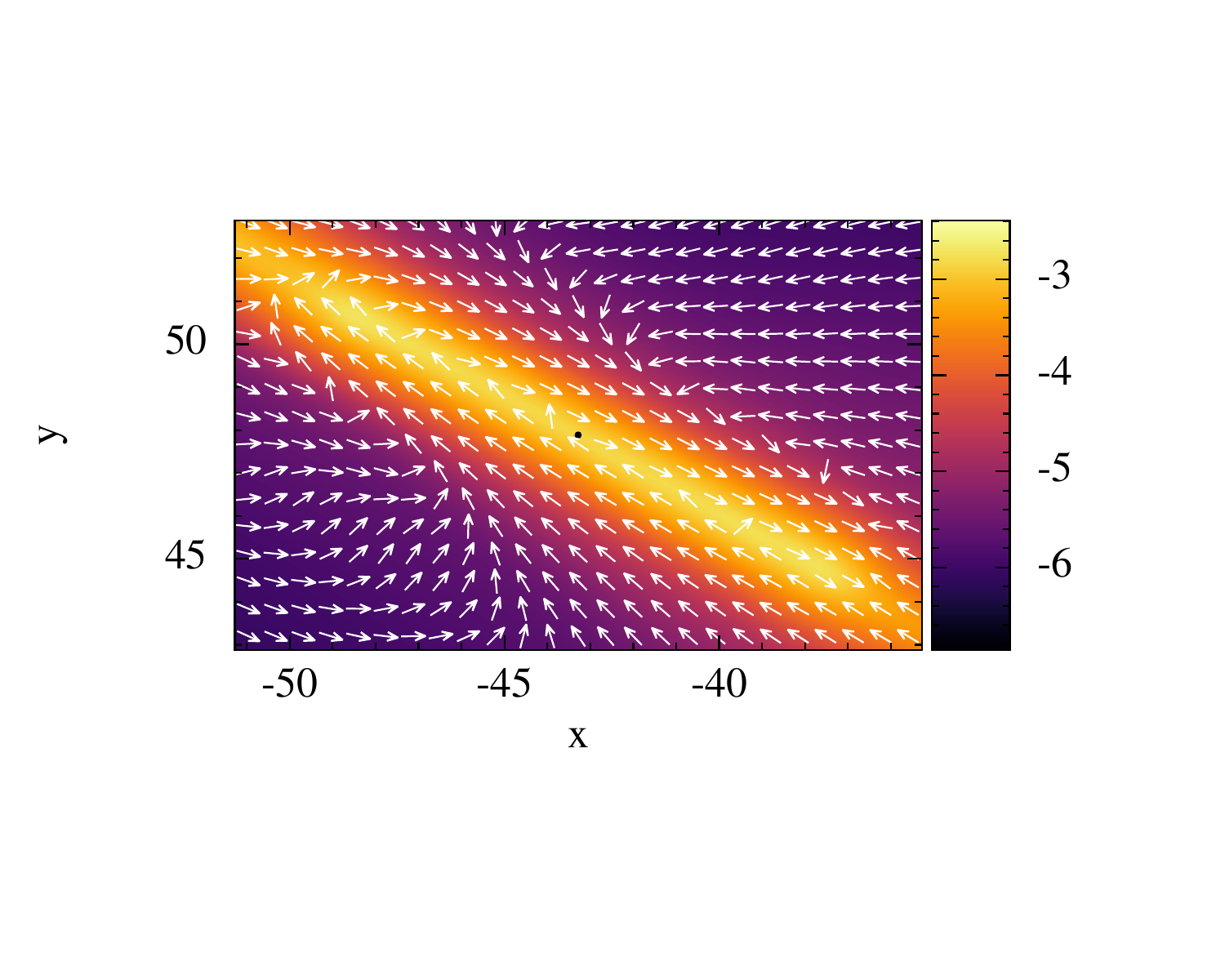}
\includegraphics[width=0.45\textwidth]{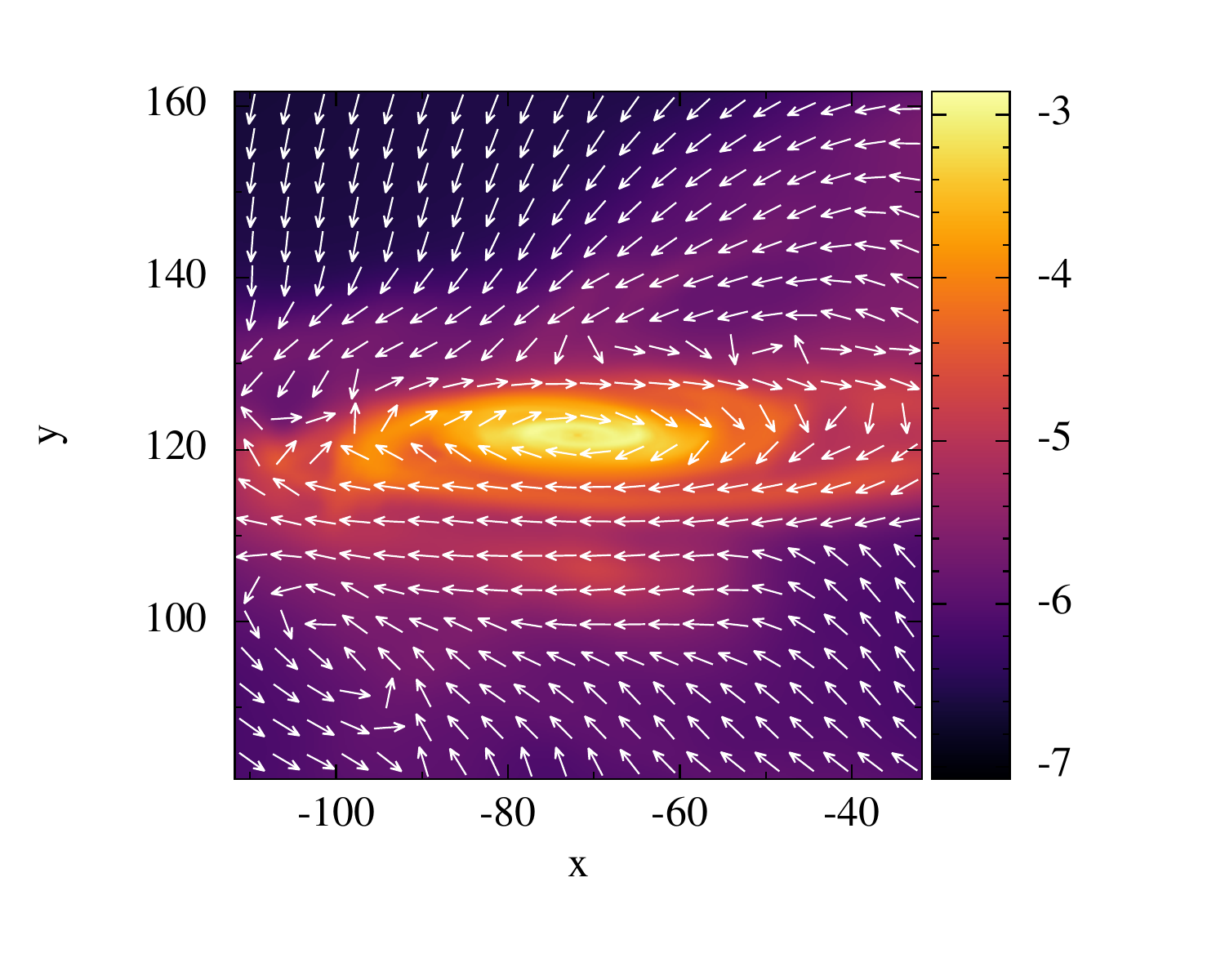}
\caption{The figure shows zoom-in views of the density (in unit of $M_\odot/\mathrm{AU}^2$) and velocity fields, focusing on sub-regions extracted from the original simulation domain displayed in Fig.~\ref{fig:general}. The left panel corresponds to 3000 years, and the right panel corresponds to 9000 years. The arrows indicate the gas velocity field.
\label{fig:vector}}
\end{figure}

\begin{figure}
\centering
\includegraphics[width=0.7\textwidth]{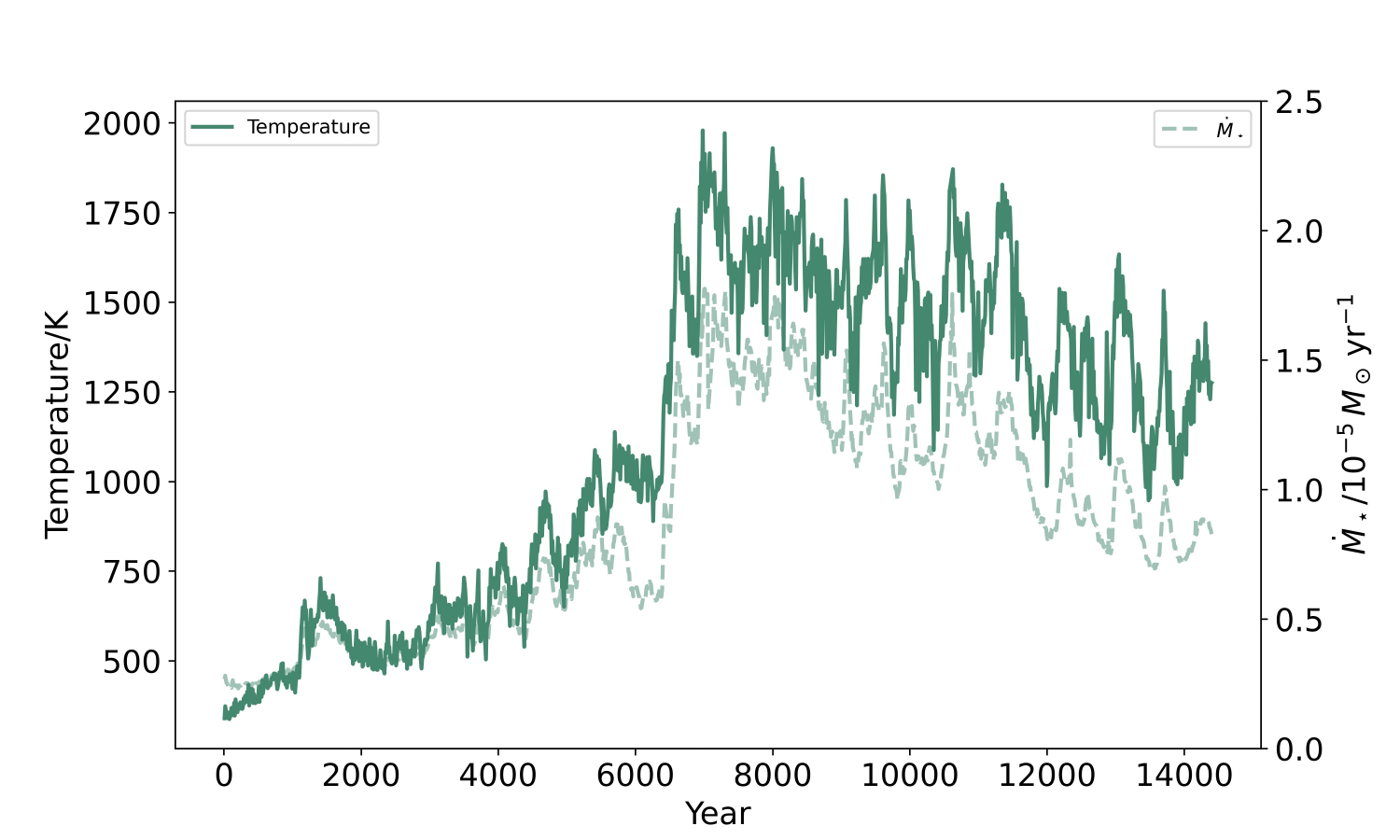}
\caption{The panel shows the time evolution of the maximum disk temperature (solid line, left axis) and stellar accretion rate $\dot{M}_\star$ (dashed line, right axis). The two quantities exhibit similar trends, with a sharp rise during the main heating phase. To mitigate the impact of numerical outliers, a representative near-maximum temperature is used instead of the absolute peak value.
\label{fig:temp53}}
\end{figure}

We begin by presenting and analyzing the results of Run 1. In the panel of Fig.~\ref{fig:general}, we show snapshots of the global evolution of the system, with the color bar representing the column density of gas in the system. Each snapshot is separated by 2000 years. The general evolution can be divided into three distinct phases: the slow infall phase, the main heating phase, and the cooling phase. The series shows the dynamic evolution of the gas, highlighting the changing density distribution and the effects of accretion and angular momentum transfer as the gas falls onto the disk. These snapshots provide insights into the gradual development of the system over time, with significant changes in the disk's structure, temperature, and density as material accretes onto the central star. During the slow infall phase (approximately 1000--6000~yr), material within the streamers surrounding the circumstellar disk gradually falls toward the disk surface under the influence of gravity. The velocity field of the gas during this stage is shown in the left panel of Fig.~\ref{fig:vector}. As infall progresses, the disk mass increases steadily at a rate of approximately $10^{-5}\,M_\odot\,\mathrm{yr}^{-1}$. The infalling material 
transports angular momentum to the disk, causing it to expand radially. Simultaneously, the central protostar gains mass through accretion at a rate of roughly $10^{-5}\,M_\odot\,\mathrm{yr}^{-1}$. As the disk continues to accrete mass, its temperature gradually increases due to viscous heating driven by turbulent stresses, as shown in Fig.~\ref{fig:temp53}.

The subsequent stage, which spans approximately 6000 to 9000~yr, corresponds to the main heating phase. As the overall disk mass increases, the infall becomes more vigorous, with a higher rate and a stronger dynamical impact. The gas carrying angular momentum from various directions accretes onto the disk surface, leading to complex redistribution of the total angular momentum within the disk. As a result, the disk undergoes a substantial reorientation, 
tilting by almost $30^\circ$ relative to its original orientation.
This change in angular momentum has a pronounced effect on the disk structure. In contrast to the first phase, where the disk radius increased steadily, the disk in this phase becomes more 
radially compact and vertically extended. The midplane density increases significantly, which in turn reduces the disk's radiative cooling efficiency. Moreover, the enhanced infall drives stronger turbulence and viscous activity within the disk. Portions of the disk material are displaced and temporarily ejected, later falling back onto the disk surface.
The right panel of Fig.~\ref{fig:vector} illustrates the gas velocity field during this phase, clearly showing both infalling streams and disk material being pushed outward. This stage corresponds to a rapid heating episode. As shown in Fig.~\ref{fig:temp53}, the maximum disk temperature rises by nearly 1000~K, with localized hot spots reaching temperatures as high as $\sim$2000~K.

The last stage, from approximately 9000 to 15,000~yr, marks the onset of the cooling phase. Since this time is a fraction of the free-fall timescale for the extended envelope, disk evolution continues, though at a less rapid rate.
As the amount of material falling from the streamer onto the disk diminishes, the disk re-expands and becomes geometrically thinner---its aspect ratio decreases from 0.18 to 0.13---due to continued angular momentum redistribution. This structural change improves the disk’s radiative cooling efficiency. Compared to the main heating phase, the maximum disk temperature gradually decreases, eventually stabilizing at around 1000~K. The stellar accretion rate also exhibits a slight decline, consistent with the reduced infall activity and overall thermal relaxation of the system.

\begin{figure}
\centering
\includegraphics[width=0.45\textwidth]{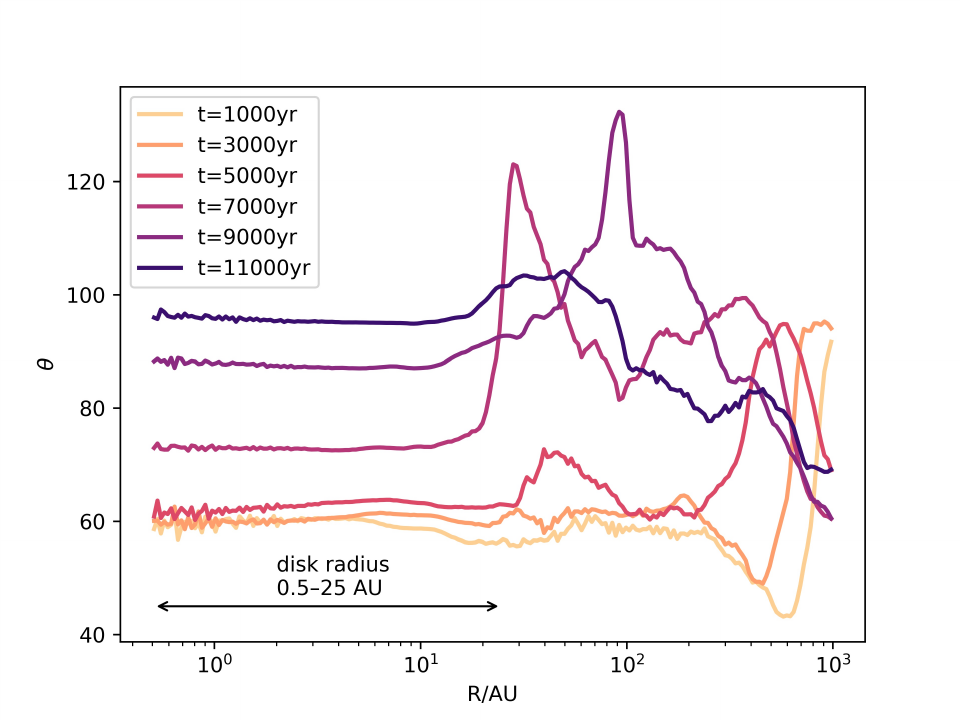}
\includegraphics[width=0.45\textwidth]{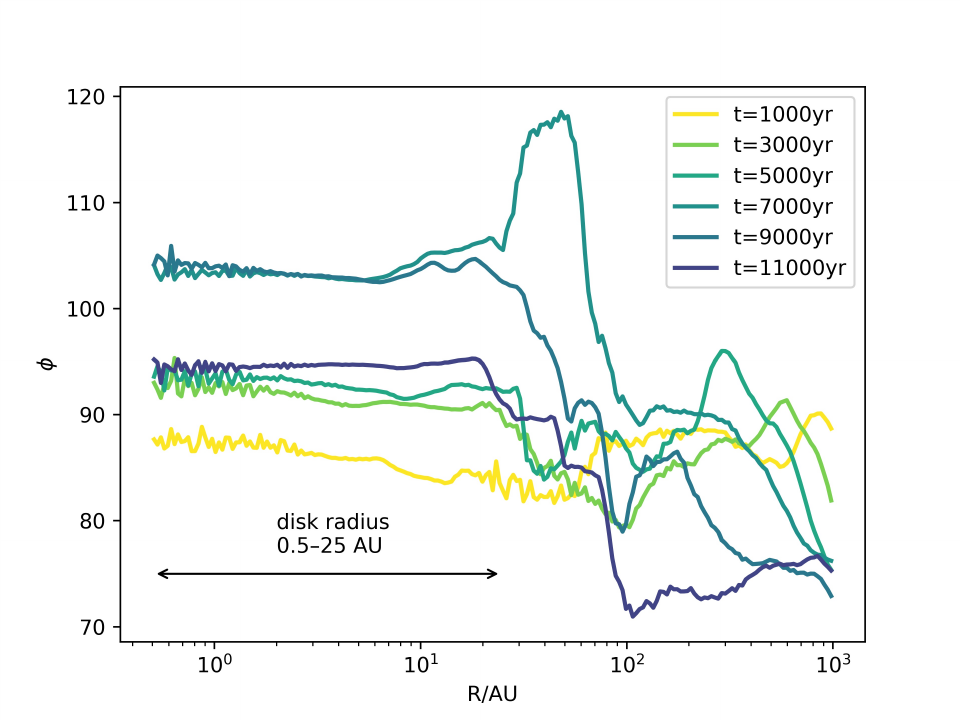}
\caption{The plots show the direction of specific angular momentum vector at various radius and times for Run 1. (Here, “radius” refers to the spherical distance from the star, the polar angle \(\theta\) is measured from the \(+\hat{\boldsymbol{z}}\) axis of the inertial frame, and the azimuth \(\phi\) is measured in the \(x\)–\(y\) plane from \(+\hat{\boldsymbol{x}}\) toward \(+\hat{\boldsymbol{y}}\).) 
The left panel represents the polar angle $\theta$, while the right panel shows the azimuthal angle 
$\phi$.   Both angles are measured relative to an arbitrary reference direction on a chosen coordinate plane. The radius refers to the spherical distance from the central stellar (i.e., the sink particle). The horizontal marker at 0.5–25 AU indicates the average disk radius across different times, the actual disk radius varies with time. Regions outside this range are the envelope.
\label{fig:angularmomentum}}
\end{figure}

We examine the evolution of the direction of angular momentum distribution throughout the simulation. 
In Figure~\ref{fig:angularmomentum}, we plot the spherical polar angles $\theta$ 
and $\phi$ of the angular momentum vector relative to the (arbitrarily defined) $z$ and $x$ axes, at 
various spherical radius $R$ and time. Non-uniform $R$ distribution of $\theta$ and $\phi$ 
corresponds to warps in the disk and their time dependence implies changes in the disk orientation.

In the early stages, the material outside the disk carries a significant amount of angular momentum 
misaligned with that of the inner disk. As time progresses, this misaligned gas gradually falls toward 
the disk. By $t = 7000$\,yr, a clear discontinuity in the angular momentum structure emerges between 
the outer ($R \gtrsim 25$ AU) disk and the inner  ($R \lesssim 25$ AU)
region, particularly in the polar angle component $\theta$. The redistribution, driven 
by the cancellation and mixing of angular momentum vectors originating from different 
infalling streams, results in a rapid contraction of the disk and a noticeable increase 
in the stellar accretion rate. During this phase, the misalignment between the inner 
and outer disk becomes more pronounced, with the polar angle difference reaching up 
to $\sim 50^\circ$. This level of misalignment reflects a significantly perturbed 
disk structure, in contrast to the more modest $\theta$ differences of 
$\sim 10^\circ$--$20^\circ$ observed in Run~2.

Later, at $t = 11,\!000$~yr, the angular mismatch between the disk and its surrounding envelope becomes significantly weaker. The $\theta$-angle of the angular momentum becomes more uniform across the disk and its outer layers, signaling the transition into the cooling phase. The angular momentum orientation within the disk shifts by nearly $35^\circ$ relative to its initial configuration.
Overall, the disk undergoes a phase of global heating over a relatively short timescale due to the impact of infall, followed by a gradual thermal relaxation, with temperatures declining again within 10,000~yr.

\subsubsection{Heating phase analysis}
\begin{figure}
\centering
\includegraphics[width=0.7\textwidth]{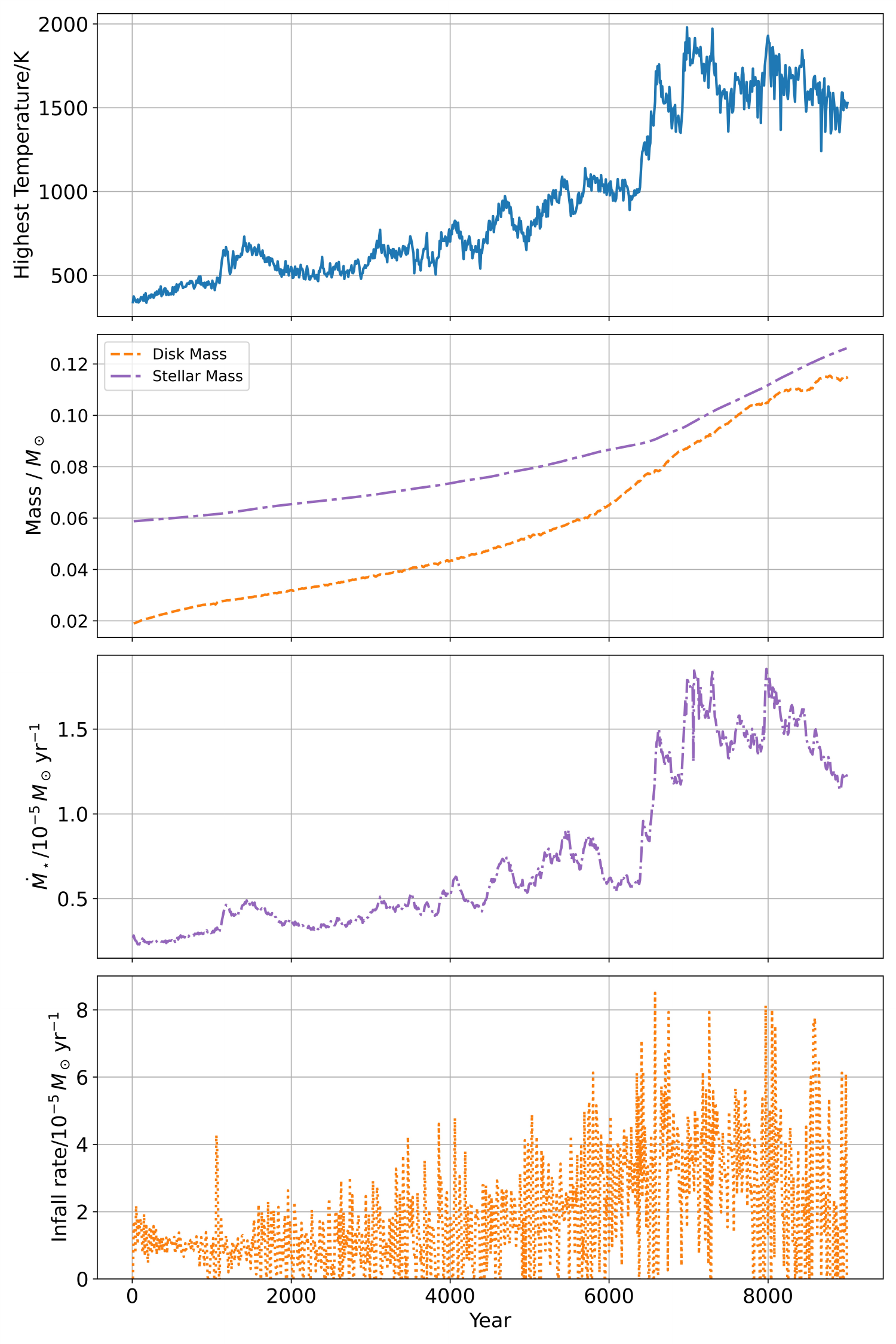}
\caption{Evolution of key disk properties over time. Panels show (top to bottom): maximum disk temperature, stellar mass, total disk mass, stellar accretion rate $\dot{M}_\star$, and infall rate from the envelope. 
Infall events correlate with accretion bursts and temperature rises, indicating that envelope feeding significantly influences disk heating and accretion activity. The disk is defined as the region where the rotational velocity satisfies $v_\phi > 0.5\,v_{\rm K}$ and the density exceeds $10^{-3}\,\rho_{\rm max}$.
\label{fig:disktime}}
\end{figure}

We now turn to a more detailed analysis of the disk heating process. Fig.~\ref{fig:disktime} shows the time evolution of several key physical quantities during the first 9000~yr of the simulation, including the maximum disk temperature, stellar mass, total disk mass, stellar accretion rate $\dot{M}_\star$, and the infall rate from the envelope. The infall rate, the mass flux of material falling from the envelope onto the disk, is calculated as the sum of the rate of change of the disk mass and the stellar accretion rate, 
\begin{equation}
    \dot{M}_{\rm infall} = \frac{dM_{\rm disk}}{dt} + \dot{M}_\star.\label{infallrate}
\end{equation}
Overall, the temperature evolution closely follows the trends of both the stellar accretion rate and the infall rate. During the first 6000~yr, all three quantities increase slowly and steadily. After $t = 6000$~yr, a marked increase occurs in each variable. The disk temperature reaches its peak value of nearly 2000~K, while the stellar accretion rate rises to approximately $1.5\times10^{-5}\,M_\odot\,\mathrm{yr}^{-1}$, and the infall rate peaks at around $4\times10^{-5}\,M_\odot\,\mathrm{yr}^{-1}$—roughly a twofold increase compared to earlier stages. Moreover, the disk mass gradually approaches the stellar mass, indicating that the disk is strongly self-gravity dominated.
Toward the end of the main heating phase, the infall rate drops sharply. However, since the disk mass remains relatively high, the stellar accretion rate decreases only slightly, and the disk temperature experiences a modest decline. This indicates that the infall process near the disk surface plays a crucial role in driving the heating. Notably, despite significant short-term fluctuations in the infall rate, the disk is able to maintain a relatively stable high temperature. This suggests that, due to reduced cooling efficiency, the heat accumulated in the disk is not radiated away immediately.

\begin{figure}
\centering
\includegraphics[width=0.45\textwidth]{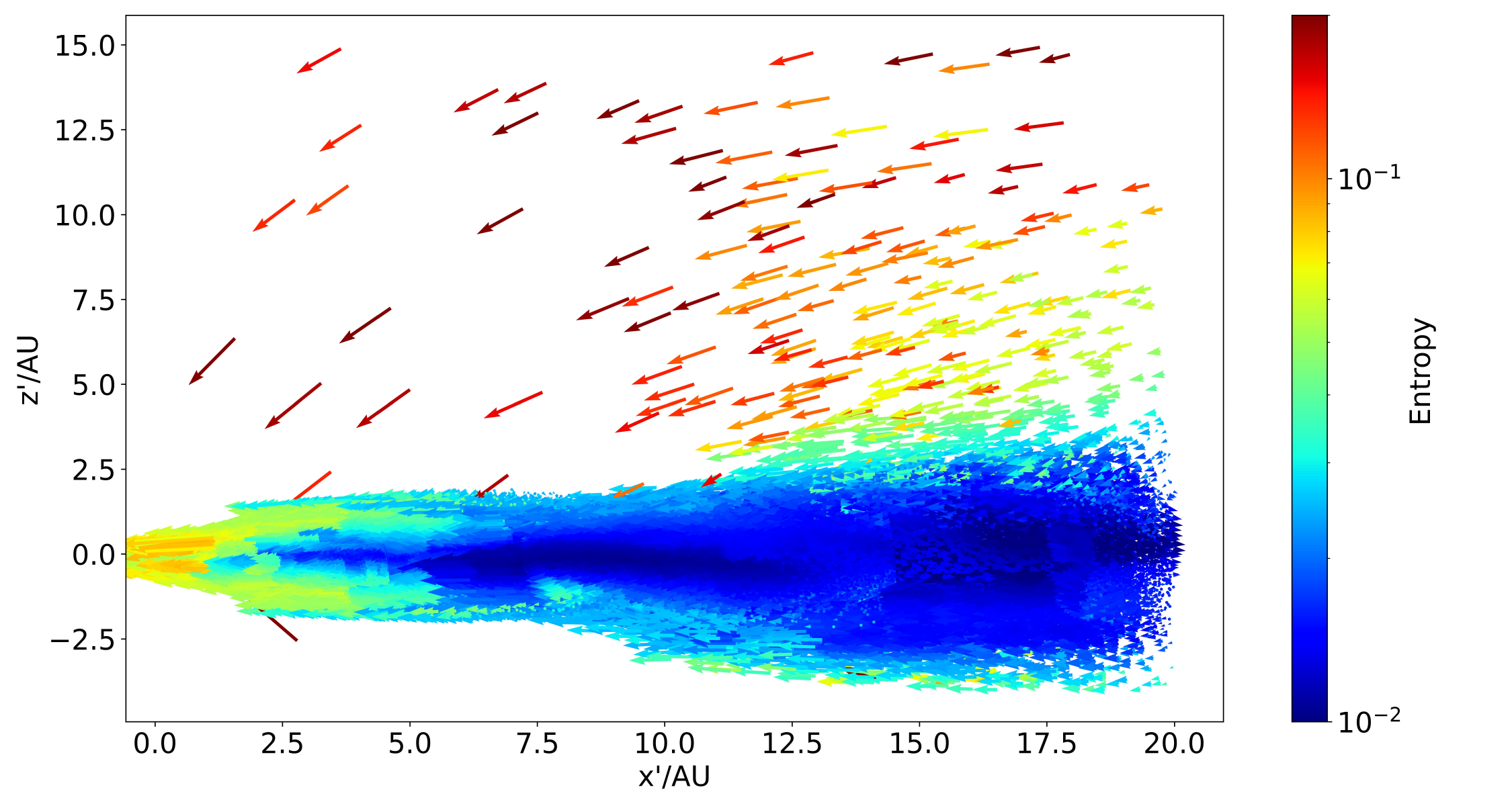}
\includegraphics[width=0.45\textwidth]{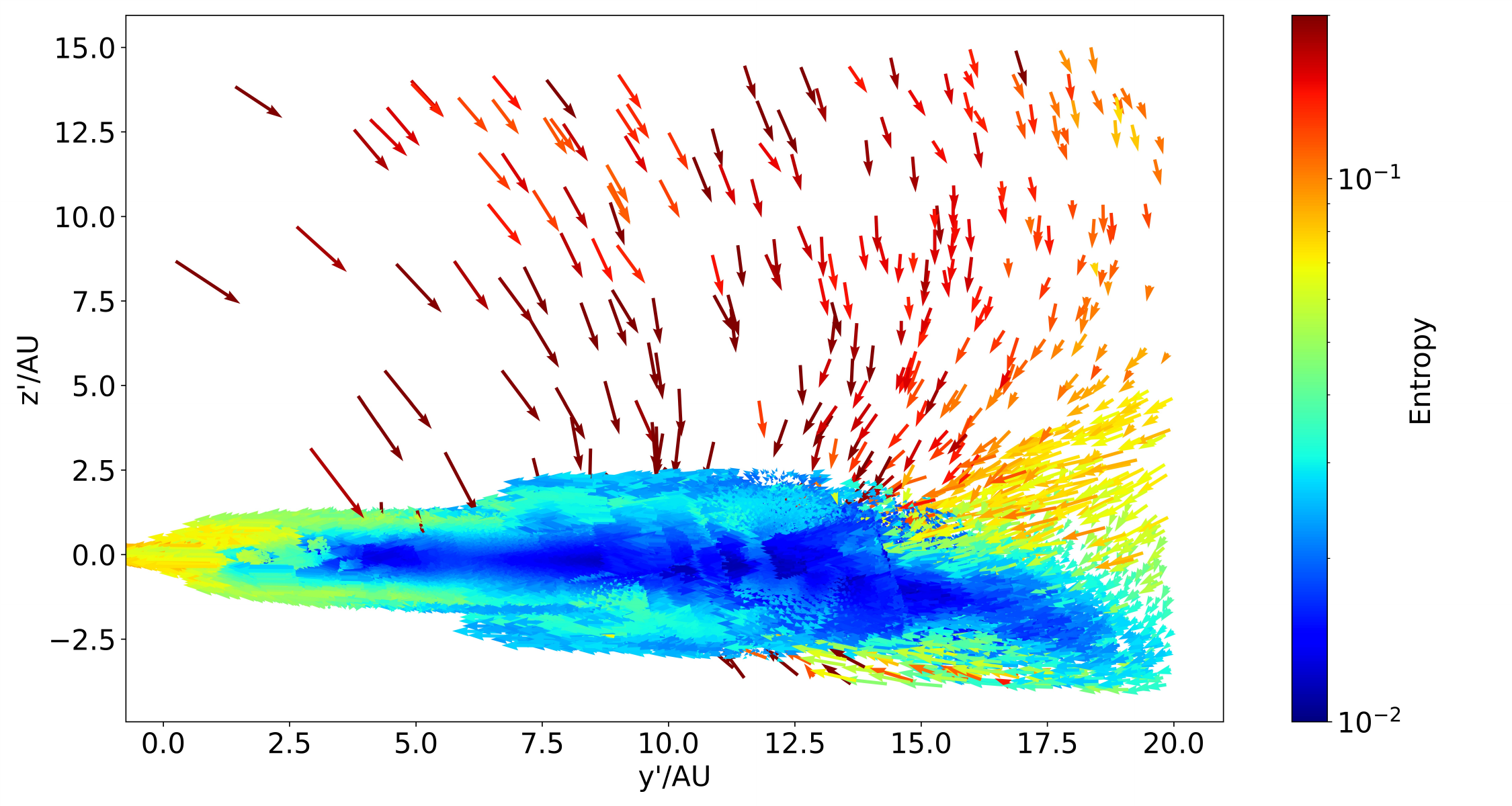}
\caption{Side view of the disk in the $x^\prime-z^\prime$ plane(left) $y^\prime-z^\prime$ plane(right), showing the distribution of entropy overlaid with velocity vectors. The arrows represent the direction and relative magnitude of the velocity field. High-entropy, shock-heated infall flows are visible above and below the disk surfaces in both directions.
\label{fig:entropyvector}}
\end{figure}

\begin{figure}
\centering
\includegraphics[width=0.45\textwidth]{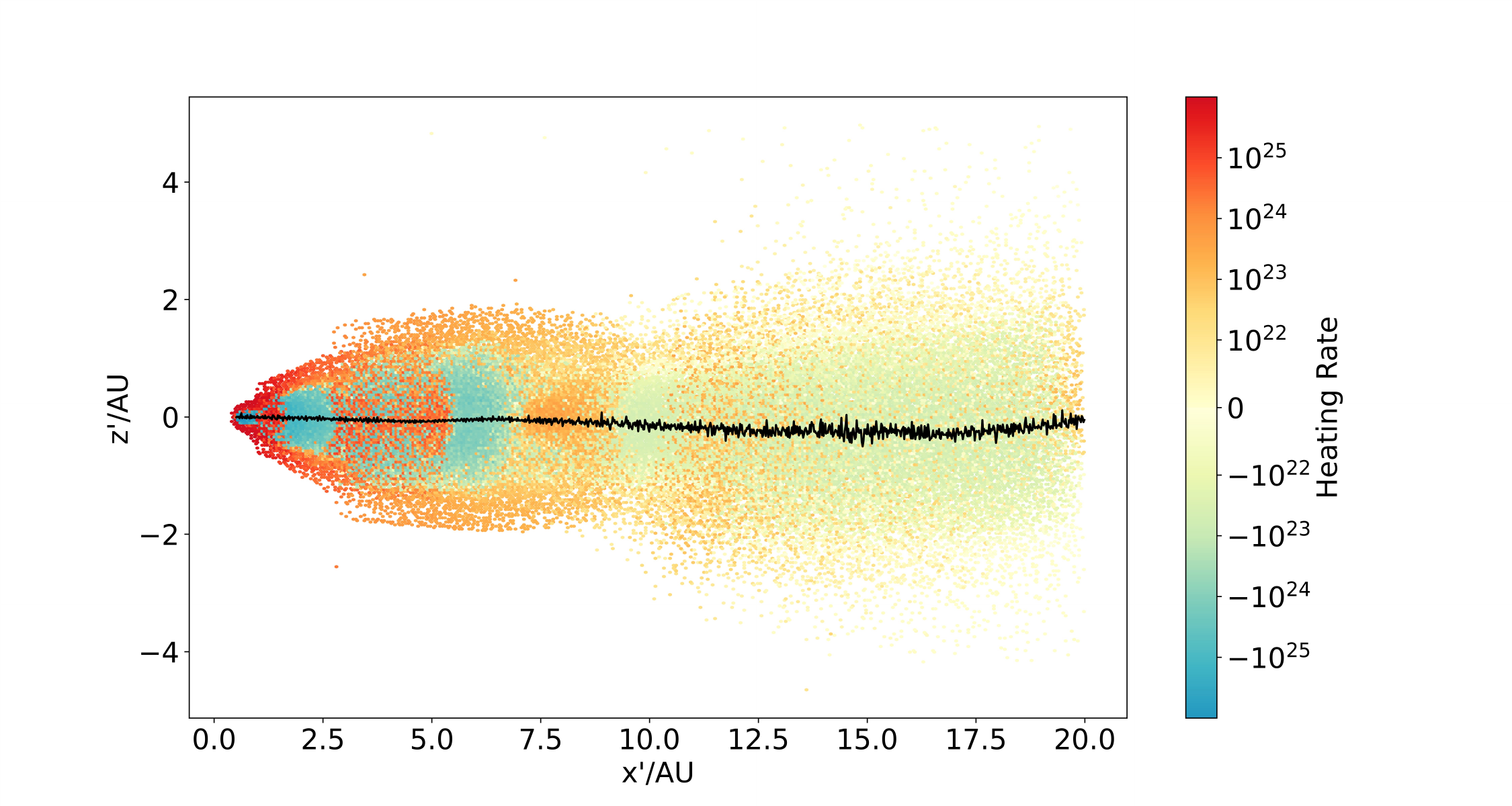}
\includegraphics[width=0.45\textwidth]{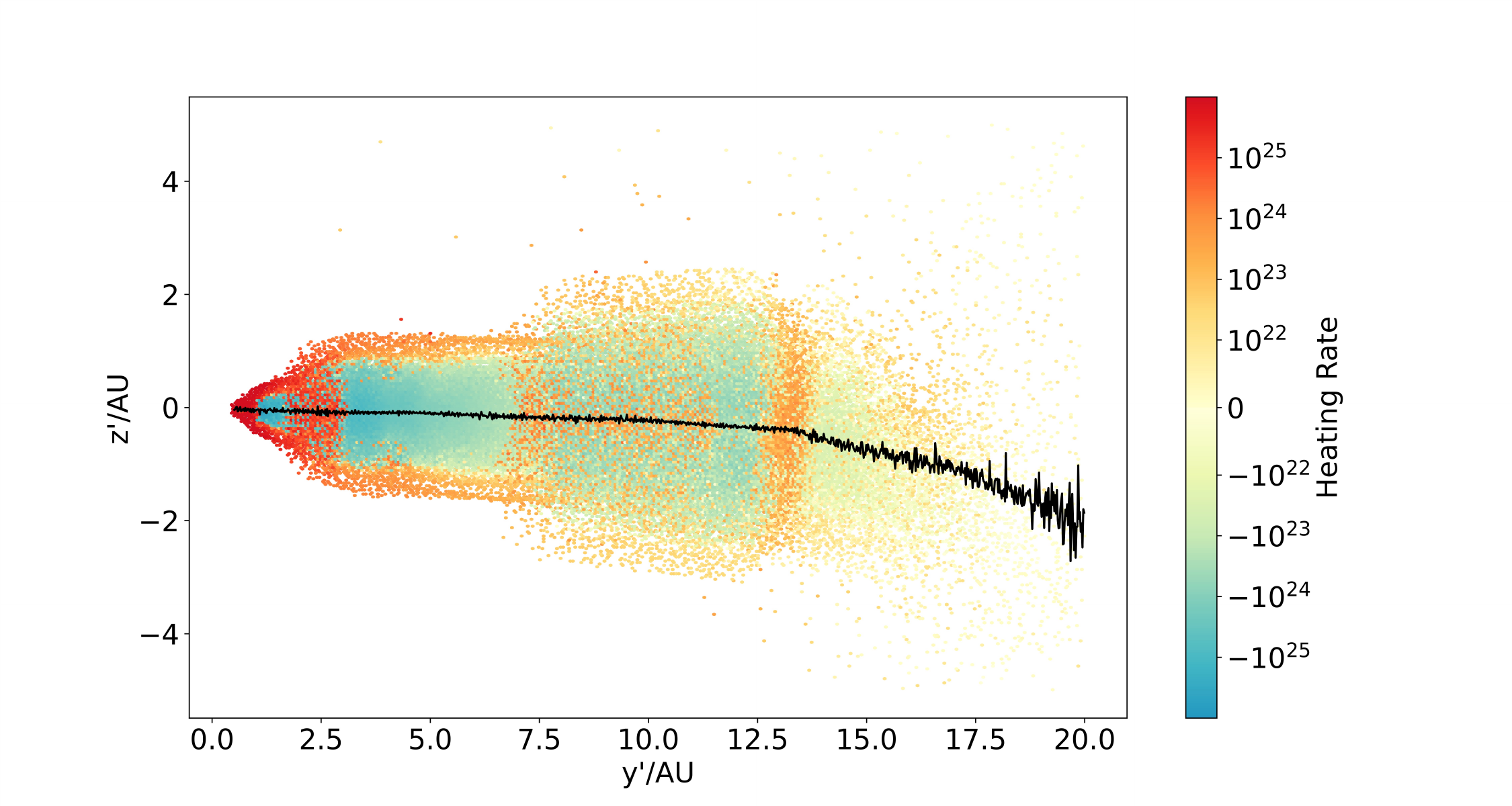}
\caption{Same views as in Fig.~\ref{fig:entropyvector} but showing the volumetric heating rate. Strong heating zones near the disk-envelope interface are seen on both sides of the midplane. The black line indicates the vertical center of mass, which highlights the disk warp in both directions.
\label{fig:heating}}
\end{figure}

\begin{figure}
\centering
\includegraphics[width=0.7\textwidth]{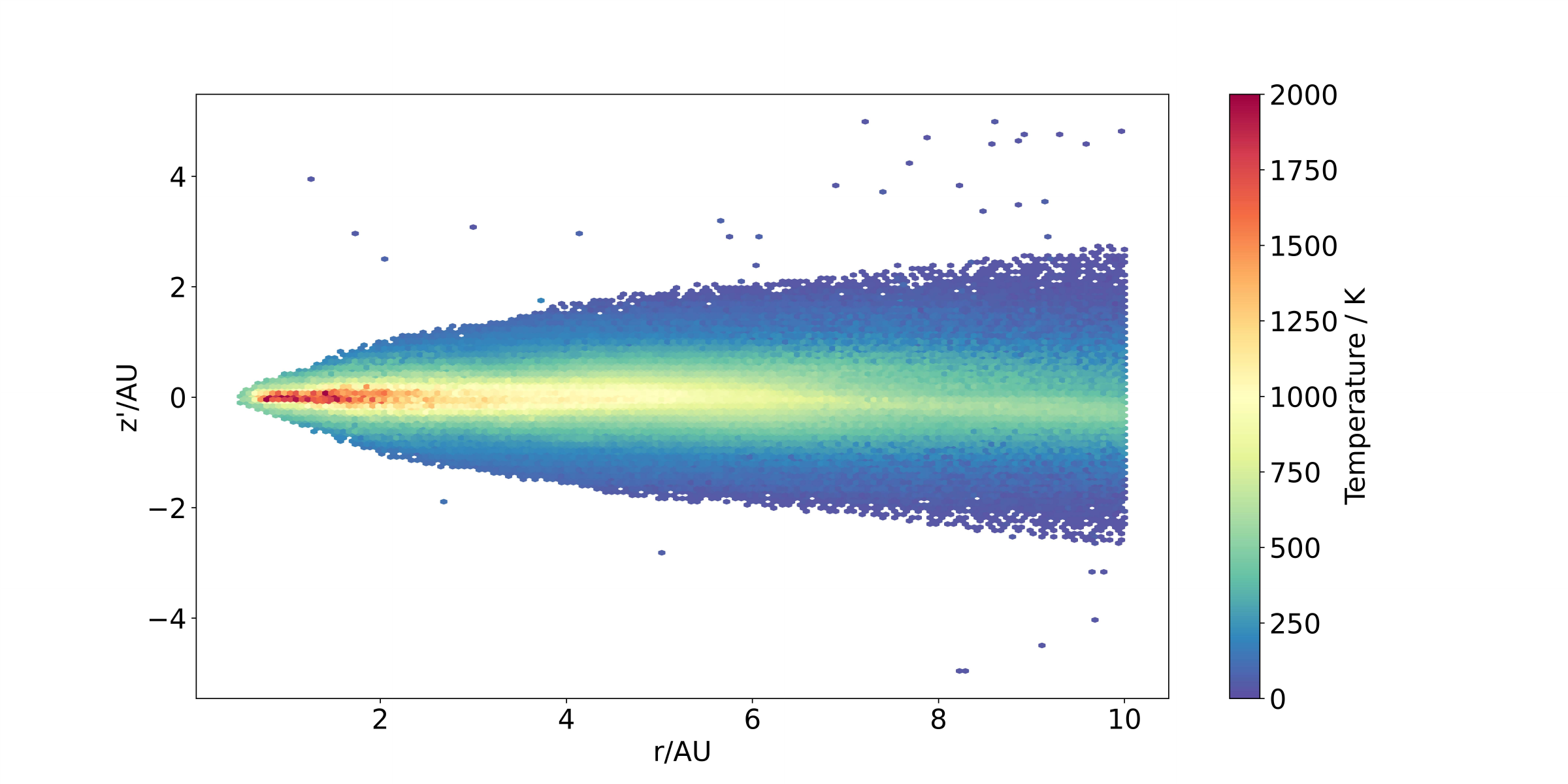}
\includegraphics[width=0.45\textwidth]{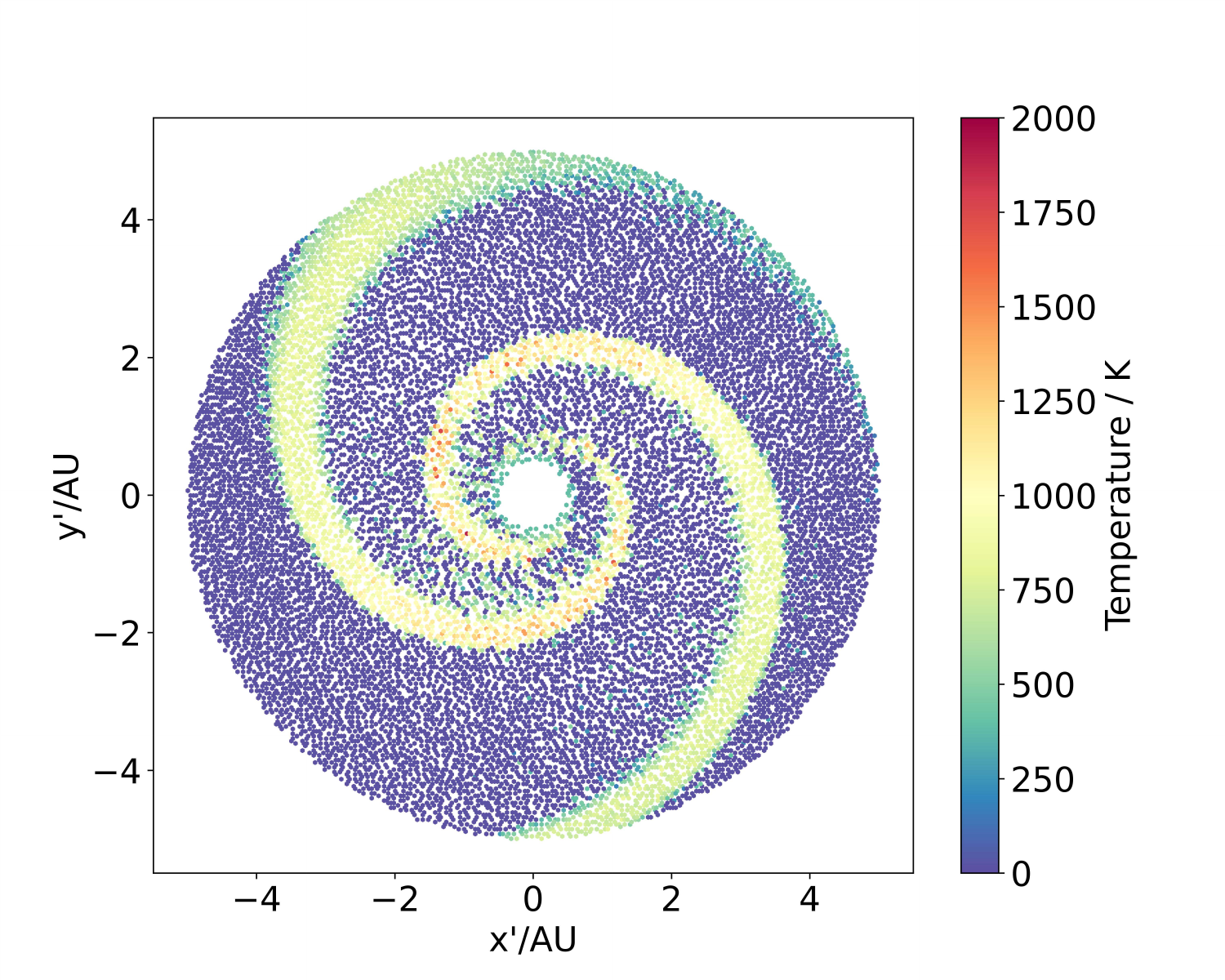}
\includegraphics[width=0.45\textwidth]{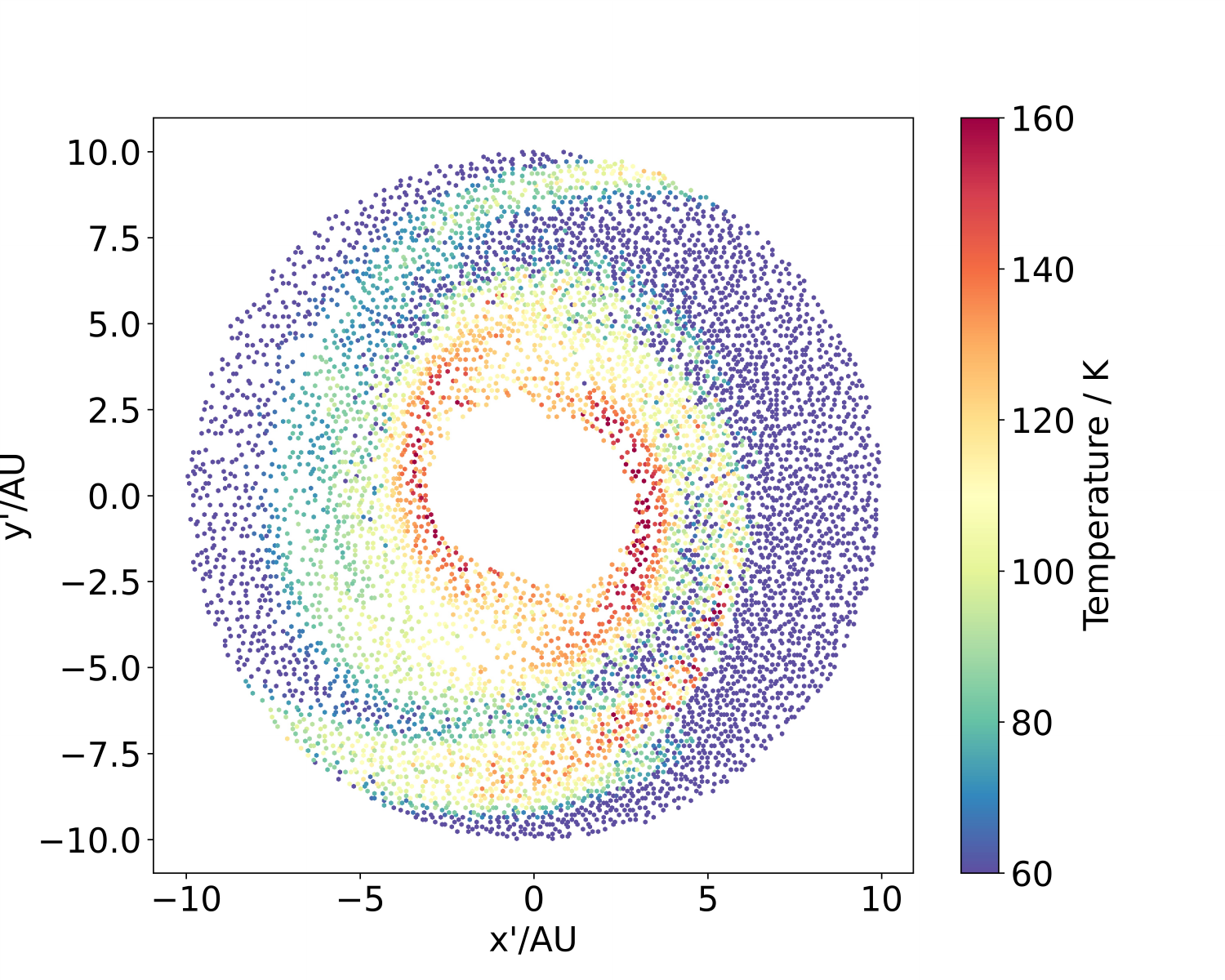}
\caption{Upper: Temperature distribution of the disk in the (r,$z^\prime$) plane. Colors represent gas temperature in Kelvin. The inner disk ($r \lesssim 5\,\mathrm{AU}$) shows elevated temperatures about $2000\,\mathrm{K}$, indicating strong heating near the midplane.
Lower: The left panels display scatter plots of the temperature distribution in the disk's midplane, within a radius of 5 AU and a disk height of 0.1 AU. Note that the temperature distribution is not axi-symmetric. The  high temperature localized around the spiral shock differs from the conventional axi-symmetric midplane temperature in viscous accretion disks.
The right panels display scatter plots of the temperature distribution in the disk's midplane, within a radius of 10 AU and a disk height of 1 $\sim$ 1.1 AU.
The temperature difference between the mid-plane and disk surface reflects  dominant radiative diffusion in the vertical direction while the more smeared out radial and azimuthal temperature distributions highlight modest radiative diffusion in the disk 
plane. The colorbar represents the temperature.
\label{fig:temp_wave}}
\end{figure}

To better investigate the infall-induced heating mechanism, we analyze a vertical slice of the disk 
during the peak heating phase, taken at a representative time 8000 years into the simulation. 
For illustration purpose, we rotate the coordinates such that the $x^\prime-y^\prime$ plane 
corresponds to the inner-disk plane.  We also adopt $r=(x^{\prime 2} + y^{\prime 2})^{1/2}$ to be
the cylindrical radius of the disk.  The left panel of 
Fig.~\ref{fig:entropyvector} presents the gas velocity field projected in the 
$x^\prime$–z$^\prime$ plane, where arrows indicate the direction of motion and their length reflects the local velocity magnitude. The color bar represents the entropy distribution, defined as $s \propto \ln(p/\rho^\gamma)$, where $p$ denotes the gas pressure, $\rho$ the density, and $\gamma$ the adiabatic index. Entropy is used here as a diagnostic to highlight regions of strong entropy variations, which are indicative of irreversible heating such as shocks in the disk.
The right panel shows the same quantities in the $y^\prime-z^\prime$ plane, providing a complementary view of the vertical infall flows. In both panels, the velocity vectors reveal converging motions toward the disk surface. These infalling streams collide with the rotating disk, leading to the formation of strong shocks due to large velocity gradients. As shown in Fig.~\ref{fig:heating}, the panels display the corresponding volumetric heating rate in the $x^\prime-z^\prime$ (left) and $x^\prime-z^\prime$
(right) planes. The black line in each panel marks the center of mass at each radius, from which we observe a noticeable warp in the outer regions of the disk. Strong heating zones are observed at the disk-envelope interface, coinciding with the shock regions inferred from the entropy and velocity structure in Fig.~\ref{fig:entropyvector}. 
As infalling streams impact the rotating disk, strong shocks—particularly infall shocks near the disk surface—form due to large velocity differences. Their spatial coincidence with regions of enhanced heating in the right panel supports the conclusion that shock heating is the primary mechanism by which infall transfers thermal energy into the disk.
In most of the disk midplane, the entropy remains low and relatively smooth, and the local heating rate is negative, indicating heat loss. 
However, we also observe localized zones of high entropy and positive heating near the midplane. The upper panel in Fig.~\ref{fig:temp_wave} shows the temperature distribution of the disk in the (r,$z^\prime$) slice. The highest temperatures are concentrated near the midplane of the inner disk. The inner disk region ($r \lesssim 3\,\mathrm{AU}$) shows elevated temperatures about $2000\,\mathrm{K}$, indicating strong heating near the midplane. This region is likely dominated by local viscous dissipation, as the mass and heat flux across the disk surface are limited in the innermost radii. However, at larger disk radii (approximately $r \gtrsim 5\,\mathrm{AU}$), the vertical temperature gradient becomes non-monotonic: it is positive near the midplane but becomes negative at higher $z^\prime$ values. This gradient reversal suggests less efficient radiative transfer. One possible explanation is that infalling material from the envelope impacts the outer disk and accumulates near the disk surface. As the surface density increases, the material gradually diffuses inward, increasing its surface residence time. This accumulation and redistribution can lead to $PdV$ work and localized heating above the midplane, thereby altering the vertical temperature gradients. Thus, the inner disk shows vertically stratified, monotonic heating, while the outer disk exhibits more complex, layered structures shaped by infall shock and accumulation effects.

\begin{figure}
\centering
\includegraphics[width=0.45\textwidth]{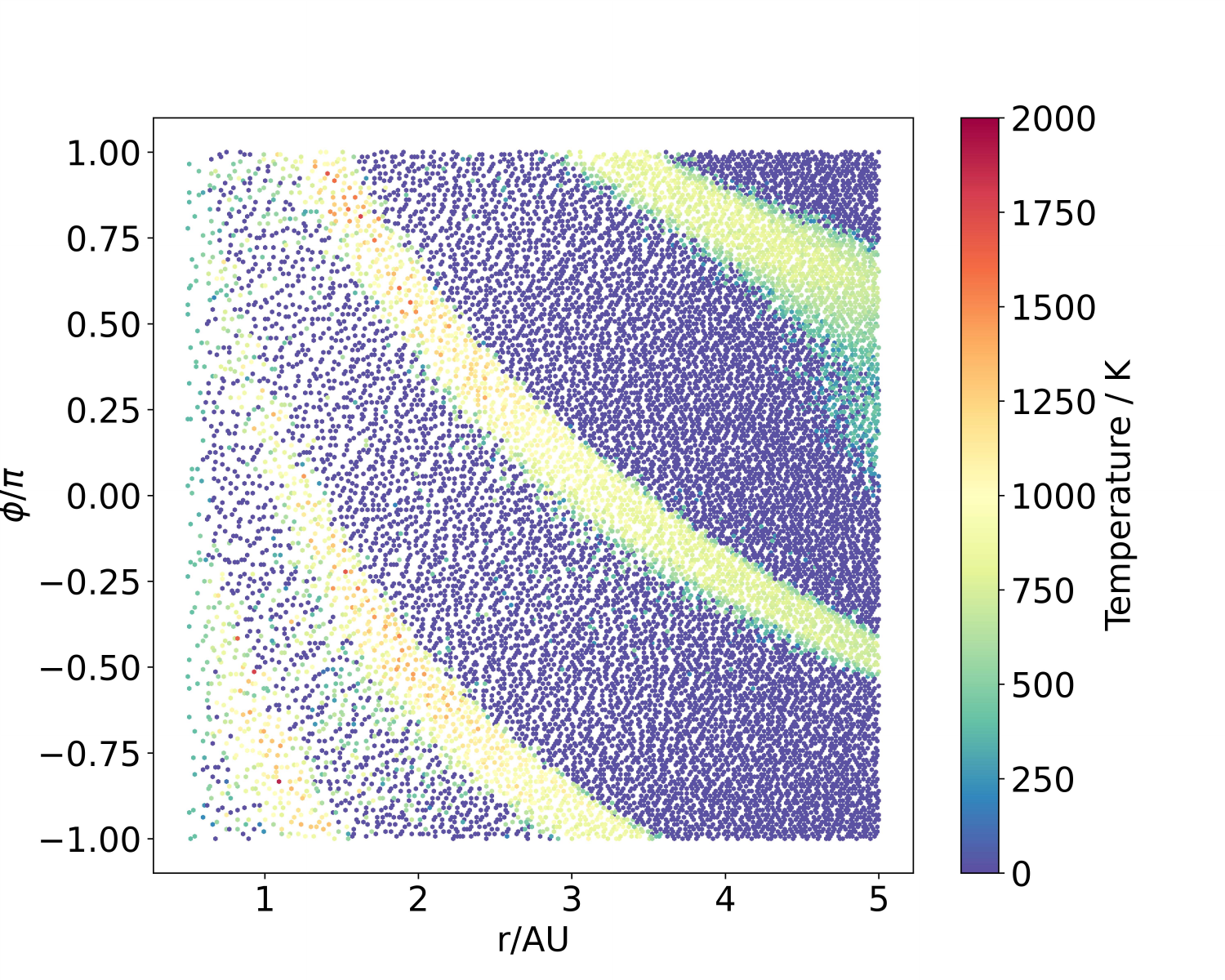}
\includegraphics[width=0.45\textwidth]{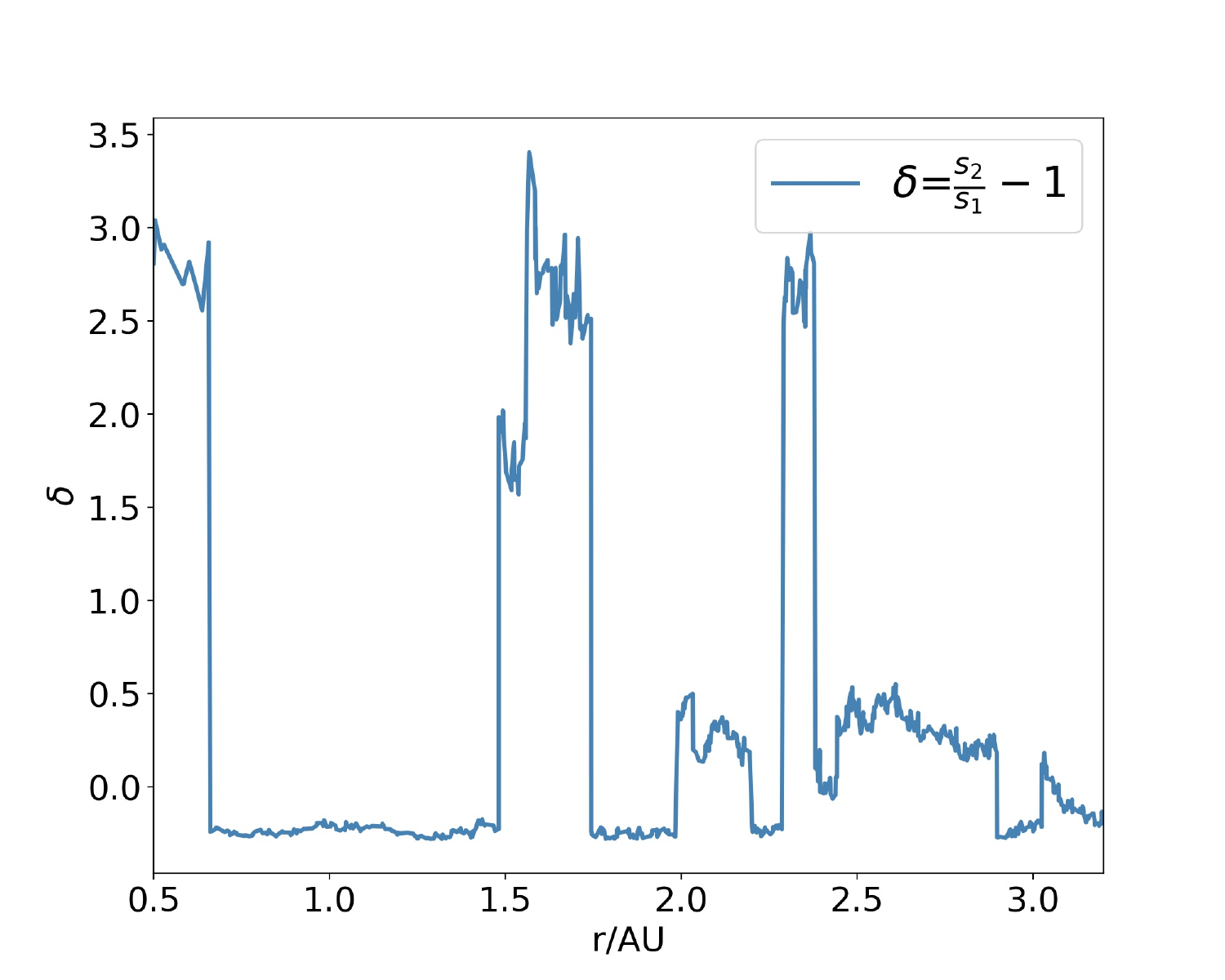}
\caption{The left panel shows the $r - \phi$ distribution of temperature in the midplane, with the same physical parameter settings as the left panel of Fig.\ref{fig:temp_wave}. The right panel shows the radial profiles of the specific entropy jump $\delta$ for an azimuthal slice (-0.05$<\phi<$0.05) of the disk.
\label{fig:entropyjump}}
\end{figure}

To further explore the thermal structure, we present midplane temperature maps in Fig.~\ref{fig:temp_wave} and \ref{fig:entropyjump}, both corresponding to the same time.
The color bars in both figures denote temperature. Fig.~\ref{fig:temp_wave} (lower left panel) shows an 
$x^\prime-y^\prime$ slice taken at a height of 0.1~AU above the midplane. Prominent spiral density waves are visible, and temperature enhancements are found along the spiral arms, with localized hotspots exceeding 1500~K at radii between 2 and 3~AU. 
The lower right panel shows a horizontal slice at 1~AU above the midplane, corresponding to $z^\prime = 1$~AU in Fig.~\ref{fig:heating}. The hottest regions here are located near $x^\prime  = 5$~AU, corresponding to shock heating at the disk surface due to infall. Outside of these regions, no strong spiral features are present, and the temperature is comparatively low.
To quantify the heating along the spiral arms, we further analyze the disk's midplane using $r$--$\phi$ slices of temperature and entropy, as shown in Fig.~\ref{fig:entropyjump}. The entropy gradient is an effective tracer of shock strength. We compute the entropy jump of gas along a narrow azimuthal sector ($-0.05 < \phi < 0.05$), focusing on regions coinciding with spiral density waves. We use $\rho_1$ and $\rho_2$ to represent the densities before and after the shock, and $P_1$ and $P_2$ are the corresponding pressures. The dimensionless entropy jump is denoted as $\delta$ 
\begin{equation}
\delta \equiv \frac{s_2}{s_1} - 1 \propto 
\frac{ \ln\left( \dfrac{p_2 \rho_1^\gamma}{p_1 \rho_2^\gamma} \right) }
     { \ln\left( \dfrac{p_1}{\rho_1^\gamma} \right) }.\label{eq7}
\end{equation}
In the right panel of Fig.~\ref{fig:entropyjump}, three distinct peaks in entropy jump are evident, which align with the locations of the spiral arms in the left panel. This confirms that shock heating also dominates within the midplane and is closely associated with spiral wave structures.

\subsubsection{Spiral structure}
To diagnose the spiral-wave structure in the disk, we performed a set of analysis measurements. First, we computed the axi-symmetric gravitational stability parameter (\citealp{safronov1960, Toomre1964})
\begin{equation}
\
Q=\frac{c_s\,\kappa}{\pi G\,\Sigma},\label{Toomre}
\
\end{equation}
where \(c_s\) is the sound speed, \(\kappa\) the epicyclic frequency, and \(\Sigma=\int \rho\,dz\) the surface density. We also quantified angular-momentum transport by measuring the Reynolds and gravitational stresses,
\[
H_{r\phi}=\rho\,\delta v_r\,\delta v_\phi,\qquad
G_{r\phi}=\frac{g_r\,g_\phi}{4\pi G},
\]
with \(\delta v_r\) and \(\delta v_\phi\) the fluctuations of radial and azimuthal velocities, and \(g_r,\,g_\phi\) the radial and azimuthal components of the self-gravitational acceleration (\citealp{Lynden1972}). The effective viscosity parameter \cite{Shakura1973}
associated with these stresses are
\begin{equation}
\
\alpha_{\rm tot} = \alpha_{\rm R} + \alpha_{\rm G} \ \ \ \ \ \ {\rm with} \ \ \ \ \ \
\alpha_{\rm R}=
\frac{\langle H_{r\phi} \rangle_{V}}{\langle P \rangle_{V}}
\ \ \ \ \ \ {\rm and} \ \ \ \ \ \ 
\alpha_{\rm G}=
\frac{\langle G_{r\phi} \rangle_{V}}{\langle P \rangle_{V}}
\label{alpha}
\
\end{equation}
where \(P\) is the gas pressure \citep[see also][]{Ni2025}, and the volume average for any variable $X$ is
\begin{equation}
\
\langle X \rangle_{V}
= \frac{\sum_{i} X_{i}/\rho_{i}}{\sum_{i} 1/\rho_{i}}\,
\
\end{equation}

\begin{figure}
\centering
\includegraphics[width=0.45\textwidth]{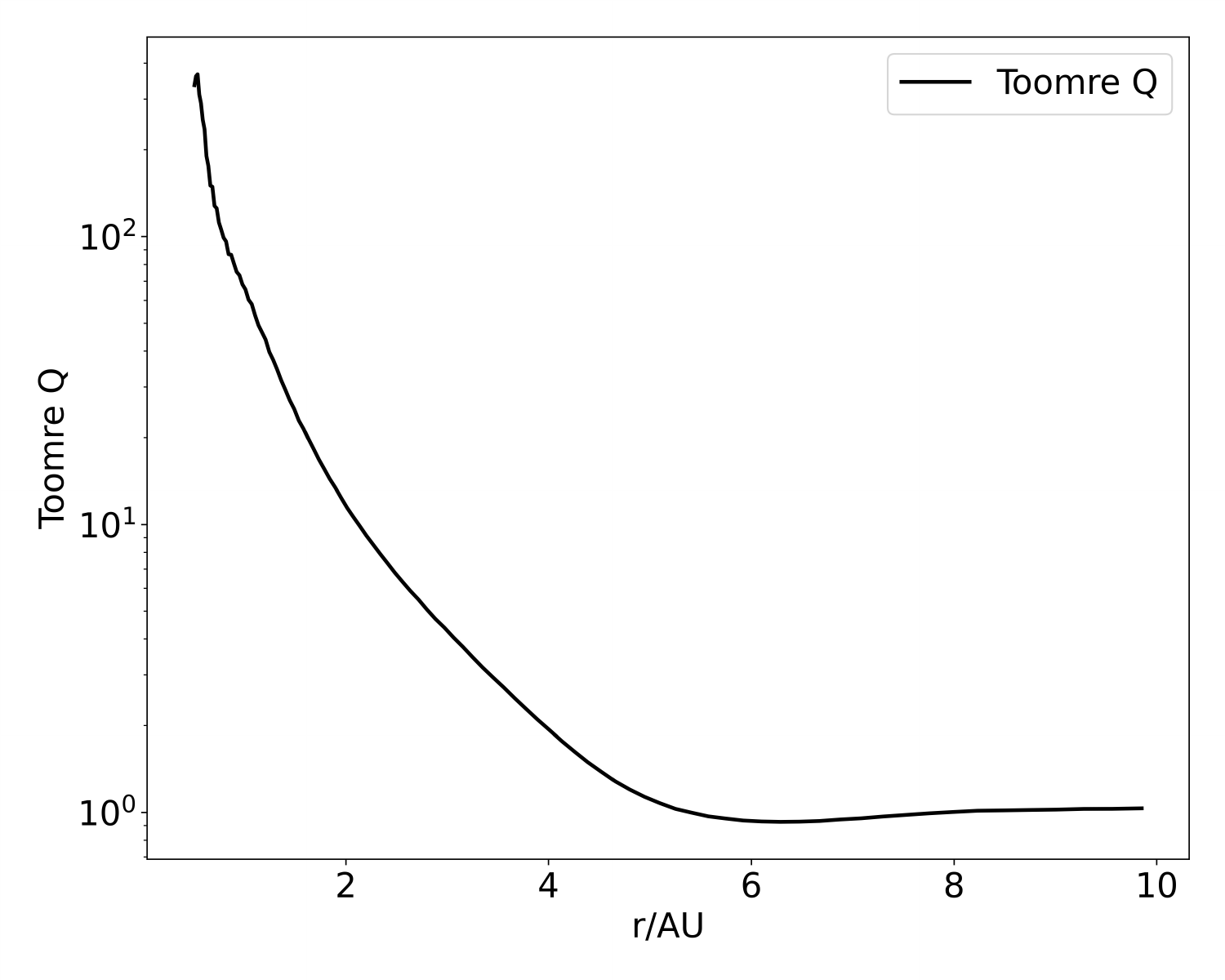}
\includegraphics[width=0.45\textwidth]{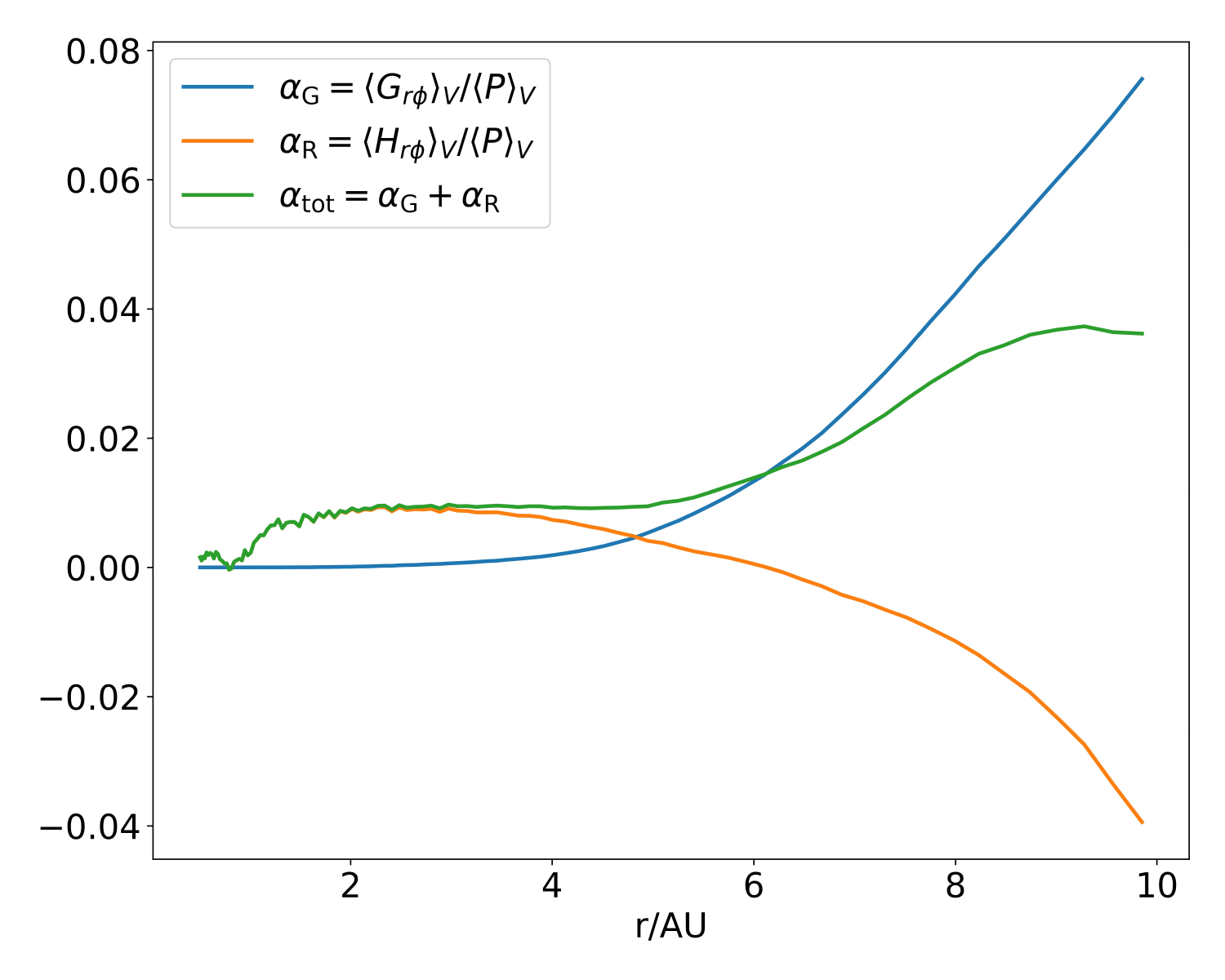}
\caption{The left panel shows the profile of the Toomre number Q maps. The right panel shows the time-averaged stress profiles \(\alpha_G\), \(\alpha_R\), and \(\alpha_{\rm tot}\).
\label{fig:spiralwave}}
\end{figure}
The left panel of Fig.~\ref{fig:spiralwave} shows the radial profile of the Toomre number \(Q\); beyond \(r\sim4~\mathrm{AU}\) the disk hovers near marginal stability with \(Q\approx1\). The right panel presents the time-averaged stress coefficients and reveals a clear transition in the transport regime: inside \(\sim5~\mathrm{AU}\) the elevated midplane temperature renders the Reynolds stress dominant, whereas outside \(\sim5~\mathrm{AU}\) the Reynolds stress declines with radius and eventually becomes negative while the gravitational stress rises rapidly and assumes the leading role. In this context, a negative Reynolds stress corresponds to inward angular momentum transport. The crossover radius coincides with the annulus where \(Q\sim1\), and the outer disk exhibits a well-defined trailing-spiral pattern that tracks the maxima of \(\alpha_G\). Despite \(\alpha_R<0\) at large radii, the total transport \(\alpha_{\rm tot}=\alpha_G+\alpha_R\) remains positive and reaches values of a few \(\times10^{-2}\), indicating a robust outward flux of angular momentum that sustains accretion. These trends are consistent with a self-regulated state, in which the disk remains near marginal gravitational stability and redistributes angular momentum primarily via gravity-wave–mediated torques in the cooler outer regions, while locally thermally driven (Reynolds) stresses dominate in the warmer inner disk, in line with gravitational self-regulation frameworks for protostellar disks \citep{xu2021formation}.

\subsection{Run 2 without infall} \label{subsec:run2}

\begin{figure*}
\centering
\includegraphics[width=0.45\textwidth]{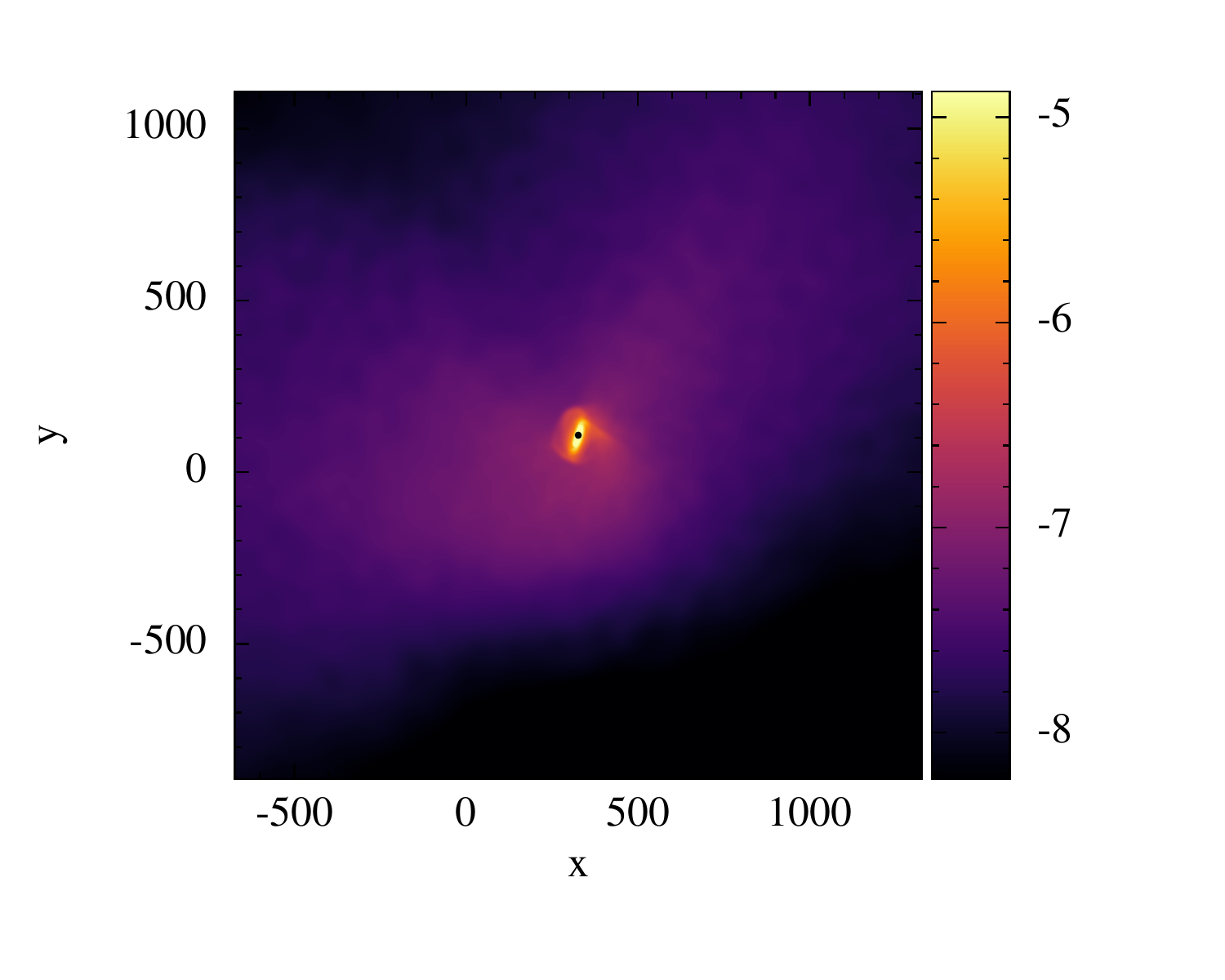}
\includegraphics[width=0.45\textwidth]{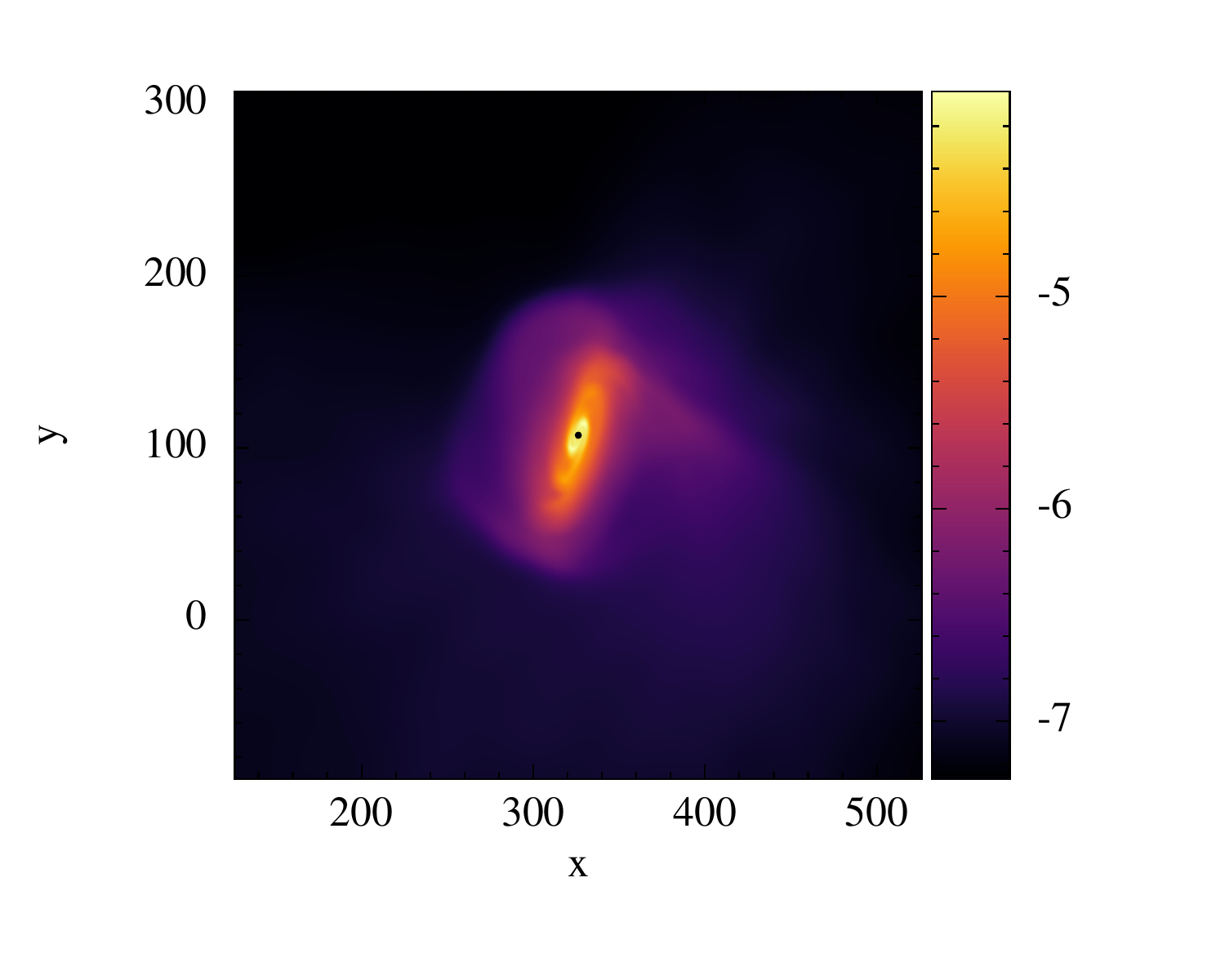}
\includegraphics[width=0.45\textwidth]{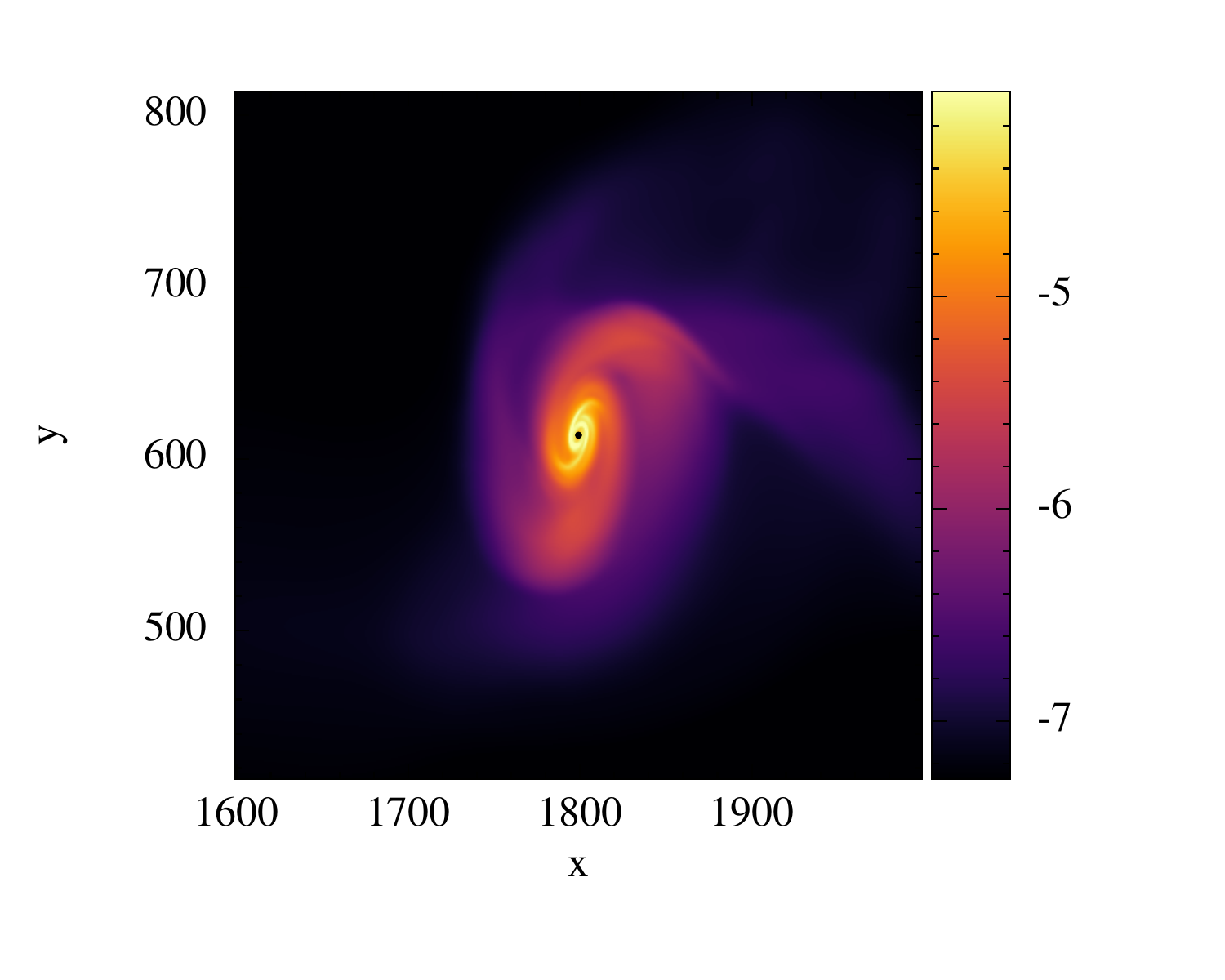}
\includegraphics[width=0.45\textwidth]{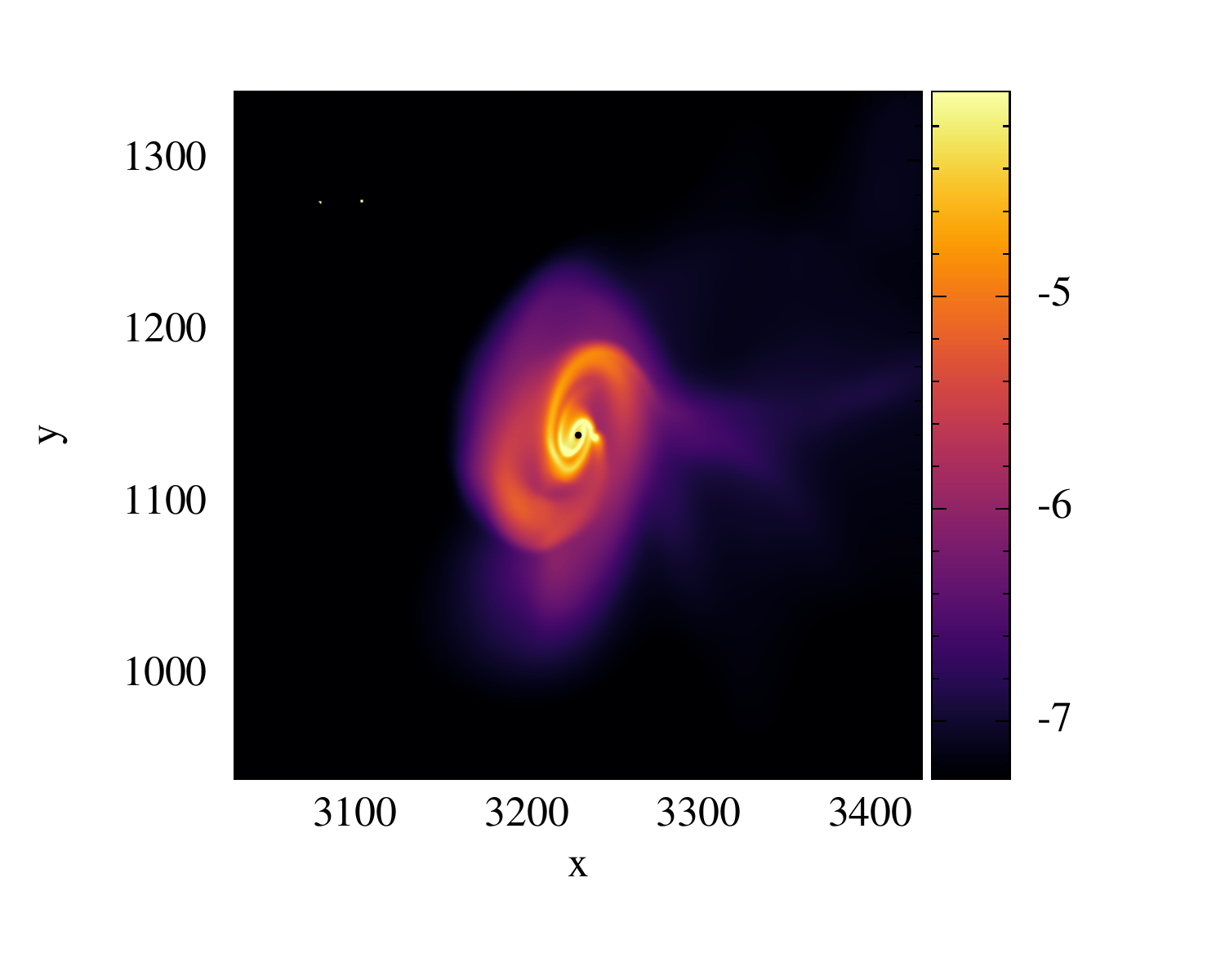}
\caption{The figures show snapshots of the system's global evolution. The first panel in the top left show the initial density distributions of Run 2 with the box size of 2000 AU, while the subsequent three images cover episodes 1,000, 8,000, and 15,000 yrs. 
The box size for the last three snapshots is 400 AU. The colorbar indicates the column density (in unit of $M_\odot/\mathrm{AU}^2$) obtained by integrating the gas density along the $z$-axis over 400 AU.
\label{fig:general2}}
\end{figure*}

\begin{figure}
\centering
\includegraphics[width=0.7\textwidth]{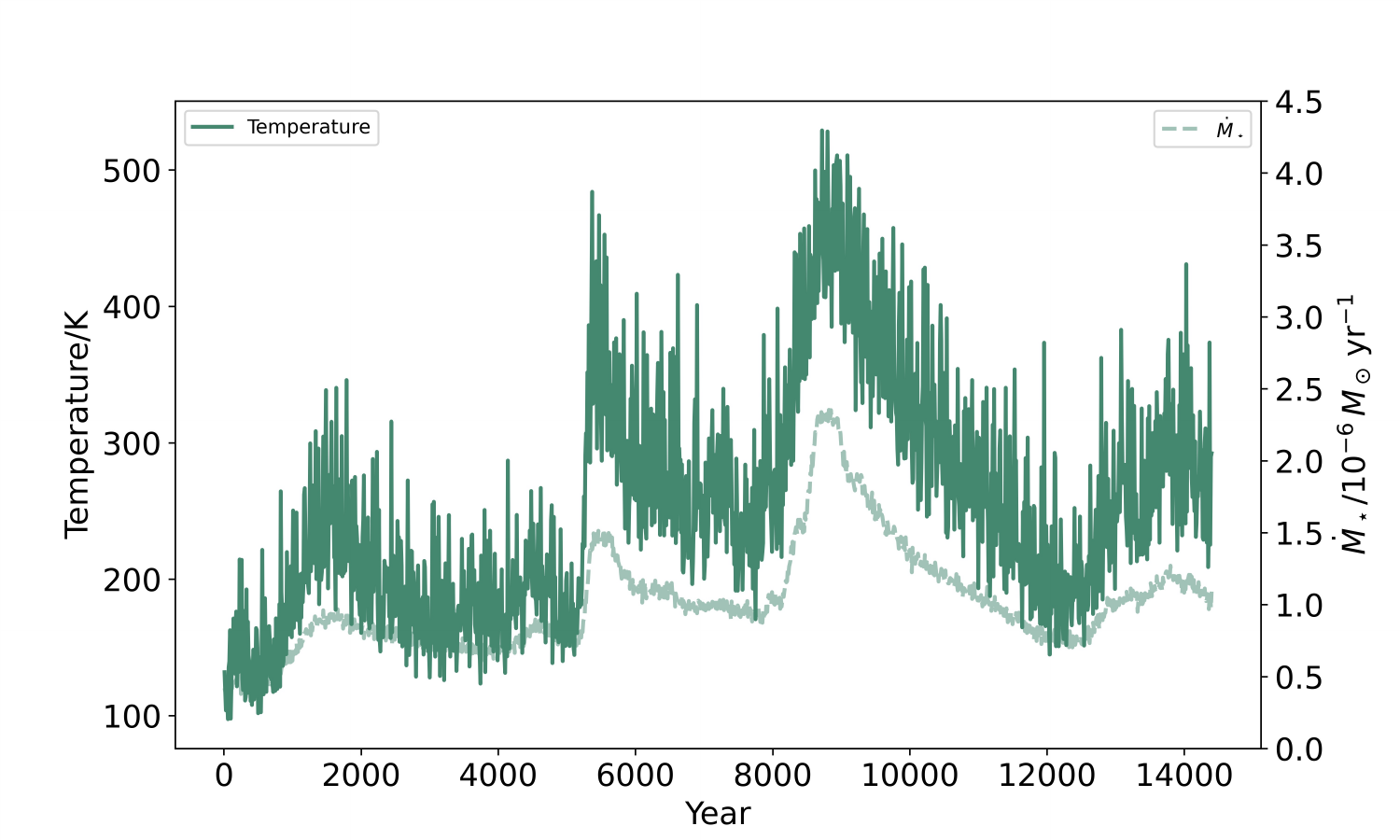}
\caption{This panel, similar to Fig.~\ref{fig:temp53}, shows the evolution for Run 2.
\label{fig:temp5}}
\end{figure}

\begin{figure}
\centering
\includegraphics[width=0.45\textwidth]{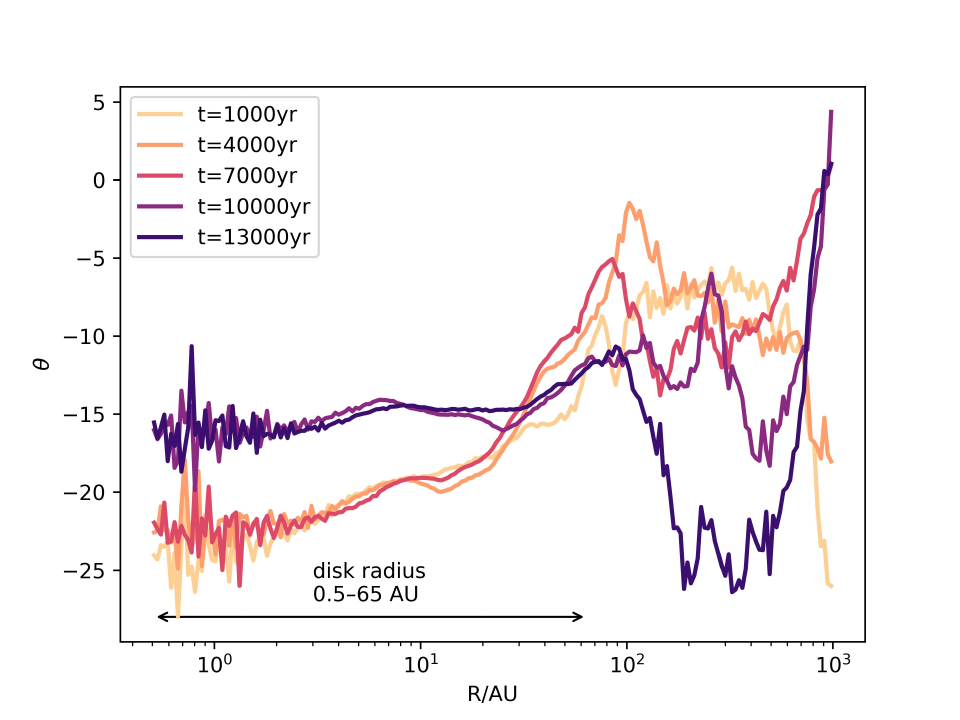}
\includegraphics[width=0.45\textwidth]{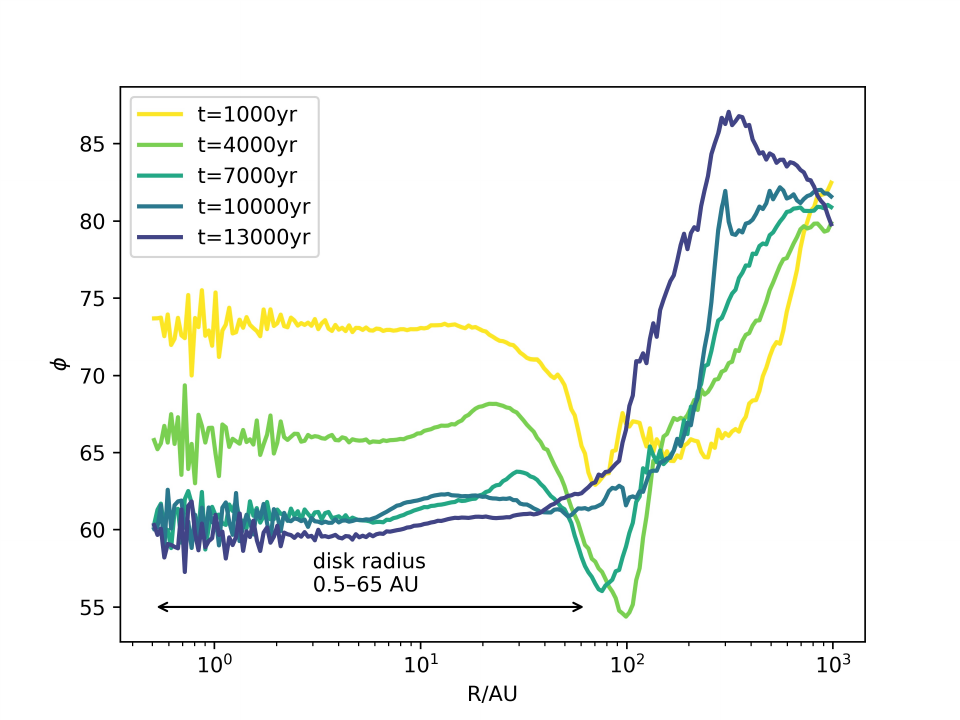}
\caption{The plots show the direction of specific angular momentum at various radius and times for Run 2. The left panel represents the polar angle $\theta$, while the right panel shows the azimuthal angle $\phi$. The horizontal marker at 0.5–65 AU indicates the average disk radius across different times; the actual disk radius varies with time.
\label{fig:angularmomentumrun2}}
\end{figure}

To further understand the role of infall in shaping the thermal and dynamical evolution of the disk, we performed a second simulation (Run 2) with two differences: a slightly smaller central stellar mass and the absence of significant vertical gas infall from the streamer. This setup serves as a comparative baseline to highlight the effects observed in Run 1.
Fig.~\ref{fig:general2} shows the time evolution of the disk structure in Run 2. Compared to Run~1, the system exhibits a more gradual evolution. No prominent tilt or large-scale reorientation of the disk is observed. As shown in Fig.~\ref{fig:angularmomentumrun2}, the radial variation in disk orientation remains moderate, with tilt angles typically confined to the range of $10^\circ$--$20^\circ$ throughout most episodes.
The density distribution evolves smoothly, without evident signs of turbulence-induced irregularities or fragmentation.
The angular momentum structure of the disk, shown in Fig.~\ref{fig:angularmomentumrun2}, further supports the overall stability of the system. Over time, the angular momentum distribution remains relatively smooth, with no significant changes in orientation or magnitude. In particular, the polar angle ($\theta$) difference between the inner and outer disk angular momentum vectors remains small—typically within $10^\circ$--$20^\circ$ over 15,000 years—and is relatively small compared to the deviations observed in Run~1. This suggests that the disk in Run~2 maintains a more stable structure with less internal warping.
There is no signature of the strong angular momentum cancellation or mixing that characterized Run~1 during the main heating phase. The thermal evolution of the disk is presented in Fig.~\ref{fig:temp5}, which shows the maximum disk temperature and stellar accretion rate over time. While both runs exhibit variations within a factor of 2, the absolute temperature change in Run~1 is more dramatic, increasing from $\sim$1000\,K to 2000\,K, whereas in Run~2 it remains in a lower range, from $\sim$300\,K to 600\,K. This suggests that the heating in Run~2 is comparatively milder in both amplitude and energetic significance.
The maximum disk temperature increases slowly to just over 400~K, and the stellar accretion rate $\dot{M}_\star$ remains about $1 \times 10^{-6}\,M_\odot\,\mathrm{yr}^{-1}$ throughout the simulation.

Overall, Run 2 serves as a control experiment demonstrating that, in the absence of vertical infall, the disk maintains a stable, low-temperature environment. The lack of strong angular momentum perturbations and external heating allows the system to evolve gradually and efficiently radiate away internal heat. This comparison strengthens the conclusion that the intense, transient heating and the prospect of efficient pre-solar grain disruption observed in Run 1 is not an inherent outcome of disk evolution but rather a direct consequence of infall events and their associated dynamical disturbances.

\subsection{Comparison of Disk Heating with and without Infall} \label{subsec:Comparison}

\begin{figure}
\centering
\includegraphics[width=0.7\textwidth]{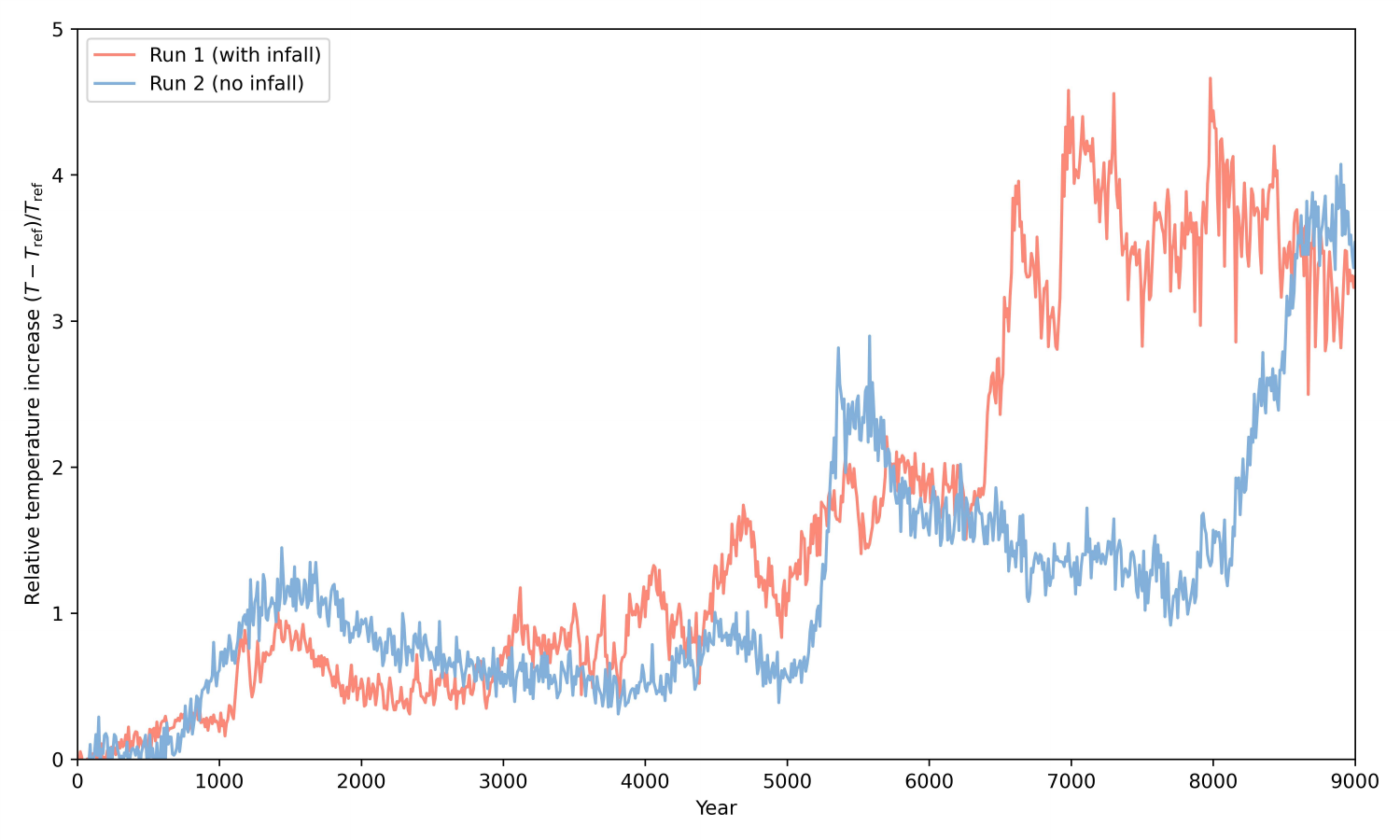}
\caption{Evolution of the relative temperature increase, $(T-T_{\rm ref})/T_{\rm ref}$, for Run 1 (with infall; red) and Run 2 (without infall; blue). The presence of infall in Run 1 leads to a distinct, sustained heating phase at $t\sim6$--$9\times10^3$ yr, while Run 2 remains comparatively cool throughout the evolution.
\label{fig:relativetemp}}
\end{figure}

To improve clarity and readability, we have added a dedicated summary and comparison in this section, where the key similarities and differences
between Run 1 and Run 2 are explicitly discussed and synthesized. This
side-by-side comparison provides a clear overview of the relative heating
efficiency in the two runs and highlights the role of infall in shaping the
thermal evolution of the disk.

We begin by comparing the temporal evolution of the disk temperature using a
relative measure, which allows a direct assessment of the heating response while minimizing uncertainties associated with the absolute temperature normalization. Fig.~\ref{fig:relativetemp} shows the relative temperature increase, $(T-T_{\rm ref})/T_{\rm ref}$, as a function of time for Run 1 (with infall) and Run 2 (without infall). $T_{\rm ref}$ is the temperature at $t = 200\,\mathrm{yr}$.

At early times ($t \lesssim 5\times10^3$ yr), both runs exhibit moderate temperature variability, indicating broadly similar thermal behaviour during the initial disk evolution. However, their subsequent thermal evolution diverges markedly. In Run 1, a pronounced main heating phase emerges between 6000 and 9000 yr, during which the relative temperature increase rises rapidly and reaches values of $(T-T_{\rm ref})/T_{\rm ref} \gtrsim 3$--4. This corresponds to a several-fold enhancement compared to the early disk state and coincides with the period of strongest mass infall.

The elevated temperature level in Run 1 is sustained over several thousand years, indicating that the heating is not driven by short-lived fluctuations
but instead reflects a prolonged, infall-dominated heating regime. Such a
persistent high-temperature phase is not observed in the absence of infall.
By contrast, Run 2 exhibits only modest temperature enhancement throughout the
simulation. Although transient increases are present, the relative temperature
increase generally remains below $\sim2$ and lacks an extended high-temperature
plateau. This behaviour is consistent with the absence of external mass infall
and suggests that intrinsic disk processes alone are insufficient to generate
the strong and sustained heating seen in Run 1.

Overall, the comparison demonstrates that infall fundamentally alters the thermal evolution of the disk by driving a distinct main heating phase. Even
when expressed in terms of relative temperature enhancement, which minimizes
the uncertainties associated with the absolute heating and cooling
prescriptions, Run~1 shows a substantially higher and longer-lasting thermal
response. This provides robust evidence that infall significantly enhances
the efficiency of disk heating and enables thermal conditions that cannot be
achieved in an isolated disk.

\section{Conclusion and Discussion} \label{sec:conclusions}
In this work, we have conducted high-resolution hydrodynamic simulations to investigate the impact of infall from streamers on the early thermal evolution of circumstellar disks. Our results demonstrate that during its formative stage, infall plays a central role in heating the disk, both globally and locally, and strongly modulates the disk’s thermal structure.

The disk undergoes a clear three early-stage evolution. During the initial infall phase 
($t \lesssim 6000$~yr), streamer material gradually accretes onto the disk, leading 
to a steady but mild increase in disk mass, temperature, and stellar accretion rate. 
In this phase, shock heating dominates, and the overall disk structure remains geometrically thin.

In the subsequent main heating phase ($6000 \lesssim t \lesssim 9000$~yr), 
infall becomes more intense, with gas carrying diverse angular momentum orientations interacting with the disk surface. This interaction leads to substantial angular momentum redistribution, disk reorientation by nearly $30^\circ$, and a temporary reduction in disk size and radiative cooling efficiency. The combined effects of strong infall and reduced cooling result in a rapid temperature rise, with local values reaching $\sim$2000~K. This global heating is primarily driven by shock dissipation at the disk surface, as evidenced by localized peaks in entropy and volumetric heating rate, coincident with infall flows in vertical slices. This is attributable to the inefficient cooling of the dense, vertically thickened disk. Furthermore, in the disk midplane, we identify spiral shock structures that also contribute to sustained localized heating via shock compression, as shown by the strong entropy jumps aligned with spiral arms.

In the final cooling phase ($t \gtrsim 9000$~yr), the infall rate declines. The disk gradually re-expands and becomes thinner, restoring more efficient radiative cooling. As a result, the maximum disk temperature slowly decreases to $\sim$1000~K, and the stellar accretion rate also declines slightly.

These results highlight that infall is not merely a mass source for disk growth but also a significant driver of thermal evolution. Through a combination of surface shock heating, angular momentum disturbance, and structural modification, infall regulates the disk's capacity to cool and thus shapes its long-term temperature profile.

Our findings also provide a theoretical framework consistent with meteoritic evidence. The formation of calcium-aluminum-rich inclusions (CAIs) in the early Solar System is thought to have occurred in a narrow temperature range around 1700–2000~K and within $\sim$1–2~AU of the Sun. These inclusions require both a transient global heating event and rapid outward transport or cooling to preserve their mineralogical integrity. The infall-driven heating mechanism identified in Run 1 of our simulations naturally produces such conditions: a rapid, system-wide temperature increase triggered by infall material, followed by gradual cooling as infall subsides and the disk re-expands. This process may provide a viable pathway for the transient thermal environments necessary to form and preserve CAIs in the early stages of disk evolution.  

Nevertheless, the results of run2 indicate that
the extensive heating in the inner disk region and sublimation of pre-solar grains out to several AU are not inherent outcomes of protoplanetary disk formation. These two models are typical outcomes of stellar cluster formation \citep{bate1995modelling, bate2010chaotic}.  They also represent the diverse observed boundary conditions including the disk fraction, proximity to 
and persistence of supernova progenitors \citep{forbes2021solar, ratzenbock2023}.  The  pre-collapse clouds in these regions contain CAIs with a wide spread of condensation age. The diverse survival probability of pre-existing refractory grains in run 1 and 2 implies that only a fraction of the emerging planetary systems and debris disks may contain single-population CAIs in chondritic meteorites out to several AUs.

It is important to emphasize that the current study primarily focuses on the initial phase of thermal evolution, particularly the rapid temperature rise associated with the early infall episode. At the end of our simulations, the central star has accreted only $\sim 0.2\,M_\odot$. Consequently, the results presented here pertain to the very early stage of disk formation, when the infall is strongest and its impact on heating is most pronounced.
In this early phase, the disk mass is comparable to that of the central star, such that self-gravity is expected to play an important role and the disk may be in a gravitationally unstable state. In such environments, turbulent transport can efficiently redistribute solid material. Previous studies of gravitationally unstable disks have shown that strong turbulent diffusion can transport a fraction of thermally processed grains from the inner disk to significantly larger radii, including distances relevant for the formation of chondritic parent bodies and comets \citep{zhou2022turbulent}. Material that is redistributed to the outer disk can remain there over longer evolutionary timescales, rather than being rapidly accreted onto the central star.
In future work, we will extend our analysis to later evolutionary stages, up to $\sim1\times10^5$ years, during which the central star grows to $\sim1~M_\odot$. This will allow us to investigate how the disk's temperature evolves over time and assess whether it remains within the $1500$--$2000~\mathrm{K}$ range required to preserve CAIs before their eventual incorporation into chondritic meteorites. Such an extended study will provide deeper insight into the long-term viability of infall-driven heating as a mechanism for both the formation and preservation of high-temperature meteoritic components in protoplanetary disks.

Acknowledgments:
We thank Mathew Bate for providing the results of an extensive set of stellar-cluster formation 
simulations, which we have adopted as initial conditions.  We also thank Joao Alves and Qingzhu Yin 
for useful conversation.

\appendix
\section{Resolution Study}

\begin{figure}
\centering
\includegraphics[width=0.3\textwidth]{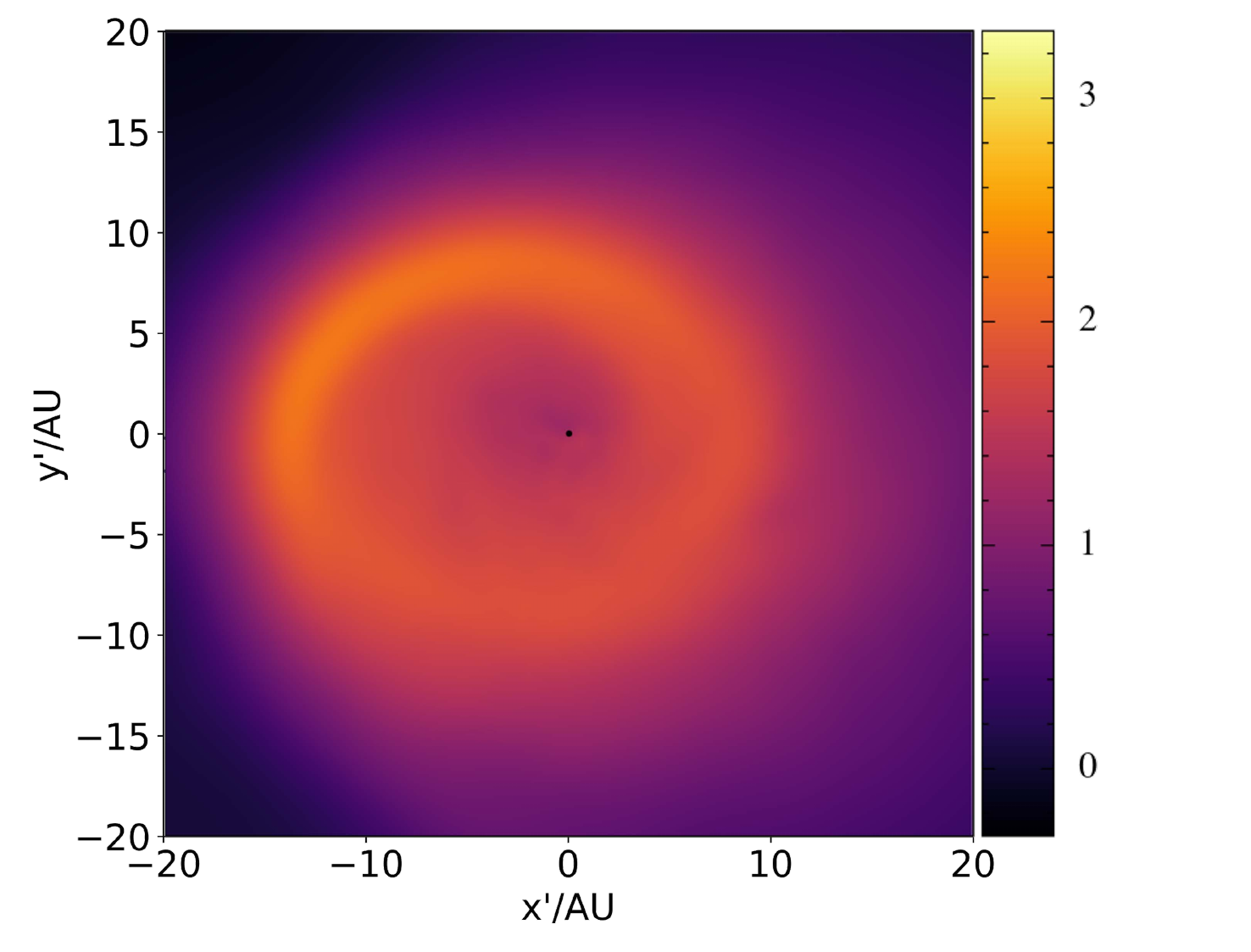}
\includegraphics[width=0.3\textwidth]{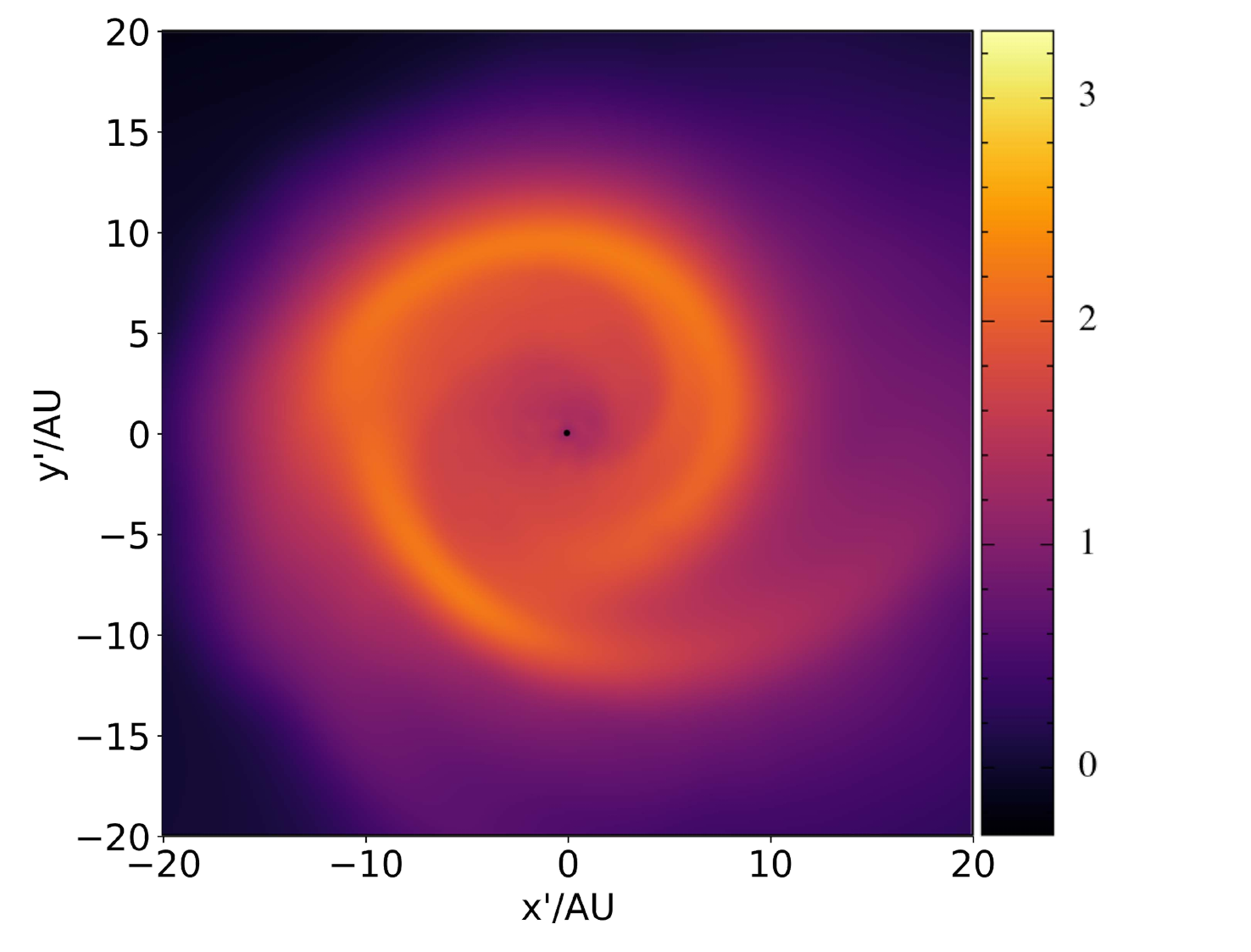}
\includegraphics[width=0.3\textwidth]{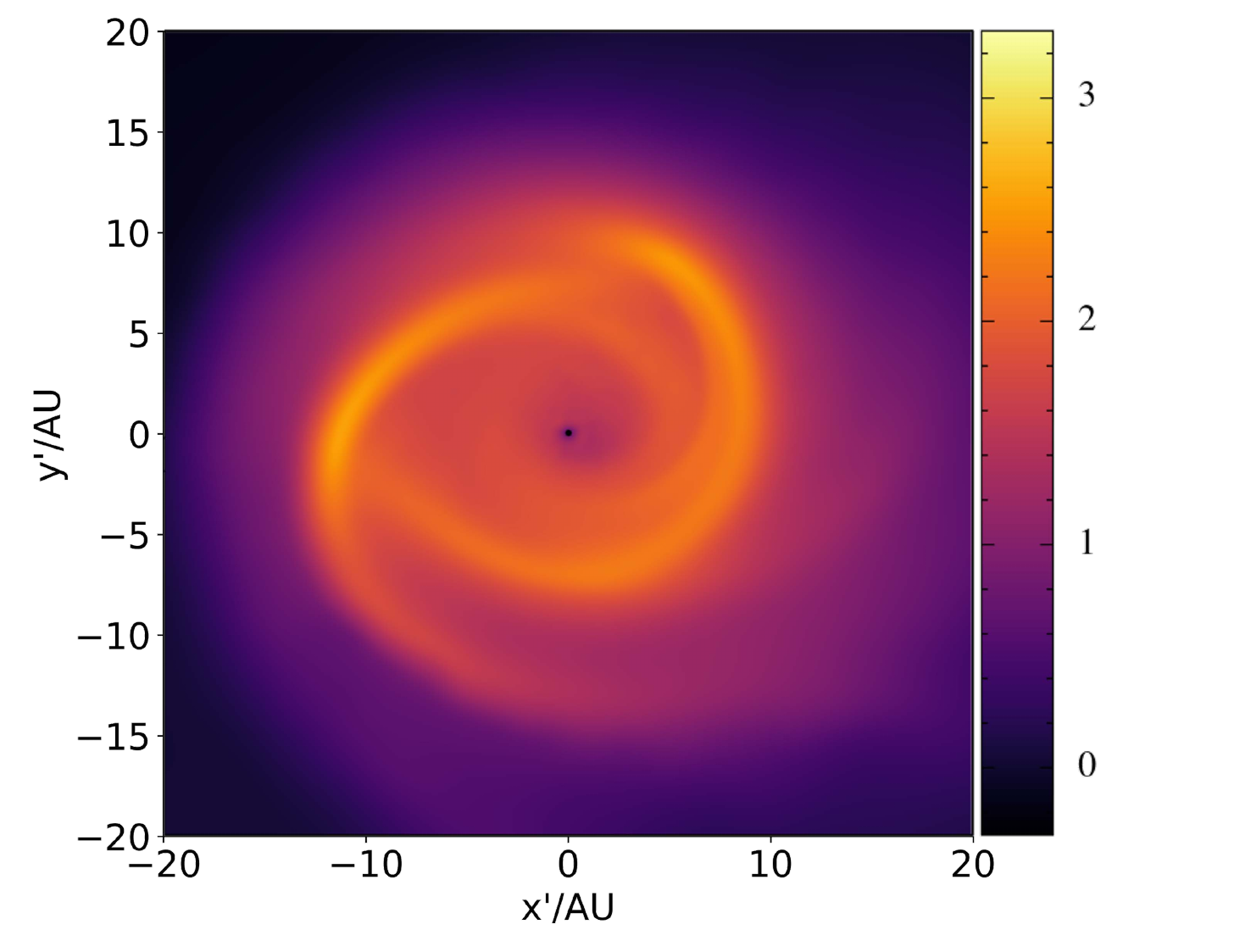}
\includegraphics[width=0.3\textwidth]{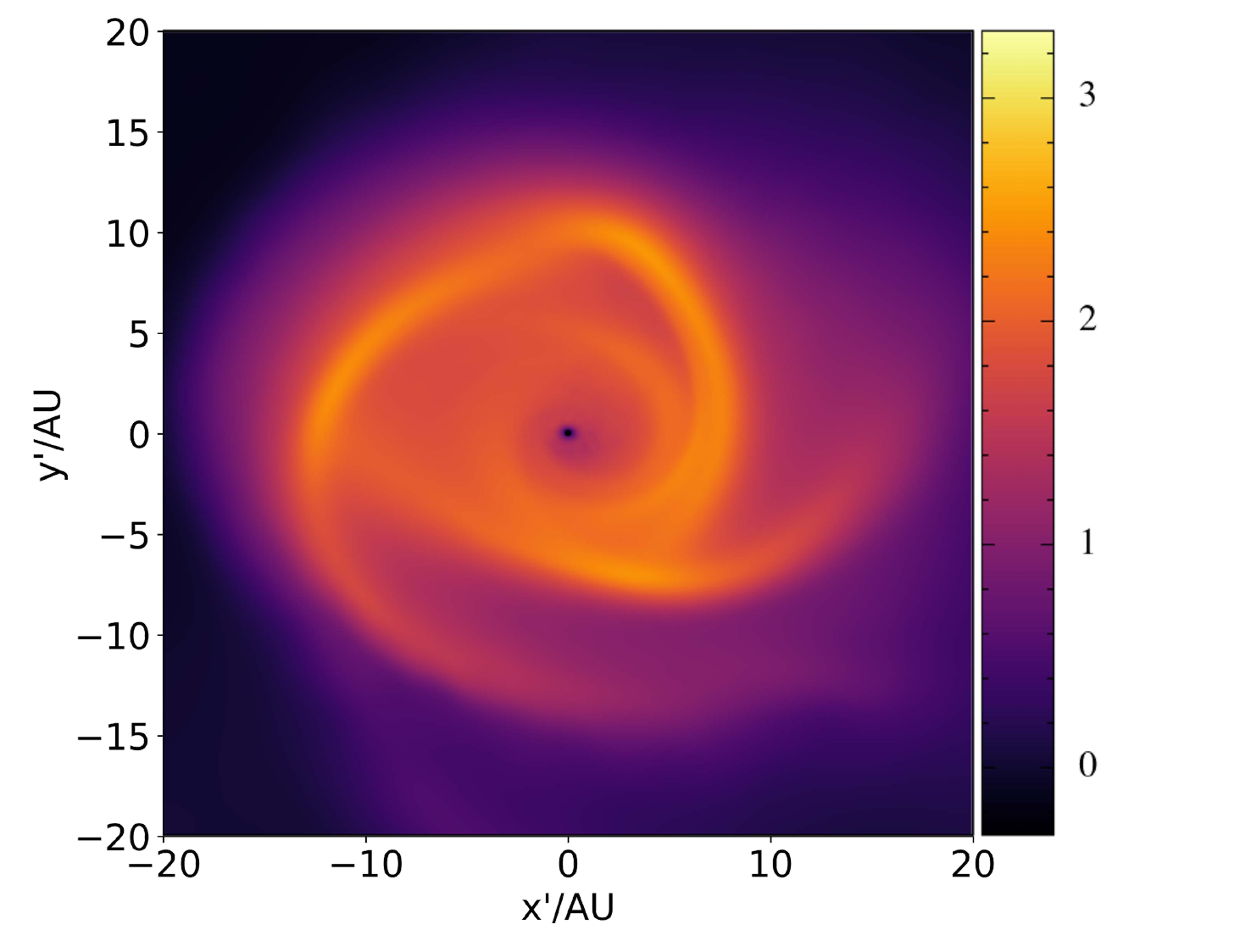}
\includegraphics[width=0.3\textwidth]{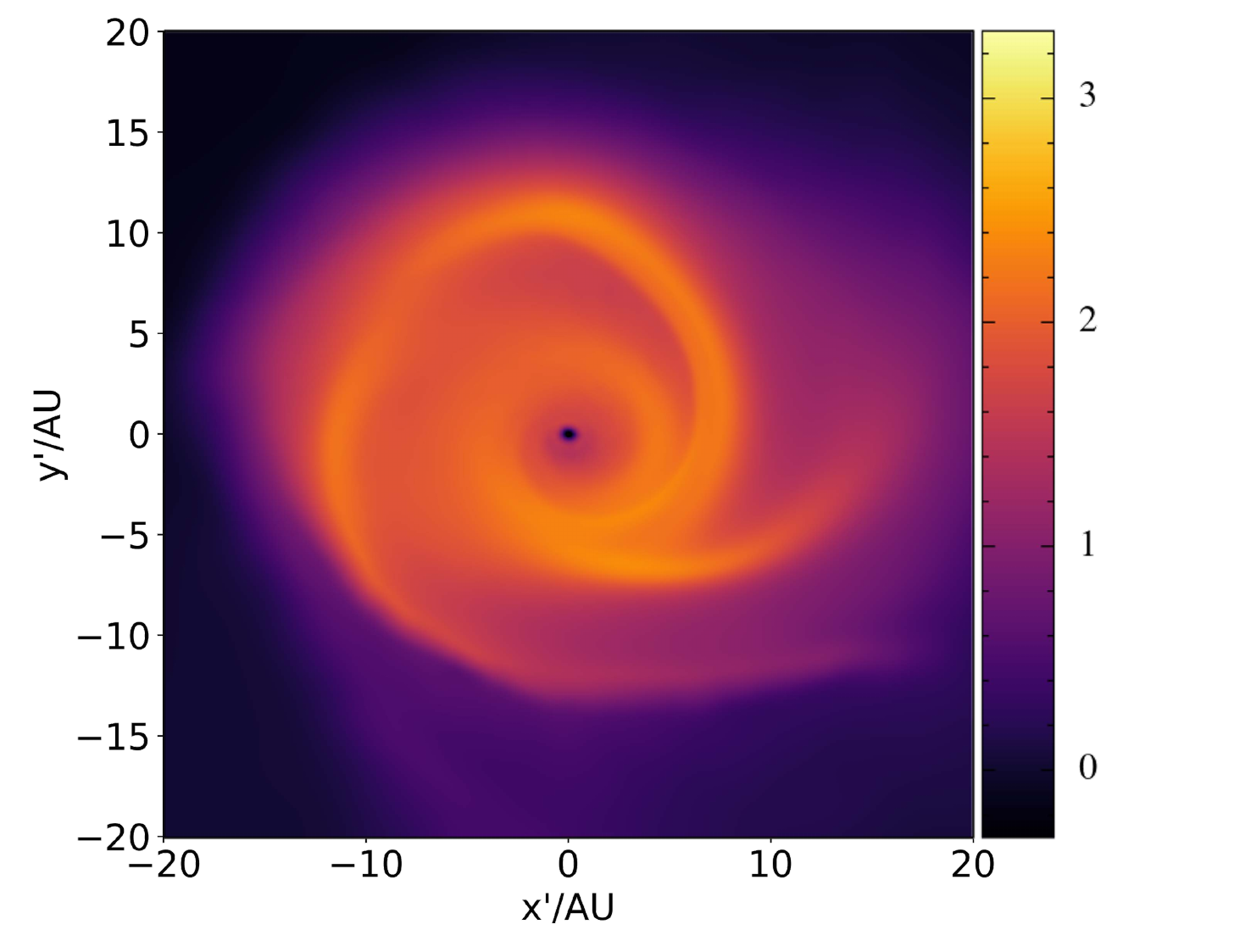}
\caption{Midplane density structure of the disk at $t = 1000\,\mathrm{yr}$ for simulations with increasingly higher numerical resolution. From left to right and continuing on the second row, the total number of particles is $1$M, $2$M, $5$M, $10$M, and $20$M.
\label{fig:resolution1}}
\end{figure}

\begin{figure}
\centering
\includegraphics[width=0.3\textwidth]{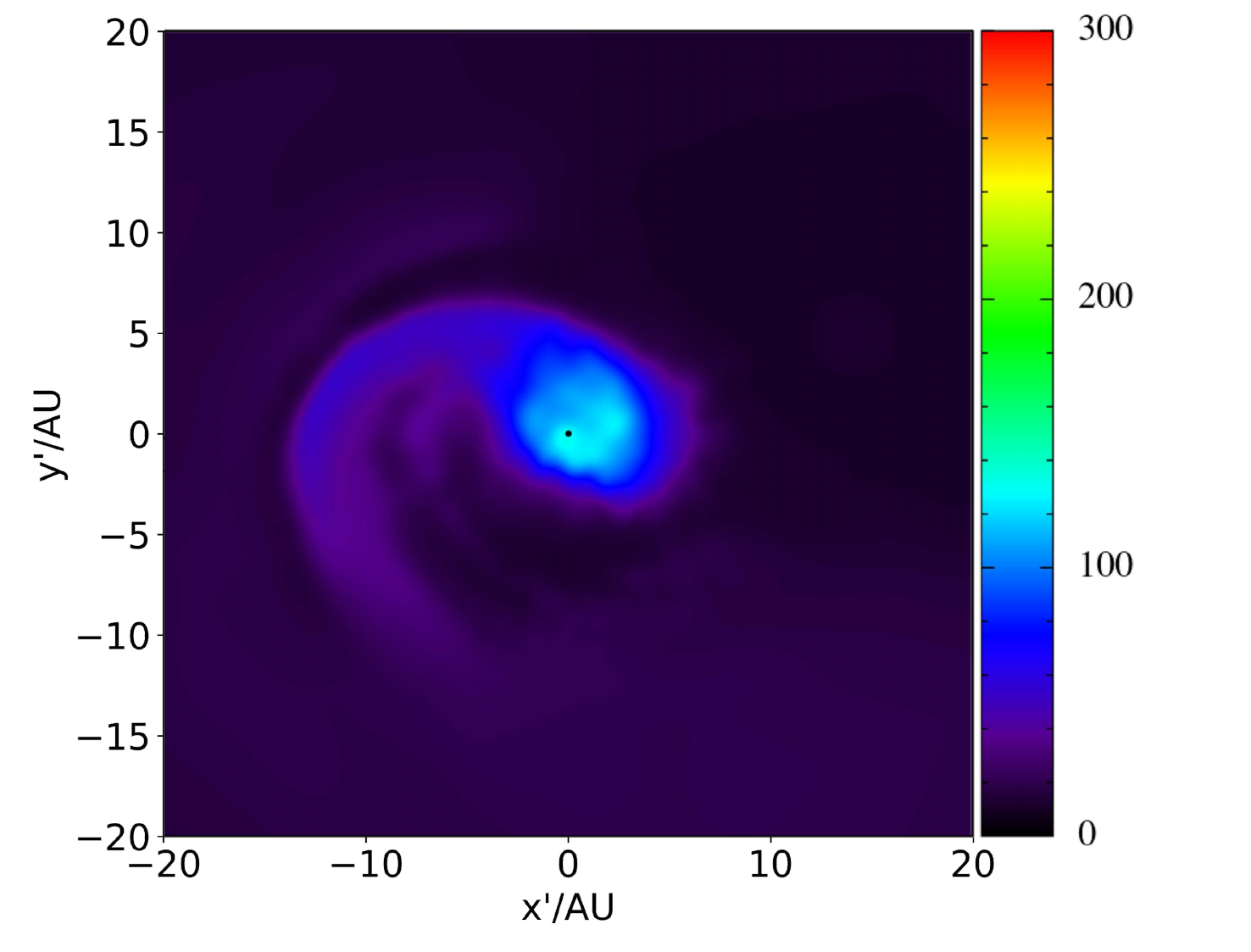}
\includegraphics[width=0.3\textwidth]{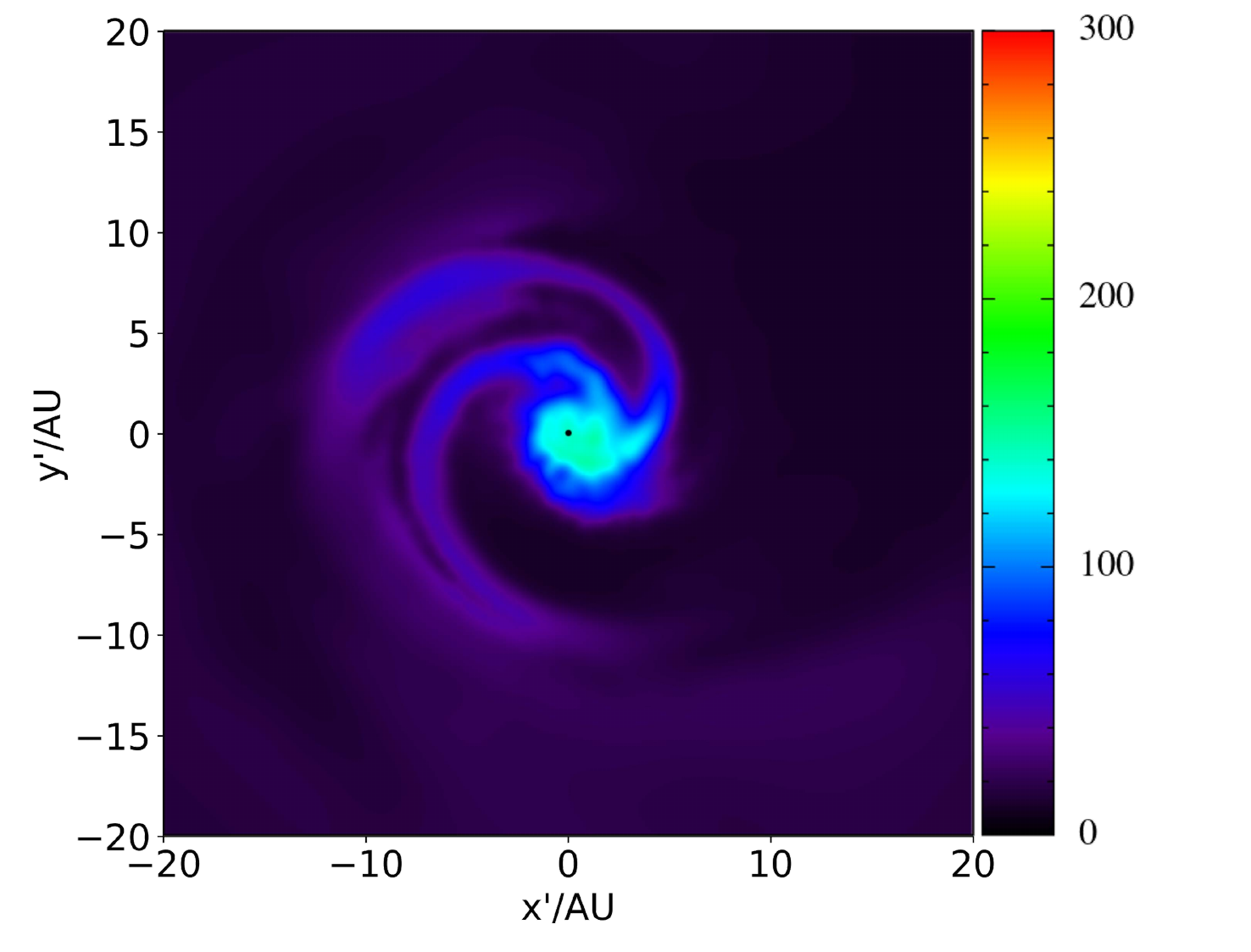}
\includegraphics[width=0.3\textwidth]{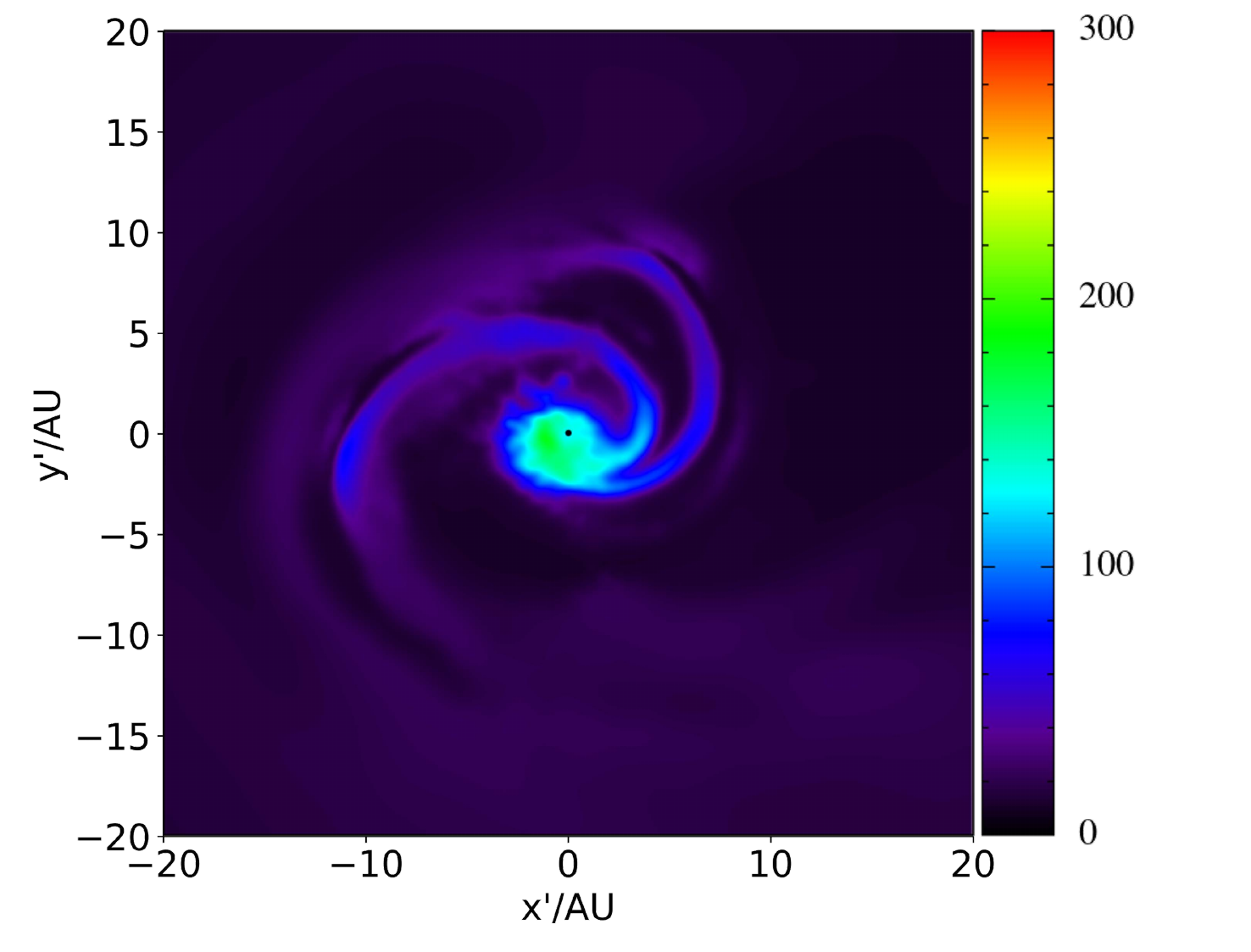}
\includegraphics[width=0.3\textwidth]{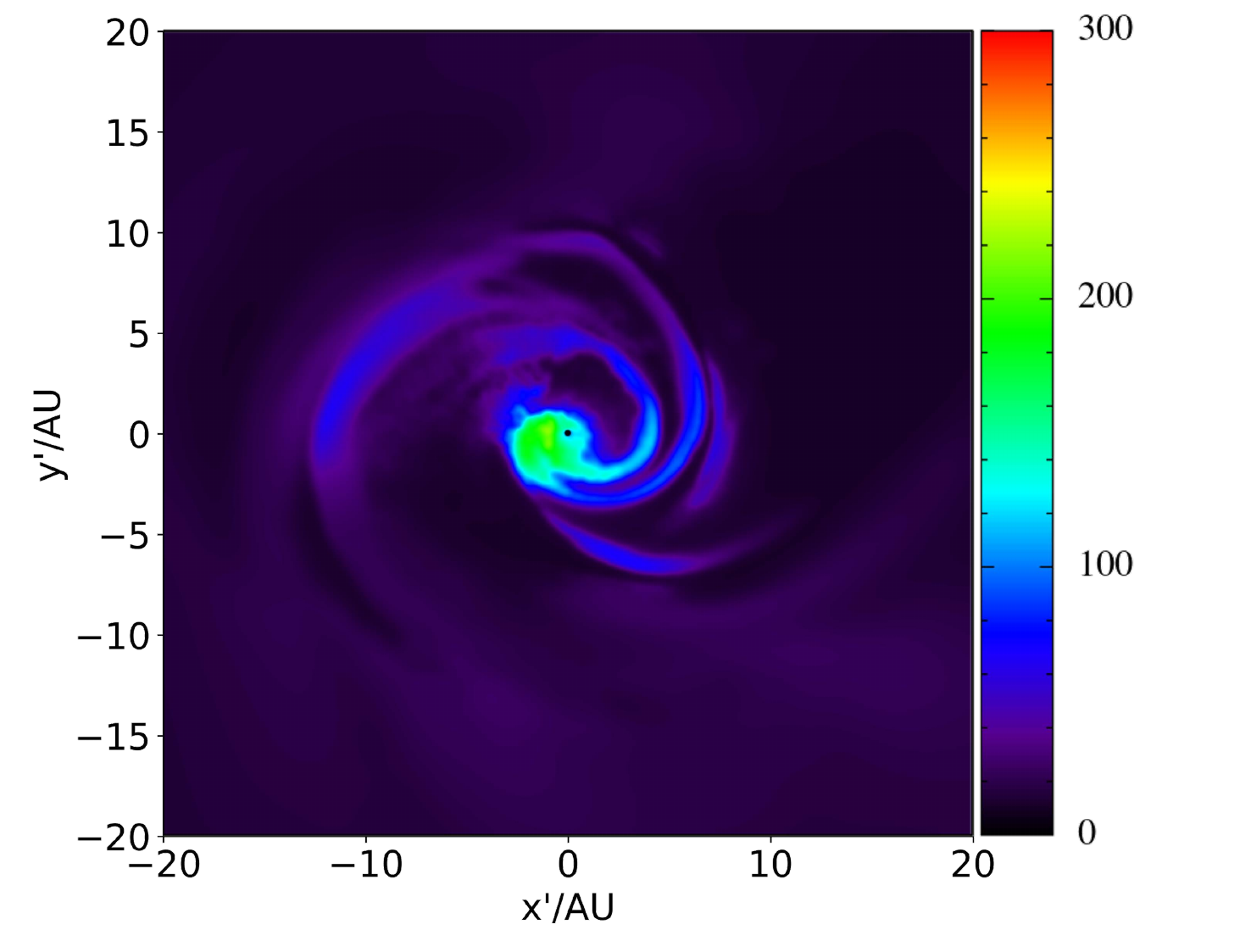}
\includegraphics[width=0.3\textwidth]{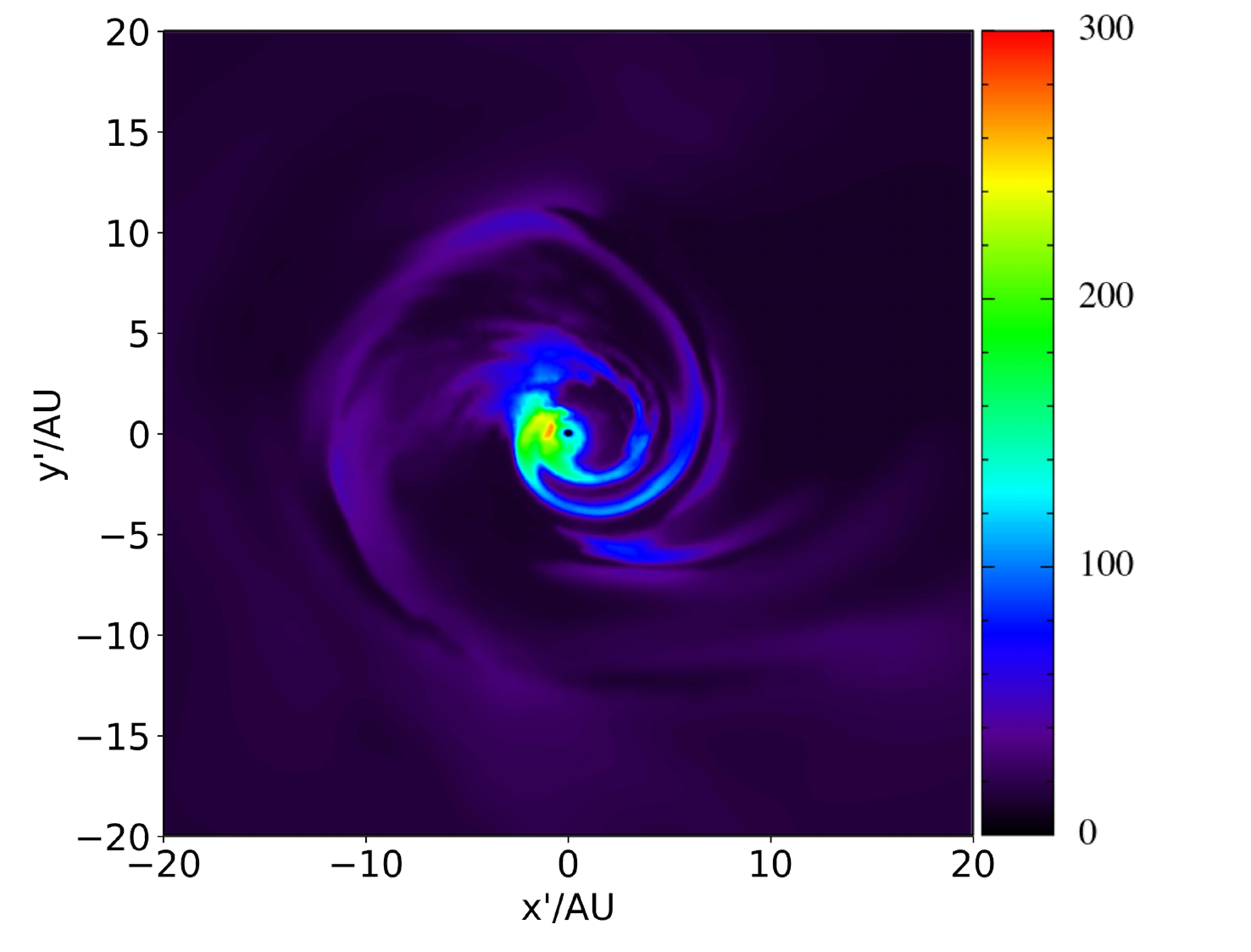}
\caption{Midplane temperature structure of the disk at $t = 1000\,\mathrm{yr}$ for simulations with increasingly higher numerical resolution. From left to right and continuing on the second row, the total number of particles is $1$M, $2$M, $5$M, $10$M, and $20$M.
\label{fig:resolution2}}
\end{figure}

To assess the effect of numerical resolution on the thermal evolution of the disk, we performed a resolution study using particle numbers ranging from 1M to 20M. We evaluate the convergence of the disk evolution by examining the disk midplane structure, which directly reflects the physical origin of the heating. Specifically, we zoom in on the disk midplane at $t = 1000\,\mathrm{yr}$ (corresponding to the first panel of Fig.~\ref{fig:general}) and compare the disk structure across different resolutions. This epoch is chosen because it represents an early stage of the simulation, when a well-defined disk structure has already formed, while avoiding the amplification of small perturbations over long integration times in this highly nonlinear system.

We find that, as the resolution increases, the spiral density wave structure in the disk midplane is consistently reproduced, as shown in Fig.~\ref{fig:resolution1} and Fig.~\ref{fig:resolution2}. The locations, geometry, and strength of the spiral arms show good agreement among the high-resolution runs, indicating that the large-scale gravitationally driven spiral structure has converged. Since shock heating in the disk is associated with these spiral density waves, the resulting shock-induced heating is therefore also consistent across resolutions. 

At resolutions of $\gtrsim 10$M particles, no qualitative differences are observed in the midplane spiral structure, and the associated heating patterns remain nearly identical. Increasing the resolution further to 20M particles does not introduce new features or systematically stronger shocks, suggesting that the relevant heating mechanisms are already well resolved at 10M. While minor differences appear in the low density region (poor resolution in our Lagrangian code) around the star (sink particle), which is not a worry as the accretion proceeds  the spiral extends close to the star (see Fig. \ref{fig:temp_wave}). Given the substantial computational cost of simulations at even higher resolutions, we adopt 10M particles as a practical compromise between accuracy and efficiency.

\bibliography{paper}{}
\bibliographystyle{aasjournal}

\end{document}